\documentclass[useAMS,usenatbib]{mn2e}
\usepackage{times}
\usepackage{graphicx}
\usepackage{txfonts}
\usepackage{subfigure}
\usepackage{color}
\pdfoutput=1

\newcommand{\dyn}{\rm\thinspace dyn}

\graphicspath{./{figures/}}
\DeclareGraphicsExtensions{.pdf}

\title[Dynamics and Metallicity of Far-Infrared Selected Galaxies]{Dynamics and Metallicity of Far-Infrared Selected Galaxies}
 
\author[R.J. Williams et al.]{R.J. Williams,$^{1,2}$ R. Maiolino,$^{1,2}$ P. Santini,$^{3}$ A. Marconi,$^{4}$ G. Cresci,$^{4}$ F. Mannucci,$^{4}$
 \newauthor D.Lutz$^{5}$ \\ \\
$^1$ Cavendish Laboratory, University of Cambridge, 19 J.J. Thomson Ave., Cambridge, UK\\
$^2$ Kavli Institute for Cosmology, University of Cambridge, Madingley Road, Cambridge, UK\\
$^3$ INAF - Osservatorio Astronomico di Roma, Via Frascati 33, I–00040, Monte Porzio Catone, Italy\\
$^4$ INAF - Obsservatorio Astrofisico di Arcetri, Largo E. Fermi 5, 50125 Firenze, Italy\\
$^5$ Max-Planck-Institut f$\ddot{u}$r extraterrestrische Physik, Postfach 1312, 85741 Garching, Germany}

\begin{document}
	\date{Accepted 2014 July 14. Received 2014 July 4}
	\pagerange{\pageref{firstpage}--\pageref{lastpage}} \pubyear{2014}
	\maketitle
	\label{firstpage}

\begin{abstract}
We present near-infrared integral field spectroscopy of ten \textit{Herschel} selected galaxies at z$\sim$1.5. From detailed mapping of the H$\alpha$ and [NII] emission lines we trace the dynamics, star formation rates, metallicities and also investigate gas fractions for these galaxies. For a few galaxies the distribution of star formation as traced by H$\alpha$ only traces a small fraction of the stellar disc, which could be tracing recent minor merging events. The rest-frame ultraviolet (UV) continuum emission often has a distribution completely different from H$\alpha$, which warns about the use of UV-SED based star formation tracers in these systems. Our analysis of galaxy dynamics shows that minor dynamical disruptions (e.g. minor merging) are generally not enough to cause a deviation from the established ``Main Sequence" relation. Most galaxies are found to follow the fundamental metallicity relation (FMR), although with large scatter. One galaxy, (a small satellite galaxy of a massive companion) is found to deviate strongly from the FMR. This deviation is in nice agreement with the correlation recently discovered in local galaxies between gas metallicity and environment, which has been ascribed to enriched inter-galactic medium (IGM) in dense environments, and therefore suggests that here the IGM was already significantly enriched by z$\sim$1.5. 
\end{abstract}

\begin{keywords}
	metallicity -- star formation rate -- dynamics.
\end{keywords}


\section{Introduction}
\label{sect:intro}

The \textit{Herschel} Space Observatory \citep{pilbratt10} has enabled major steps forward in our understanding of galaxy evolution throughout the cosmic epoch. In particular, the various extensive extragalactic surveys have allowed astronomers to resolve the bulk of the cosmic far-infrared background \cite[hence detecting the bulk of the star formation that has occurred in the Universe, e.g.][]{berta10, berta11,magnelli13} and to trace in detail the evolution of star formation of different classes of galaxies across the cosmic epoch. Within this context an important result has been the finding that the bulk of star forming galaxies at high redshift (z$\sim$2) is located on the so-called ``Main Sequence'' \citep{noeske07,rodighiero_oct11}. The latter is a relation between the star formation rate (SFR) and stellar mass. This relation was initially identified in the local Universe \cite[e.g.][]{brinchann04, salim07, peng_sep10} and its small scatter ($\sim$0.3 dex), over several orders of magnitude, suggests that galaxies along this sequence are evolving in a secular, quiescent mode. Instead, starbursting galaxies, which are forming stars violently likely as a consequence of merging/interactions, are located above such Main Sequence \citep{stott13b}. Such relation is also in place at high redshift \cite[e.g.][]{elbaz_jun07, daddi_nov07, daddi09}, but offset relative to the local relation, with galaxies at a given stellar mass having higher star formation rates. This evolution of the Main Sequence has been ascribed to the larger amount of gas in high-z galaxies, which implies higher star formation. The \textit{Herschel} surveys have revealed that the fraction of ``starburst galaxies'', i.e. those deviating from the Main Sequence, is small even at z$\sim$2 \citep[e.g.][]{rodighiero_oct11, elbaz_11}, hence indicating that the bulk of galaxies are forming stars through secular, quiescent processes even at the epoch when the cosmic star formation reaches its maximum.

Near-infrared (near-IR) integral field spectroscopic observations (probing the main optical nebular lines at high redshift) have also been successful in providing independent evidence for secularly evolving galaxies in place at  z$\sim$1-3, although with vigorous star formation. Indeed, near-IR integral field unit (IFU) spectroscopic observations of star forming galaxies at z$\sim$1-3 have revealed that a significant fraction of them are characterized by regular rotation curves typical of disc galaxies (although dynamically ``hot'', i.e. with high $\sigma _V/V$ ratio), without evidence of major disturbances associated with mergers \cite[e.g.][]{forster_jul06,forster_schreiber09, stark08,cresci_may09, law09, gnerucci_apr11, genzel13,newman_apr13}. Furthermore, the investigation of the metal content in z$\sim$1-3 galaxies, (enabled by the mapping of metallicity diagnostics through near-IR IFU spectroscopy) has revealed regular metallicity gradients suggestive of smooth evolutionary processes in these systems \citep{cresci_oct10,jones10, yuan11,jones12, swinbank12}.

In this paper we combine the two techniques discussed above to shed light on the physics of star forming galaxies at z$\sim$1.5. We have selected a small sample of FIR sources in GOODS-S, detected within the Herschel-PACS GTO survey ``PEP'' \citep{lutz_aug11}, with redshifts carefully selected so that the main optical nebular lines avoid strong airglow emission lines and atmospheric absorption. We have observed these galaxies with the Spectrograph for INtegral Field Observations in the Near Infrared (SINFONI), at the ESO-VLT, targeting the main optical nebular lines redshifted into the near-IR. The SINFONI spectra have enabled us to investigate various key properties of these galaxies, in particular: the galaxy dynamics and how star formation is distributed, which has been compared with the distribution of the old stellar population (traced by the rest-frame optical light) and with the (rest-frame) UV tracers of star formation. We also exploit the nebular line ratios to investigate the metallicity and metallicity gradients in these galaxies. These observational results are used to shed light on the evolutionary processes of the bulk of the galaxy population at these redshifts. Throughout this paper we adopt the \cite{chabrier_jul03} initial mass function (IMF) and a cosmological model with ($\Omega_{\Lambda}$, $\Omega_{m}$, $h$) = (0.685, 0.315, 0.673) \citep{planck_res13}.


\section{Sample selection, observations and data reduction}
\label{sect:sample}

As previously mentioned in the Section~\ref{sect:intro}, the targets of our observations have been extracted from the PEP Herschel GTO survey. The PEP survey provides deep observations taken with the PACS instrument \citep{pacs} at wavelengths $\lambda = 70, 100~\&\ 160~\rmn{\mu m}$ in several extragalactic fields. Galaxies for our IFU spectroscopic follow-up have been extracted from the GOODS-S field (which is the deepest of the PEP fields). We selected 12 galaxies which have at least 1 detection in 1 of the 2 PACS bands at 110~$\rm{\mu}$m and/or 160~$\rm{\mu}$m, and with a spectroscopic identification in the redshift range 1.2$<$z$<$1.7, which is around the peak of cosmic star formation and which allows the H$\rm{\alpha}$+[NII] lines to be observed in the H-band and H$\rm{\beta}$+[OIII] lines to be observed in the J-band. The specific redshift was accurately selected so that none of these emission lines are affected by strong OH airglow lines.

We have accurately discarded AGNs both by using the ultra-deep Chandra 4Ms observations (none of our sources has hard X-ray emission in excess of what is expected from their star formation rate) and the ultra-deep MIPS observations \cite[none of our sources has excess 24~$\mu$m emission which would suggest the presence of Compton-thick AGN, according to the criteria of ][]{fiore08}.

As we will show later on, all galaxies but one are located on the Main Sequence at z$\sim$1.5. The list of sources observed is given in Table~\ref{tab:sources}. Note that the total number of observed sources is 13, since 1 field serendipitously includes an additional galaxy, as discussed in more detail in the following.

We also mention that all of these galaxies benefit from extensive photometric and multi-band Hubble Space Telescope (HST) imaging information, which has been used, for instance to infer the stellar masses and extinction towards the stellar continuum.

Observations were performed at the ESO-VLT with the near-IR integral field spectrometer SINFONI in 2011 and 2012. We used both the J and H
gratings, delivering a spectral resolution of R=2000 and R=3000, respectively. We adopted the coarse pixel scale, giving pixels with angular size of $0.250'' \times 0.125''$ on the sky. Integration times were typically between 1.5 and 3 hours on source for both observations in H \& J bands (depending on scheduling constraints). The seeing during the observations was generally around or better than 0.6$''$.

Data were reduced using the ESO-SINFONI pipeline (version 2.3.9). The pipeline subtracts the sky from the temporally contiguous frames, flat-fields the images, spectrally calibrates each individual slice and then reconstructs the cube. Within the pipeline the pixels are re-sampled to a symmetric angular size of $0.125\arcsec \times 0.125\arcsec$. The atmospheric absorption and instrumental response were taken into account and corrected for by using a telluric standard star.

\begin{table}
	\caption{Our targeted galaxies in the GOODS-S field extracted from the PEP Herschel GTO survey listing their position (R.A. and Dec.), optical redshift and the redshift inferred from our H$\rm{\alpha}$ detection. The last three were only marginally detected in H$\alpha$ therefore we cannot reliably constrain the spectroscopic redshift.} 
	\label{tab:sources} 
	\begin{tabular}{@{}lcccc@{}}  
	\hline
	Galaxy  & R.A. $^\circ$ & Dec. $^\circ$ & z (optical) & z (H$\rm{\alpha}$)\\
	&  (J2000)&  (J2000)& &\\
	\hline
	CDFS13844 & 53.0494 & -27.7370 & 1.325 & 1.327\\
	CDFS13784 & 53.1408 & -27.7382 & 1.299 & 1.298\\
	CDFS6758 & 53.0727 & -27.8341 & 1.616 & 1.616\\
	CDFS14224 & 53.1326 & -27.7323 & 1.550 & 1.548\\
	CDFS16485 & 53.1297 & -27.7014 & 1.550 & 1.549\\
	CDFS2780 & 53.2106 & -27.8883 & 1.611 & 1.615\\
	CDFS11583 & 53.1655 & -27.7697 & 1.552 & 1.553\\
	CDFS15753* & 53.0923 & -27.7121 & 1.610 & 1.612\\
	CDFS15764 & 53.0917 & -27.7120 & 1.613 & 1.612\\
	CDFS10299 & 53.0759 & -27.7858 & 1.381 & 1.382\\
	CDFS12664 & 53.1409 & -27.7552 & 1.391 & -\\
	CDFS12910 & 53.1572 & -27.7515 & 1.607 & -\\
	CDFS16661 & 53.1336 & -27.6986 & 1.393 & -\\
	\hline
	\end{tabular}\\
	{  * Detected in the same field of view as CDFS15764}
\end{table}


\section{Data analysis}
\label{sect:data}
\begin{figure}
	\includegraphics[trim=0cm 1.2cm 0cm 2cm, clip=true, width=84mm]{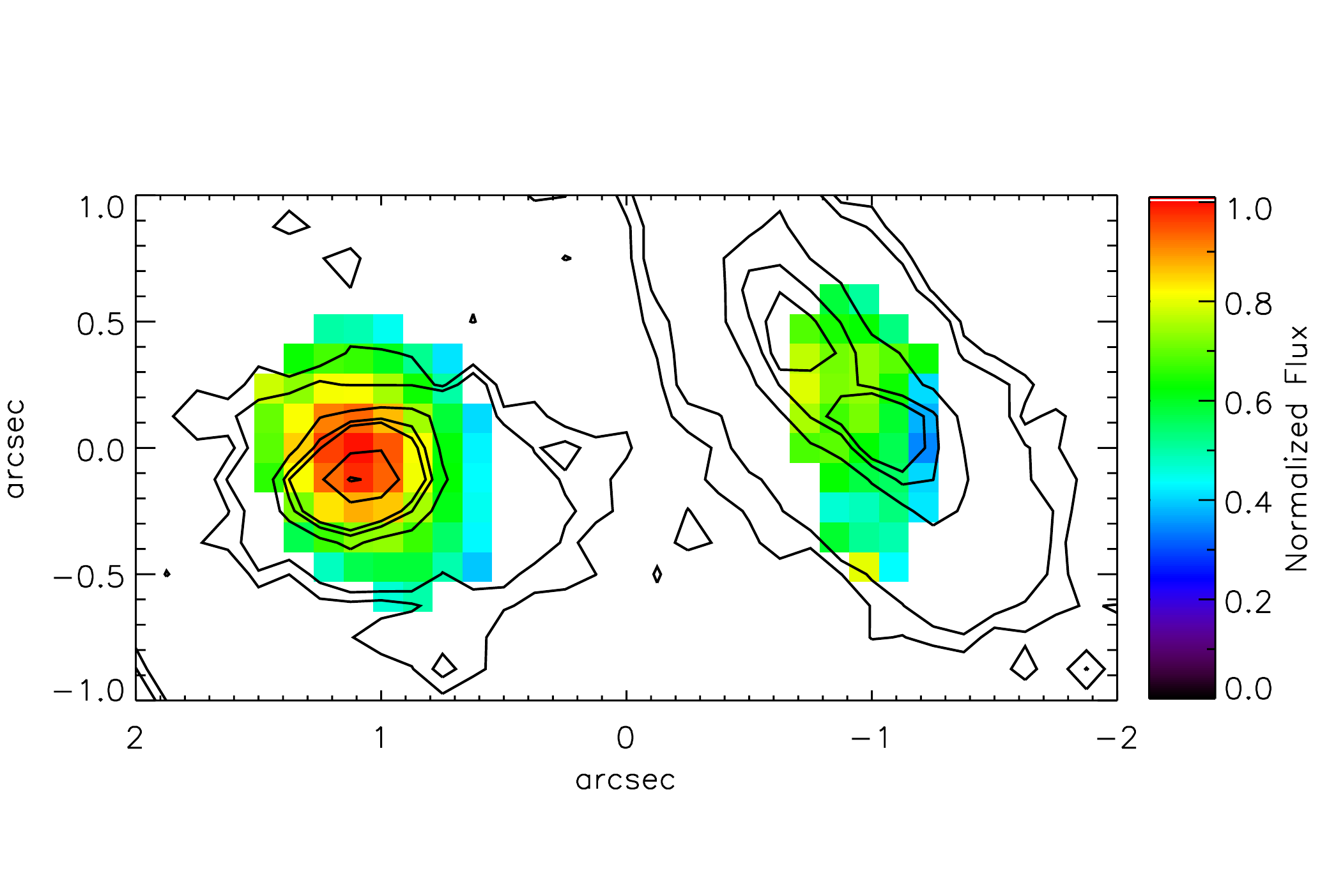}
	\caption{Flux map of the interacting system with the serendipitous galaxy, CDFS15753 (left) which was not selected from the \textit{Herschel} survey but was detected in the same field of view as CDFS15764 (right). Color shows the H$\alpha$ flux (normalized to the peak emission), while contours indicate the emission in the near-IR \textit{HST} band. The central position (0,0) corresponds to R.A. = 53.0907$^\circ$, Dec. = -27.7122$^\circ$. The close proximity ($\sim17$~kpc) suggests that they are strongly interacting and possibly in the process of merging.}
	\label{fig:merger}
\end{figure}

H$\alpha$ and [NII] are clearly detected in nine of the twelve galaxies. In the other three H$\rm{\alpha}$ is only marginally detected and so we can only place upper limits on the fluxes, given in Table~\ref{tab:fluxes}. However they are not useful to perform any spatially resolved analysis or to infer metallicity hence these three sources (CDFS12664, CDFS12910, CDFS16661) will not be discussed further. Within the SINFONI field of view of CDFS15764, the H$\rm{\alpha}$+[NII] of the companion galaxy CDFS15753 is also detected (see Fig.~\ref{fig:merger}) and so included in the sample (the former is actually the most massive galaxy in a small group including CDFS15753).

For most galaxies [OIII] and H$\rm{\beta}$ are not detected and this is the case even when stacking all spectra, indicating very high extinction. The lower limit on the Balmer decrement H$\rm{\alpha}$/H$\rm{\beta}$ (given in Table~\ref{tab:fluxes}) is generally 3.5, indicating a visual extinction $\rm{A_V}$ generally higher than 2.3~magnitudes (assuming a Calzetti extinction curve). Only in the case of CDFS2780 are the [OIII] and H$\rm{\beta}$ lines marginally detected (as will be discussed in Section~\ref{sect:results}).

The H$\alpha$ and [NII] lines in the H-band are fitted at each spatial pixel with three Gaussians, linked to have the same velocity dispersion and same velocity shift. Furthermore the [NII]$\lambda6548$,6584 doublet was forced to have a flux ratio of 1:3, as given by the Einstein coefficients. The continuum was fitted with a linear function with wavelength within the H-band (avoiding spectral regions contaminated by OH sky lines).
 
We have complemented our data with measurements of the stellar mass in these galaxies obtained by exploiting the extensive multi-band photometric data available in this field. These data are also used to infer the extinction towards the stellar continuum. Details on the spectral energy distribution (SED) fitting method to derive these quantities are given in \cite{santini_09}. The stellar masses and continuum reddening inferred for the galaxies in our sample are given in Table~\ref{tab:properties}. It should be noted that the lower limit on the extinction inferred from the Balmer decrement is consistent with the extinction inferred from the continuum, assuming $\rm R_V=A_V/E_{B-V}=4.05$ and assuming a differential line to continuum extinction of the form $\rm A_V(nebular) = A_V(line)/0.44$, as suggested by \cite{calzetti_apr00}, although the latter assumption has been recently revised, as discussed in the next section.

\begin{table*}
	\caption{Emission line properties. The second and third columns give the H$\alpha$ and [NII] fluxes (not corrected for extinction). The forth gives the SFR from the H$\alpha$ emission corrected for dust extinction using the recipe from \citet{wuyts_13}. The last two columns give the Balmer decrement H$\alpha$/H$\beta$ (generally lower limits) and the implied extinction A$_V$ applying the Calzetti extinction law and method (i.e. by applying a differential extinction factor of $0.44$ between continuum and nebular lines.} 
	\label{tab:fluxes} 
	\begin{tabular}{@{}lcccccc@{}}  
	\hline
	Galaxy  & Flux(H$\alpha$) & Flux([NII]) & logSFR(H$\alpha$) & H$\alpha$/H$\beta$ & A$_{\rm v}$\\
	&  $\times 10^{-16} ~\rmn{erg~s^{-1}~cm^{-2}}$ &  $\times 10^{-16} ~\rmn{erg~s^{-1}~cm^{-2}}$ & $\rmn{M_{\odot}~yr^{-1}}$ & &mag\\
	\hline
	CDFS13844 & 1.21 $\pm$ 0.10 & 0.35 $\pm$ 0.08 & 1.96 $\pm$ 0.18 & $>3.6$ & $>1.02$\\
	CDFS13784 & 0.54 $\pm$ 0.06 & 0.09 $\pm$ 0.05 & 1.58$\pm$ 0.18 & $>3.6$ & $>1.02$\\
	CDFS6758 & 0.62 $\pm$ 0.07 & 0.24 $\pm$ 0.07 & 2.27 $\pm$ 0.21 & $>4.1$ & $>1.51$\\
	CDFS14224 & 0.61 $\pm$ 0.06 & 0.25 $\pm$ 0.05 & 0.94 $\pm$ 0.13 & $>5.1$ & $>2.34$\\
	CDFS16485 & 0.29 $\pm$ 0.03 & 0.12 $\pm$ 0.03 & 1.61 $\pm$ 0.19 & $>4.7$ & $>2.02$\\
	CDFS2780 & 1.08 $\pm$ 0.07 & 0.24 $\pm$ 0.05 & 1.89 $\pm$ 0.16 & $3.5$ & $0.86$\\
	CDFS11583 & 0.16 $\pm$ 0.02 &  0.04 $\pm$ 0.02 & 1.90 $\pm$ 0.21 & $>3.7$ & $>1.13$\\
	CDFS15753 & 0.19 $\pm$ 0.02 & 0.03 $\pm$ 0.01 & 0.88 $\pm$ 0.12 & $>1.0$ & $>0$\\
	CDFS15764 & 0.10 $\pm$ 0.02 & 0.07 $\pm$ 0.02 & 1.21 $\pm$ 0.19 & $>0.7$ & $>0$\\
	CDFS10299 & 0.35 $\pm$ 0.05 & 0.08 $\pm$ 0.05 & 1.95 $\pm$ 0.21 & $>0.6$ & $>0$\\
	CDFS12664 & $<0.006$ & $<0.0001$ & $<0.60$ & - & - \\
	CDFS12910 & $<0.085$ & $<0.001$ & $<0.62$ & - & - \\
	CDFS16661 & $<0.039$ & $<0.0003$ & $<0.63$ & - & -  \\
	\hline
	\end{tabular}\\
	
\end{table*}

\section{Results}
\label{sect:results}
\subsection{SFR distribution and main sequence}

For these targets we can compare the SFR inferred from the H$\rm{\alpha}$ with the SFR inferred from the FIR emission. This test has also been performed by other authors \cite[e.g.][]{kewley_02, sanchez12}, but either at lower redshifts and/or by using single slit spectroscopy, hence subject to strong aperture effects and slit losses on the H$\rm{\alpha}$ luminosity measurement. Our integral field spectra allows us to recover all of the H$\rm{\alpha}$ flux in these sources.

\begin{figure}
	\includegraphics[width=80mm]{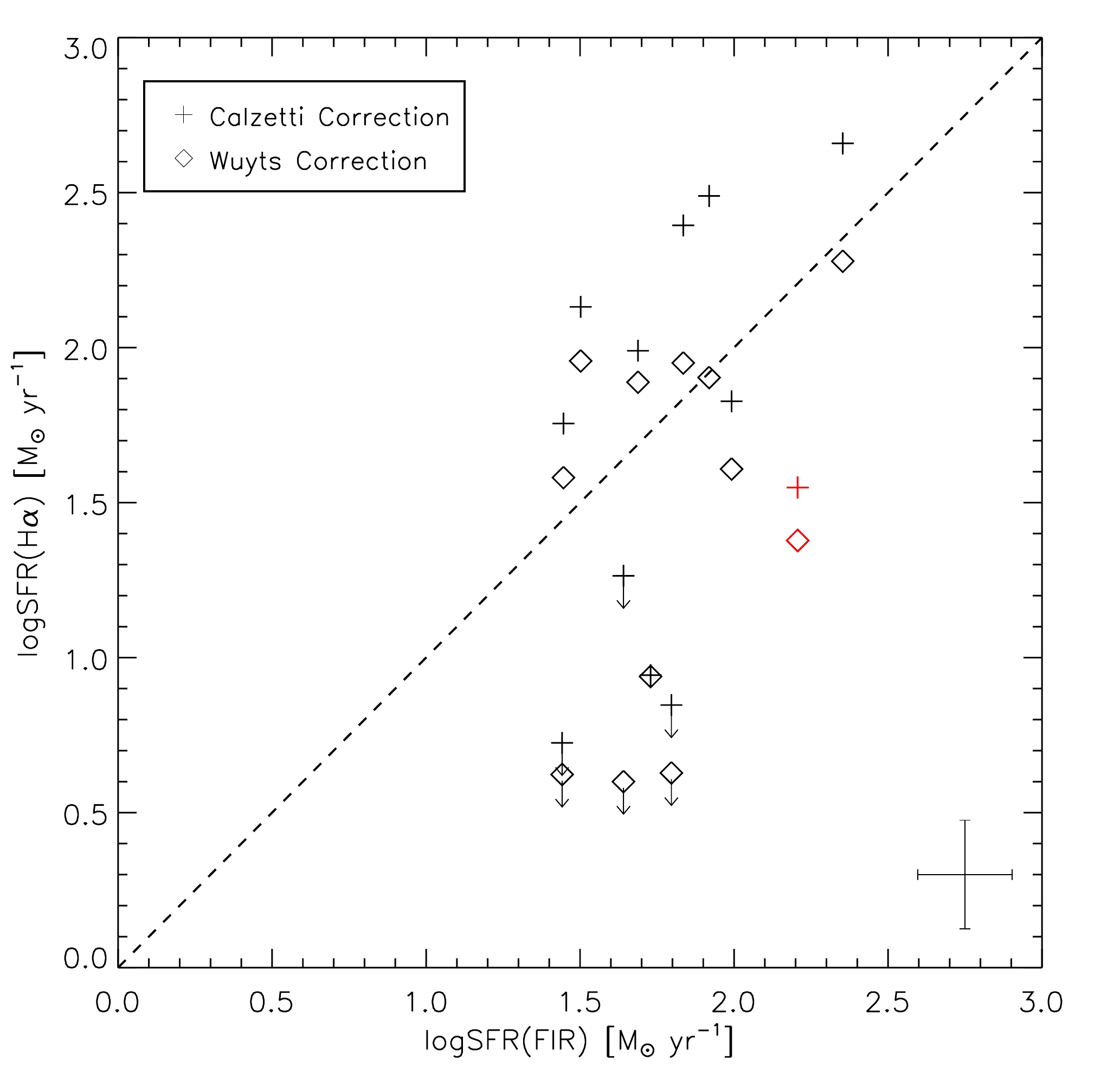} 
	\caption{SFR inferred from H$\alpha$ versus SFR inferred from the FIR luminosity. Here we compare two different corrections for dust extinction, i.e. using the \citet{calzetti_apr00} correction (plus symbols) and that from \citet{wuyts_13} (diamond symbols). The dotted line represents the expected 1:1 relation between the two quantities and suggests that the recipe given by \citet{wuyts_13} is appropriate here. Included are the three galaxies where H$\alpha$ is only marginally detected, giving upper limits on the SFR(H$\alpha$). The average uncertainty in the extinction corrected SFR(H$\alpha$) and SFR(FIR) is shown in the bottom right corner. In red we give just one point for the joint system (CDFS15764 and CDFS15753) by summing the SFR(FIR) and corrected SFR(H$\alpha$) to give a combined total as in this case the FIR emission is unresolved. }
	\label{fig:fir}
	
\end{figure}
The SFR from the FIR emission is obtained by fitting the monochromatic 100~$\rm{\mu}$m and 160~$\rm{\mu}$m fluxes to the \cite{dale_sep02} templates \cite[see Appendix A of ][ for details on how to associate a SFR estimate to these templates]{santini_09}. This was done for each individual galaxy except for the joint system, CDFS15753 and CDFS15764, as the FIR fitting was unresolved across these two. To attempt to separate them out, the total SFR(FIR) for the system was distributed according to the level of extinction-corrected SFR(H$\rm{\alpha}$) for each of the two galaxy. Here the assumption is made that, although there might be an offset between the two SFR tracers, the two quantities should correlate linearly (once extinction corrections are applied).

We infer the SFR(H$\rm{\alpha}$) from the observed H$\rm{\alpha}$ luminosity and correct for dust extinction using the reddening towards the continuum inferred from the global SED fitting and then correcting for the differential extinction between continuum and nebular lines both with the recipe given by \cite{calzetti_apr00} and with the method given in \cite{wuyts_13}. Both recipes account for extra extinction towards HII regions; the former assumes this to be proportional to the attenuation by diffuse dust in the galaxy, while the latter finds that a more complex
(luminosity-dependent) relation provides a better fit with observations. Then we apply the $\rm{H\rm{\alpha}}$ luminosity to SFR conversion factor given in \cite{kennicutt_98}.

Fig.~\ref{fig:fir} shows the comparison between SFR(FIR) and SFR(H$\rm{\alpha}$). Plus symbols indicate the SFR inferred from H$\rm{\alpha}$ and extinction corrected using the \cite{calzetti_apr00}, where we use E(B-V)/0.44 which accounts for the larger obscuration of the star forming regions emitting H$\alpha$. The diamond symbols indicate the SFR inferred when applying the reddening correction given by \cite{wuyts_13}. The relation between SFR(FIR) and SFR(H$\rm{\alpha}$)$_{\rm{corrected}}$ is close to 1:1 for most targets when using the \cite{wuyts_13} recipe, suggesting that this is the most appropriate extinction correction. However, there are six galaxies for which the SFR(H$\alpha$) is still significantly lower (by even more than an order of magnitude) than SFR(FIR), implying that in these galaxies the H$\alpha$ optical emission associated with the bulk of star formation is completely absorbed by dust (or not even produced as large amounts of dust may directly absorb the ionizing photons) and is therefore undetectable. In Table~\ref{tab:fluxes} we show the H$\alpha$ and [NII] fluxes obtained for each galaxy and also give the inferred SFR from the \cite{kennicutt_98} relation using the \cite{wuyts_13} extinction corrected fluxes. We also include upper limits for the three galaxies only marginally detected in H$\alpha$.

In the following we use the SFR inferred from the FIR (i.e. from \textit{Herschel}), but we use the H$\rm{\alpha}$ maps to obtain spatially resolved information on the SFR by simply scaling the SFR(H$\rm{\alpha}$) maps to obtain a total SFR consistent with the one measured in the FIR.

\begin{figure}
	\includegraphics[trim=0.9cm 0.2cm 0.2cm 1.5cm, clip=true,width=84mm]{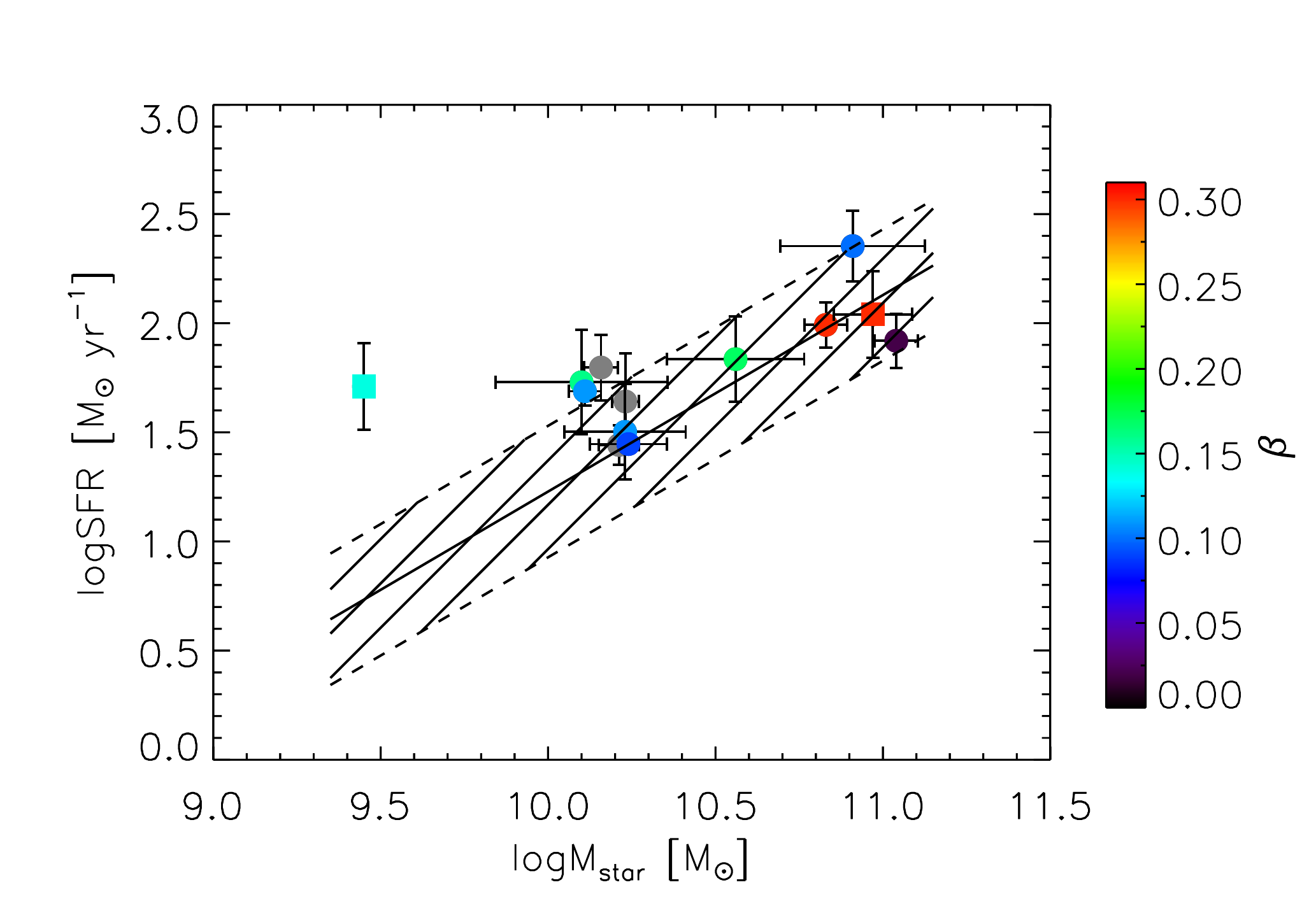}
	\caption{The mass--SFR digram where the solid line and hatched area show the Main Sequence along with the dispersion at $z\sim 1.5$. We color code each galaxy with respect to the galaxy dynamics represented by the parameter $\beta$ (Eq.~\ref{eq:beta}). The three galaxies where H$\alpha$ is only marginally detected are also shown using the SFR(FIR) and stellar mass from the SED fitting, but are color-coded in grey as we are not able to map them through the H$\alpha$ emission. In addition we highlight the location of the two galaxies in the interacting system (CDFS15764 and CDFS15753) with square symbols. }
	\label{fig:MS}
\end{figure}

The SFR inferred from the FIR emission can be used to investigate how these objects are located relative to the Main Sequence on the stellar mass--SFR diagram. Fig.~\ref{fig:MS} shows the stellar mass--SFR diagram, in which the solid line and hatched area show the Main Sequence, along with its dispersion, at z$\sim$1.5, as interpolated from \cite{elbaz_jun07} and \cite{daddi_nov07}. The colored symbols show the location of the galaxies in our \textit{Herschel} sample (the color coding will be discussed later in Section~\ref{sect:dynamics}), most of which are clearly located on the Main Sequence. A notable exception is the interacting system CDFS15764-CDFS15753: the former (one of the most massive galaxies in the sample) appears to lie on the Main Sequence relation; while the latter (which is the least massive galaxy in the sample) could have a SFR about 10 times higher than the Main Sequence, likely as a consequence of a starburst possibly induced by the interaction with the massive companion.

\begin{table*}
	\caption{List of derived properties for the galaxies in our sample (see text for details).}
	\label{tab:properties} 
	\begin{tabular}{@{}lccccccccc@{}}  
	\hline
	Galaxy & $\beta$ & $V/\sigma$ & 12+log(O/H) &  E(B-V) & log($\rm M_{*}/M_{\odot}$) & log($\rm M_{\dyn}/M_{\odot}$) & log(SFR\_FIR) & F$_{\rm{gas}}$ & $\rm M_{\rm{gas}}/M_{\odot}$ \\
	& & & & & & & $\rm M_{\odot}$yr$^{-1}$ & & \\
	\hline
	CDFS13844 & 0.11 & 3.73 & $8.86 \pm 0.08$ & 0.45 & $10.23 \pm 0.18$ & $10.83 \pm 0.16$ & $1.50 \pm 0.22$ & $0.46 \pm 0.04$ & $10.16 \pm 0.28$\\
	CDFS13784 & 0.09 & 4.05 & $8.68 \pm 0.14$ &  0.45 & $10.24 \pm 0.11$ & $10.91 \pm 0.10$ & $1.45 \pm 0.02$ & $0.41 \pm 0.04$ & $10.09 \pm 0.28$\\
	CDFS6758 & 0.10 & 1.70 & $8.99 \pm 0.06$ & 0.65 & $10.91 \pm 0.21$ & $11.09 \pm 0.14$ & $2.35 \pm 0.16$ & $0.38 \pm 0.08$ & $10.69 \pm 0.28$\\
	CDFS14224 & 0.16 & 2.01 & $9.01 \pm 0.08$ & 0.10 & $10.10 \pm 0.25$ & $11.17 \pm 0.20$ & $1.73 \pm 0.24$ & $0.61 \pm 0.06$ & $10.29 \pm 0.28$\\
	CDFS16485 & 0.30 & 1.01 & $8.95 \pm 0.15$ & 0.50 & $10.83 \pm 0.06$ & $11.12 \pm 0.11$ & $1.99 \pm 0.10$ & $0.27 \pm 0.06$ & $10.41 \pm 0.28$\\
	CDFS2780 & 0.11 & 2.17 & $8.79 \pm 0.10$ & 0.35 & $10.11 \pm 0.05$ & $10.57 \pm 0.31$ & $1.68 \pm 0.07$ & $0.56 \pm 0.06$ & $10.22 \pm 0.28$\\
	CDFS11583 & 0.02 & 3.24 & $8.86 \pm 0.06$ & 0.80 & $11.04 \pm 0.06$ & $11.31 \pm 0.11$ & $1.92 \pm 0.12$ & $0.15 \pm 0.04$ & $10.31 \pm 0.28$\\
	CDFS15753 & 0.14 & 0.90 & $8.68 \pm 0.06$ & 0.25 & $9.45 \pm 0.03$ & $10.41 \pm 0.19$ & $1.71 \pm 0.20$ & $0.83 \pm 0.04$ & $10.15 \pm 0.28$\\
	CDFS15764 & 0.30 & 0.63 & $9.06 \pm 0.05$& 0.50 & $10.97 \pm 0.11$ & $11.39 \pm 0.19$ & $2.02 \pm 0.20$ & $0.18 \pm 0.05$ & $10.33 \pm 0.28$\\
	CDFS10299 & 0.17 & 0.71 & $8.99 \pm 0.08$ & 0.70 & $10.56 \pm 0.20$ & $11.12 \pm 0.10$ & $1.84 \pm 0.19$ & $0.33 \pm 0.06$ & $10.26 \pm 0.28$\\
	CDFS12910 & - & - & - & 0.85 & $10.23 \pm 0.04$ & $ - $ & $1.64 \pm 0.22$ & $-$ & $-$\\
	CDFS12664 & - & - & - & 0.35 & $10.21 \pm 0.06$ & $ - $ & $1.44 \pm 0.09$ & $-$ & $-$\\
	CDFS16661 & - & - & - & 0.50 & $10.16 \pm 0.05$ & $ - $ & $1.80 \pm 0.15$ & $-$ & $-$\\
	\hline
	\end{tabular}
\end{table*}

\begin{figure*}
	CDFS13844\\
	\includegraphics[width=0.88\textwidth]{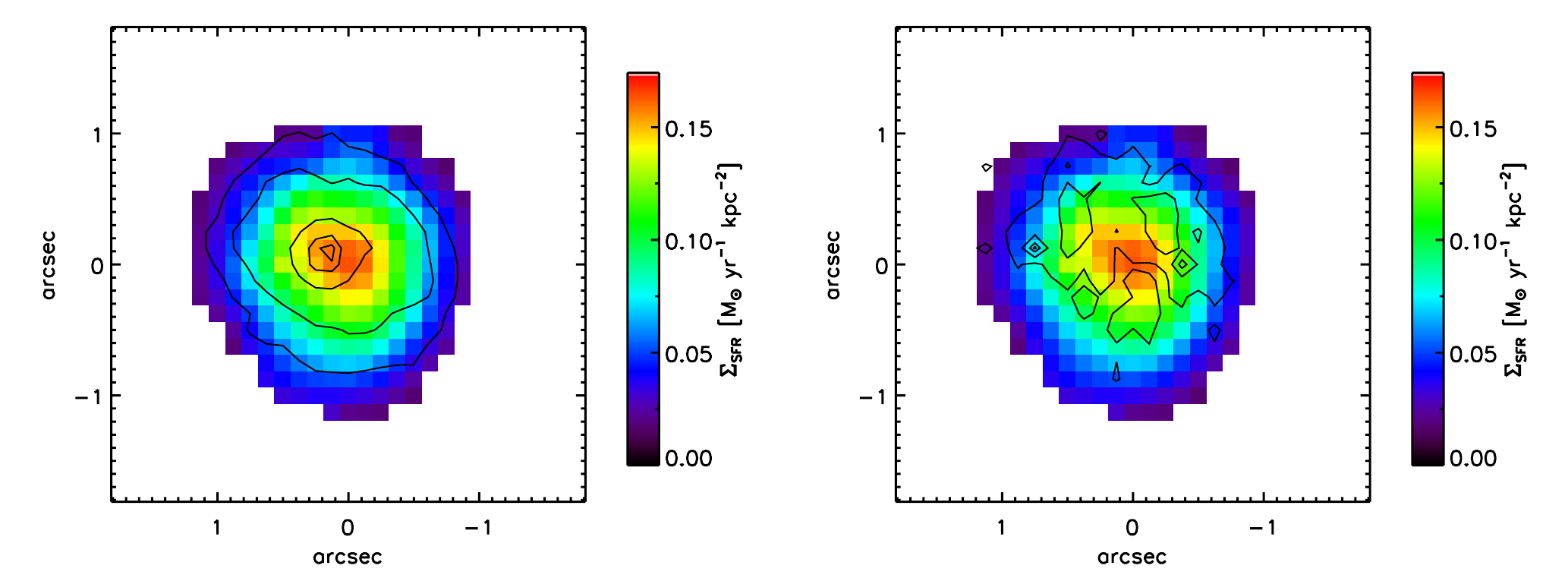}\\
	CDFS11583\\
	\includegraphics[width=0.88\textwidth]{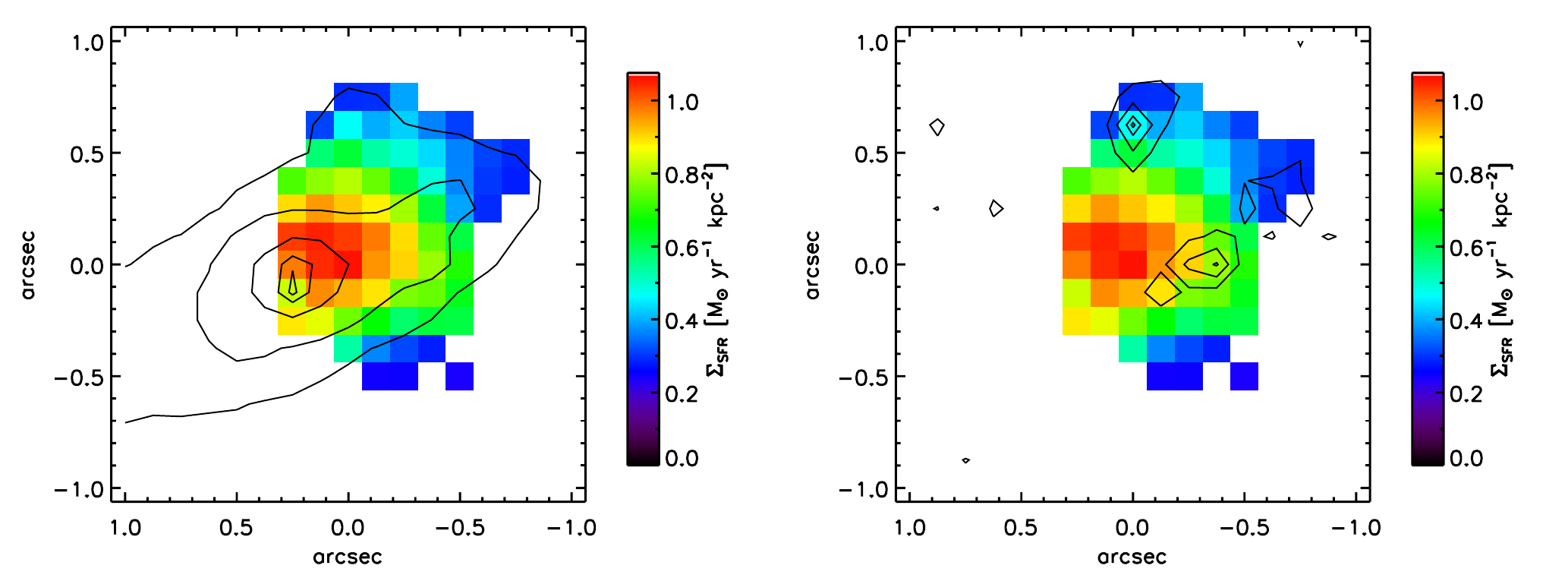}
	\caption{Star formation rate surface density map inferred from the H$\rm{\alpha}$ surface brightness distribution, scaled to match the total 	SFR inferred from the FIR, shown for two galaxies from the sample. The $\rm \Sigma _{SFR}$ map is compared with ({\it left}) the near-IR (160W) HST images (contours) and ({\it right}) the optical V-band HST image. The central positions (0,0) corresponds to: (top) R.A. = 53.0481$^\circ$, Dec. = -27.7371$^\circ$ and (bottom) R.A. = 53.16425$^\circ$, Dec. = -27.77698$^\circ$.  Note: the star formation density maps for the full sample are available online.}
	\label{fig:sfr}
\end{figure*}

Fig.~\ref{fig:sfr} shows the map of star formation rate surface density $\rm \Sigma _{SFR}$ inferred from the H$\rm{\alpha}$ surface brightness distribution, scaled to match the total SFR inferred from the FIR (which is nearly equivalent to correcting for extinction following the steps discussed above) for two galaxies in the sample. The maps for the entire sample are available online. For the H$\alpha$ fluxes, we impose a signal-to-noise (S/N) cut of $\sim 4$. On the left-hand panels the $\rm \Sigma _{SFR}$ map are compared with the near-IR (160W) HST image (contours), sampling the rest frame light at about 0.6$~\mu$m, hence roughly tracing the distribution of the stellar mass, while on the right-hand panels the maps are compared with HST images taken with the optical V-band F606W filter (contours), which trace the rest-frame UV light. Note that the SINFONI and HST images are aligned using the galaxy continuum detected in the SINFONI observations in the H-band.

The first interesting result to note is that the star formation rate generally peaks in the central region of the galaxy, where the stellar mass distribution peaks. Here the $\rm \Sigma _{SFR}$ is typically in the range  $\rm 0.2-2.0 ~M_{\odot}yr^{-1}kpc^{-2}$. This is certainly much higher than in local star forming disc galaxies ($\rm \Sigma _{SFR} \sim 0.01 ~M_{\odot}yr^{-1}kpc^{-2}$), but well below the extreme values observed in local ULIRGs and in some high-z hyperluminous systems, which are characterized by $\rm \Sigma _{SFR} \sim 100-1000~M_{\odot}yr^{-1}kpc^{-2}$. These results are consistent with the fact that most of these galaxies are located on the Main Sequence, hence likely evolving secularly, not through violent processes (unlike ULIRGs and merging systems), although at an accelerated rate relative to local galaxies because of larger gas content. However, we cannot exclude that the limited angular resolution of our data is smearing out peaks of high $\rm \Sigma _{SFR}$. This may be the case for the strongly interacting small companion CDFS15753 that, despite being well above the Main Sequence, has a peak $\rm \Sigma _{SFR}$ of ``only'' $\rm 0.6 ~M_{\odot}~yr^{-1}~kpc^{-2}$. However \cite{troncoso_13} found that the SINFONI beam smearing had a negligible effect on the measurements made of $\rm \Sigma _{SFR}$ for simulated galaxies at $\rm z \sim 3$.

It should also be noted that the SFR does not always peak on the location of the stellar light, especially in massive galaxies. The most remarkable case is CDFS11583, whose star formation (with S/N(H$\rm \alpha$ $>4$)) is active only on half of the stellar disc, and whose peak of SFR is clearly offset with respect to the peak of stellar light. To be certain this distribution is not a consequence of the S/N cut, we also directly collapse the reduced SINFONI cube for this galaxy around the H$\rm \alpha$ wavelength and find that this asymmetry is still present. A similar effect is seen in CDFS15764, the massive companion of CDFS15753 (see Fig.~\ref{fig:merger}). In these massive systems the distribution of star formation is likely tracing the most recent minor merger event of a gas rich galaxy, or massive infalls from the IGM, hence the non-axisymmetric morphology. The distribution being far from axisymmetric is however a warning about using SFR or gas spectroscopic tracers to map the dynamics of high-z massive galaxies, since these may only trace part of the whole galaxy. 

It is also very interesting to compare the distribution of SFR traced by H$\rm{\alpha}$ with the morphology of the rest-frame UV light ($\sim$3000~\AA), which should trace similar regions in the absence of dust extinction (Fig.~\ref{fig:sfr}). In many cases the UV light is either totally undetected or tracing outer regions, in which dust extinction is probably low enough that UV radiation can escape. If these galaxies are representative of Main Sequence galaxies, our results implies that the UV studies of star forming galaxies at high-z provide a very distorted and very partial view of the star formation process due to the effects of dust extinction. The same is true for H$\alpha$ detection, although this is less affected. Therefore by combining UV, H$\alpha$ and FIR observations as in the present work, we can get a more complete view of the star formation at high-z.

\subsection{Galaxy dynamics}
\label{sect:dynamics}
\begin{figure*}
	\includegraphics[width=0.9\textwidth]{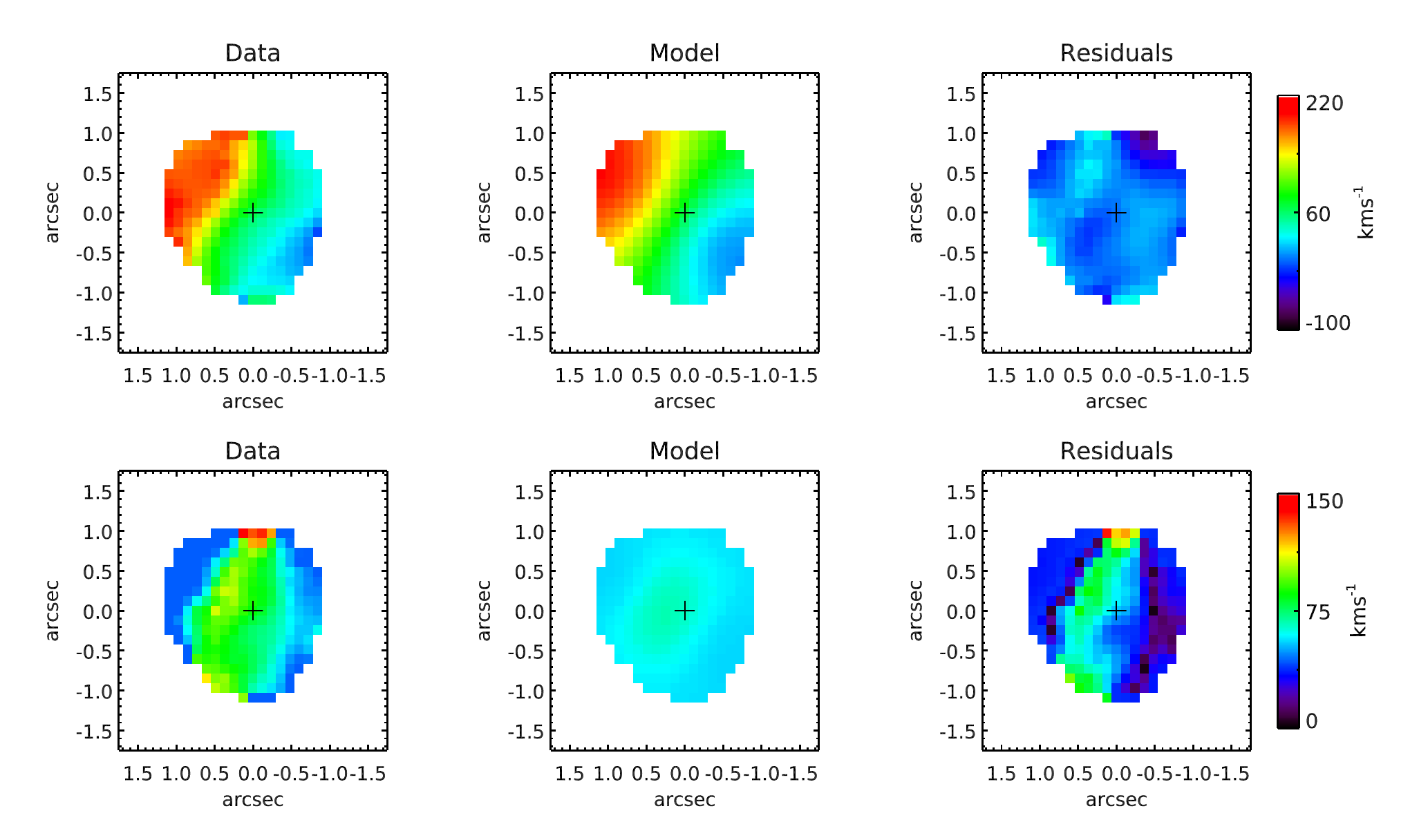}
		\caption{Velocity field and associated modelling inferred from the H$\alpha$+[NII] lines (using a S/N cut of 4) for one galaxy in the sample (CDFS13844). The top row shows the velocity field, where the left panel shows the observed velocity pattern, the central panel shows the best fit with a rotating disc model and the right panel shows the residuals. The bottom row shows the velocity dispersion, where the left panel shows the observed velocity dispersion, the central panel the velocity dispersion due to the PSF smearing of the velocity field and the right panel shows the intrinsic velocity dispersion obtained by the difference in quadrature of the previous two quantities. The cross marks the position of the peak H$\alpha$ emission and the central position (0,0) corresponds to R.A. = 53.0481$^\circ$, Dec. = -27.7371$^\circ$.  Similar maps for the entire sample are available online. } 
	\label{fig:vel}
\end{figure*}

The leftmost panel of Fig.~\ref{fig:vel} show the observed velocity field (top) and observed velocity dispersion (bottom) inferred from H$\alpha$+[NII], using an average S/N cut of 4, for one galaxy in the sample. Velocity maps for the full sample are available online. Most of the velocity fields show a clear rotation pattern. In terms of the velocity dispersion, one should bear in mind that the peak of dispersion around the central region may be a consequence of the steep rotation field within the seeing resolution element, which makes the line artificially broader.

The velocity field was modelled with a rotating disc assuming an exponential surface brightness. Firstly we fit the flux distribution to obtain the best fitting values for the luminosity and the exponential scale radius, which we then fix in the velocity modelling. We also fix the inclination of the disc along our line of sight to the value inferred from the axial ratio observed in the H-band HST images. All other parameters are allowed to vary to produce the best fitting by minimizing the $\chi^{2}$ probability. The model is convolved with the point spread function (PSF) and compared to the observed velocity to calculate the residuals from the fitting. Further details of the method are given in \cite{gnerucci_apr11} and specifics of the model in \cite{carniani_13}. 

In the following we define $\Delta V$ as the maximum projected rotation curve obtained by the model:
\begin{equation} \label{eq:vmod}
	\Delta V = \frac{1}{2}(V_{\rmn{max}}-V_{\rmn{min}})
\end{equation}
where $V_{\rmn{max}}$ and $V_{\rmn{min}}$ are the maximum and minimum velocities from the model within the observed region.\\
To correct for the orientation of the galaxy with respect to our line of sight we use the inclination inferred from the HST images to de-project the modelled velocity and give the rotational velocity as:
\begin{equation} \label{eq:vreal}
	V_{\rmn{rot}}=\frac{\Delta V}{\cos(\frac{\pi}{2} -i)}
\end{equation}
where $i$ is the inclination angle of the disc.

\begin{figure}
	\includegraphics[trim=0.9cm 0.4cm -0.2cm 1.5cm, clip=true, width=90mm]{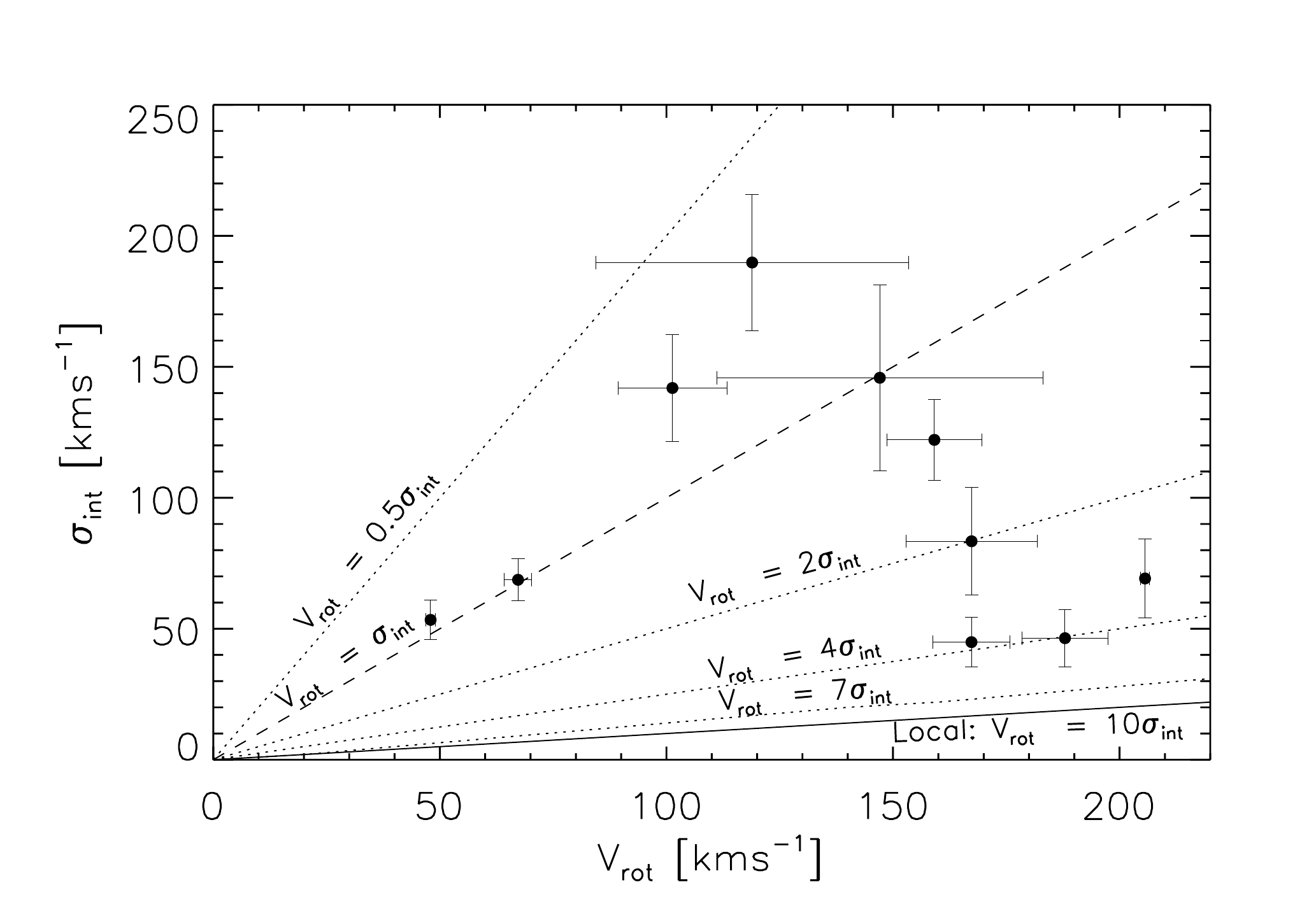}
	\caption{De-projected rotational velocity versus intrinsic velocity dispersion. The solid black line shows the relationship found in local galaxies which are rotation dominated with $V/\sigma \sim 10$. We also show with dotted lines the case where $V$ is 7, 4, 2, 0.5 times $\sigma$, while the dashed line shows the case where $V=\sigma$. All of our galaxies have $V/\sigma$ less than that of local galaxies showing that these are more turbulent systems with some approaching $\sigma = 2V$. }
	\label{fig:vsig}
\end{figure}

The central top panel of Fig.~\ref{fig:vel} shows the velocity model for one galaxy, while the top-right model shows the residuals. Clearly the residuals are very low, indicating that indeed the overall velocity field is well described by a simple rotating disc. The bottom-central panel shows the velocity dispersion resulting from the model. Since the model has no intrinsic velocity dispersion, the dispersion seen in this panel (central) is simply ``artificial'' velocity dispersion introduced in the data as a consequence of the velocity field smeared by the seeing. This is essentially the spread in the velocity of each pixel about the mean value, within the PSF, rebinned to the correct pixel size and weighted by the flux of each pixel and also taking into account the spectral resolution. By subtracting in quadrature this component from the observed velocity dispersion it is possible to obtain the intrinsic velocity dispersion (i.e. $\sigma_{\rmn{int}}= (\sigma_{\rmn{obs}}^2 - \sigma_{\rmn{mod}}^2)^{1/2} $), which is shown in the bottom-right panel. Most of the galaxies have significant velocity dispersion compared to the rotational velocity. This is better illustrated in Fig.~\ref{fig:vsig}, which shows the intrinsic velocity dispersion as a function of rotational velocity. Clearly, the bulk of galaxies have $\rm V/\sigma <4$, some of them approaching
$\rm V/\sigma \sim 0.5$, i.e. they are highly turbulent relative to local discs, for which $\rm V/\sigma \sim 10$ \cite[e.g][]{downes98}. This finding is in line with other works at similar and higher redshift \citep{forster_jul06, cresci_may09, gnerucci_apr11, swinbank_nov11} and it has been ascribed to the large gas content characterizing high-z star forming galaxies \cite[e.g.][]{ daddi10, tacconi_may13}, which makes gaseous discs self-gravitating and gravitationally unstable. In Section~\ref{sect:fgas} we will discuss the gas content in these galaxies in more detail.

\begin{figure}
	\includegraphics[trim=1.3cm 0.4cm 0cm 0.8cm, clip=true,width=95mm]{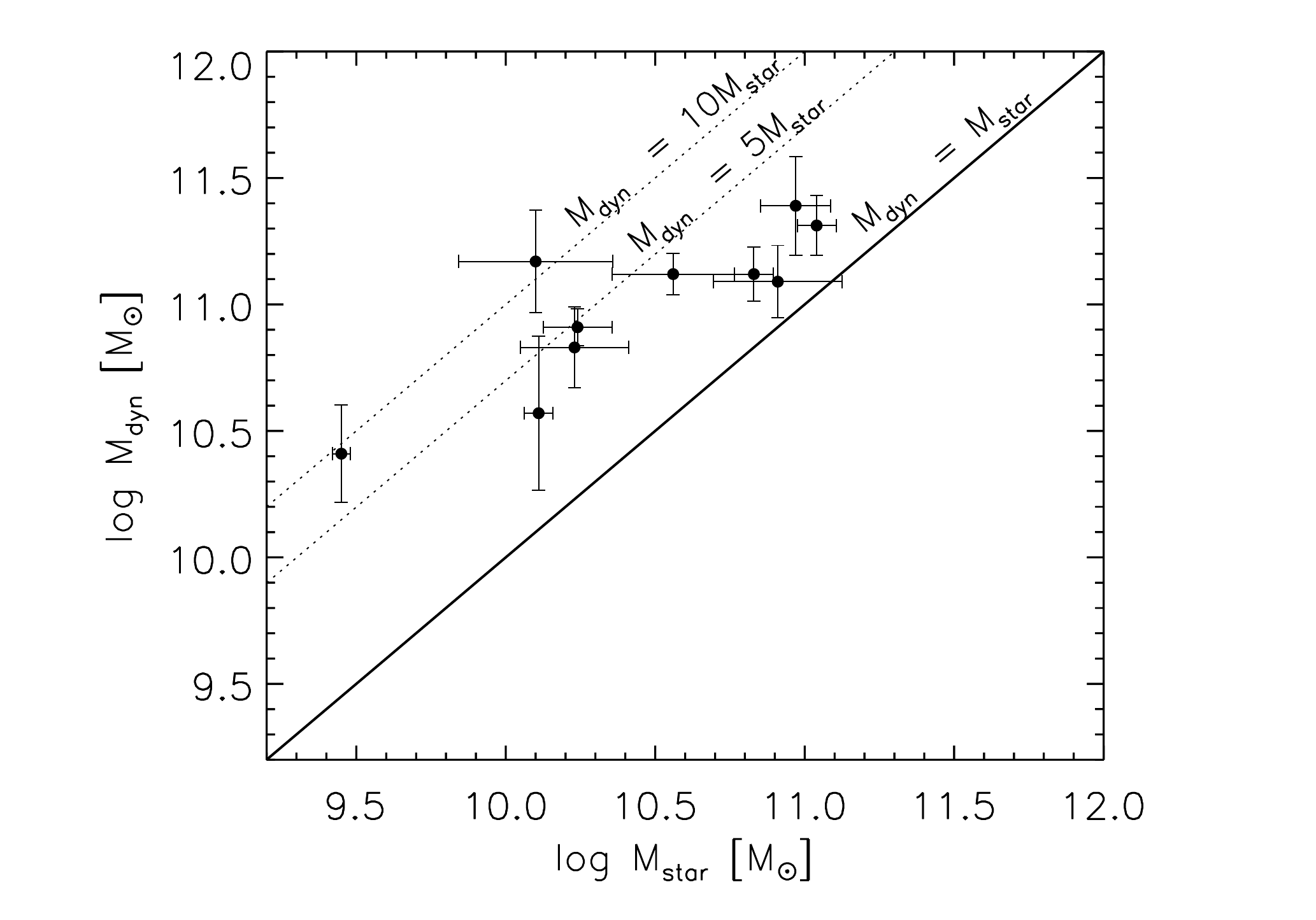}
	\caption{The stellar mass versus the dynamical mass for each galaxy. The solid line shows the 1:1 ratio between the stellar and dynamical mass, whereas the dotted lines indicate dynamical masses five and ten times the stellar.}
	\label{fig:mdyn}
\end{figure}

\begin{figure}
	\includegraphics[trim=1.3cm 0.4cm 0cm 0.8cm, clip=true,width=95mm]{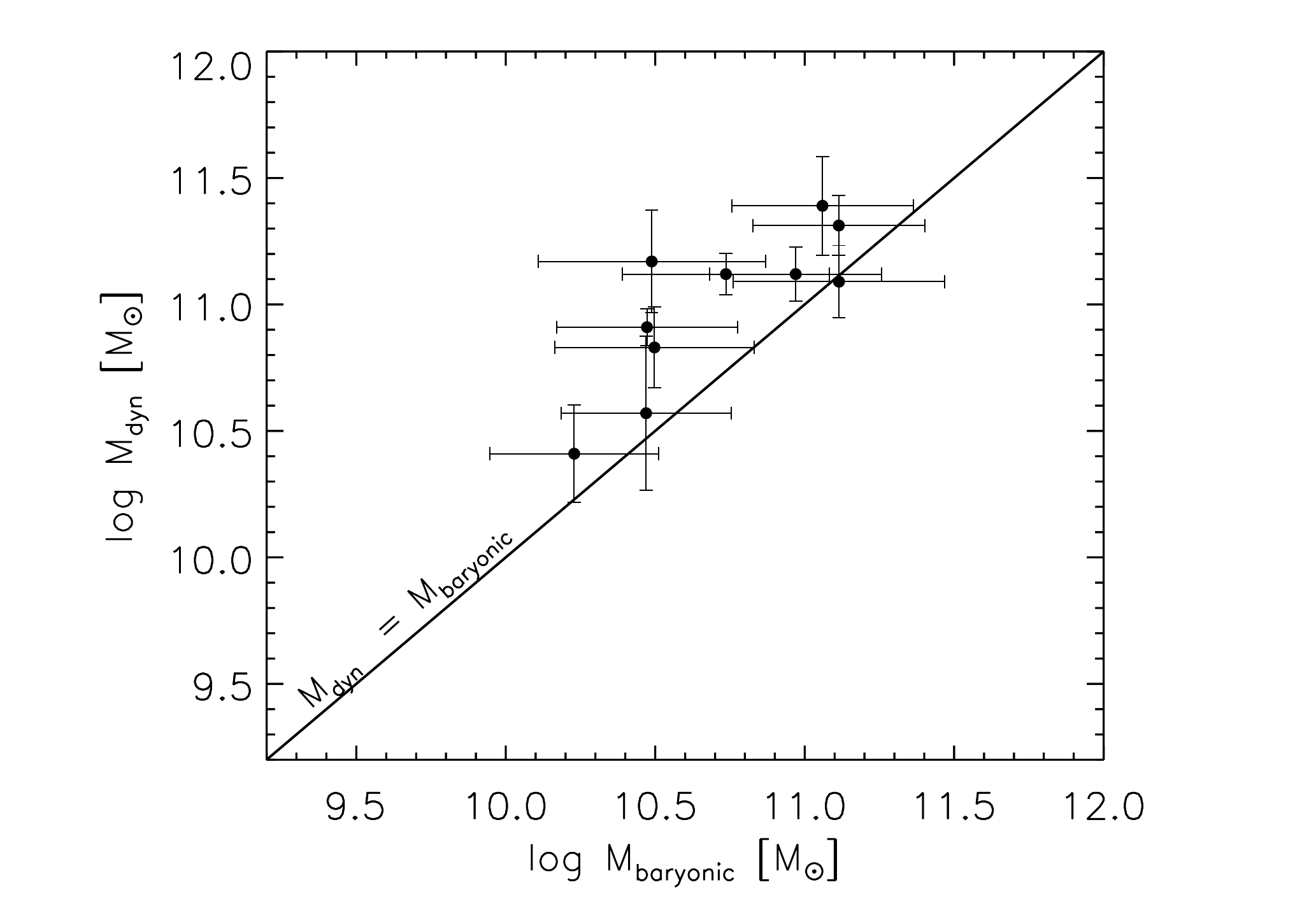}
	\caption{Dynamical mass versus total baryonic mass ($\rmn{M_{star} + M_{gas}}$). The solid black line represents the condition when $\rmn{M_{dyn} = M_{baryonic}}$.	}
	\label{fig:mbar}
\end{figure}

In Fig.~\ref{fig:mdyn} we compare the dynamical mass given by the velocity model to the stellar mass. The model assumes that the gravitational potential is that of a thin disc with an exponential profile and computes the dynamical mass as the mass within 10~kpc. More details on the method are given in \cite{gnerucci_apr11}. We find that on average the dynamical mass is about three times the stellar mass, which is comparable to the results found by \cite{gnerucci_apr11}, for galaxies at $\rm z\sim 3$, and with \cite{cresci_may09}, who find $\rm {M_{dyn}}\sim3.3 M_{star}$ for galaxies at $\rm z=2$. This excess of dynamical mass relative to the stellar mass is mostly ascribed to the large masses of gas which are expected to be hosted in these systems and is consistent with the large gas fractions observed in these galaxies, as inferred by various studies \cite[e.g][]{tacconi_feb10, santini14} and also by us. Indeed, in Section~\ref{sect:fgas} we attempt to infer the gas mass in our sample of galaxies by inverting the Schmidt-Kennicutt (S-K) relation \citep{kennicutt_may98}.
In Fig.~\ref{fig:mbar} we show the comparison of the dynamical mass and the baryonic mass, which is obtained by summing the stellar mass inferred from the SED fitting and the gas mass computed by inverting the S-K relation.

For most of our galaxies the dynamical mass is consistent with the total baryonic mass within the error bars suggesting that dark matter has a negligible contribution to the total mass within the region probed by our observations. However there is some tension for some of the galaxies that show a dynamical mass significantly higher than the baryonic mass, and the two masses are not consistent with a significance of 2--3$\sigma$. For these few galaxies dark matter may contribute significantly within the aperture.

In the following we investigate whether the dynamical properties of our galaxies correlate with their star formation activity and, in particular, with the deviation from the Main Sequence. We quantify the amount of dynamical disturbance with a {\it dynamical disturbance index} $\beta$ defined as the average residuals from the velocity fit relative to the maximum velocity gradient, hence gives a measure of the quality of the fit to the data with a pure regular rotating disc model, i.e.
\begin{equation} \label{eq:beta}
	\beta = \frac{\langle \rmn{|V_{obs}-V_{model}|} \rangle}{\Delta V}
\end{equation}

\begin{figure}
	\includegraphics[trim=0.9cm 0.4cm 0.9cm 0.5cm, clip=true,width=90mm]{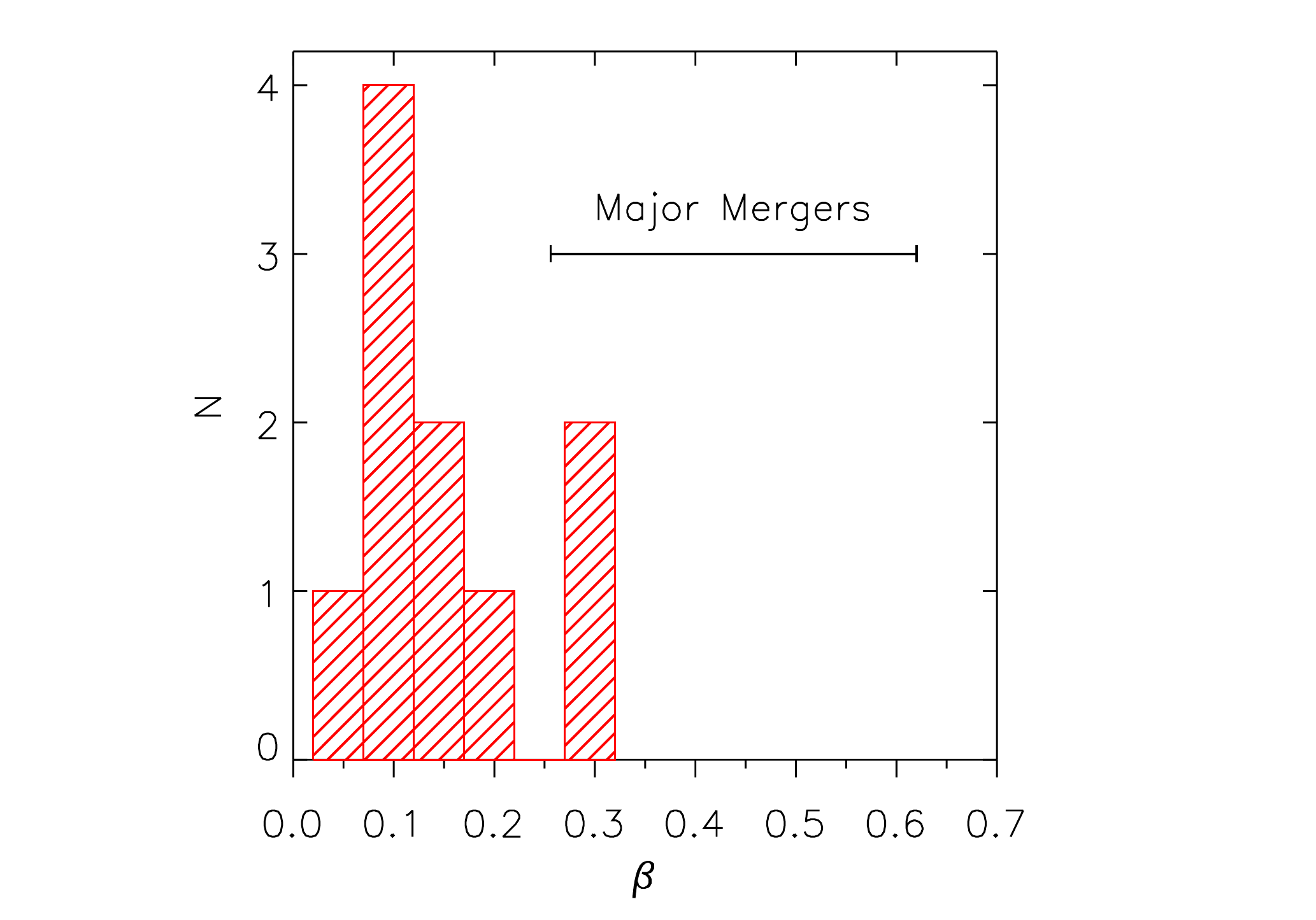}
	\caption{Distribution of the dynamical disturbance index $\beta$ (Eq.~\ref{eq:beta}) which quantifies the level of dynamical disturbance. This shows that in most cases the deviations from pure rotation are at most a few 10\%. We also indicate the range of the $\beta$ parameter for a small sample of local major mergers (see Appendix~\ref{appen1} for further details).}
	\label{fig:beta}
\end{figure}

The distribution of $\beta$ for the galaxies in our sample is shown in Fig.~\ref{fig:beta}, illustrating that in most cases the deviations from pure rotation are below 20\% of $\Delta V$. This result supports, in a quantitative way, our initial statement that the dynamics of most of these galaxies are well represented by pure rotation. To further quantify our result, we compare our $\beta$ values with those obtained for a sample of local major mergers artificially redshifted to z=1.5 and convolved with our resolution (see Appendix~\ref{appen1}). In general we find major mergers to have significantly higher values of $\beta$ ($0.37<\beta <0.62$) whereas our galaxies have in most cases much lower values of $\beta$ ($0.02<\beta<0.17$). There are two exceptions in our sample, the galaxies CDFS15764 and CDFS16485 which have $\beta \sim 0.3$, i.e. approaching the values observed in local major mergers (although not yet so high), however there is evidence to suggest that these galaxies could be in the initial phases of a major merger. Indeed, we have shown that CDFS15764 is undergoing interactions with CDFS15753 (see Fig.~\ref{fig:merger}) and for CDFS16485 there is also a hint of an interaction with a close neighbouring galaxy shown in the HST H-band contours (available online). 

We note that the index $\beta$ does not correlate with the deviation from the Main Sequence. Indeed, in Fig.~\ref{fig:MS} galaxies are color-coded with the value of the $\beta$ index and no correlation is observed, although our galaxies do not probe a broad range of deviations from the Main Sequence. Most interestingly, mild deviations ($\beta \gtrsim 0.2$), possibly associated with minor mergers or mild interactions, are apparently not enough to move galaxies outside the Main Sequence. Even more interestingly, the low mass system CDFS15753 (indicated with a solid green square), interacting with the massive system CDFS15764 (located at a projected distance of about $17$~kpc), does not show strong deviations from the rotational velocity field. The finding that the central dynamics of interacting systems at distances of a few $10$~kpc is not strongly perturbed, even if one of the companions is very massive, has been found also in other high-z systems \cite[e.g.][]{carniani_13}. Therefore, disturbed central dynamics as a consequence of the interaction with the massive companion is probably not at the origin of the strong deviation of CDFS15753 from the Main Sequence.

\begin{figure}
	\includegraphics[trim=0.0cm 0.0cm 0.0cm 1.8cm, clip=true,width=80mm]{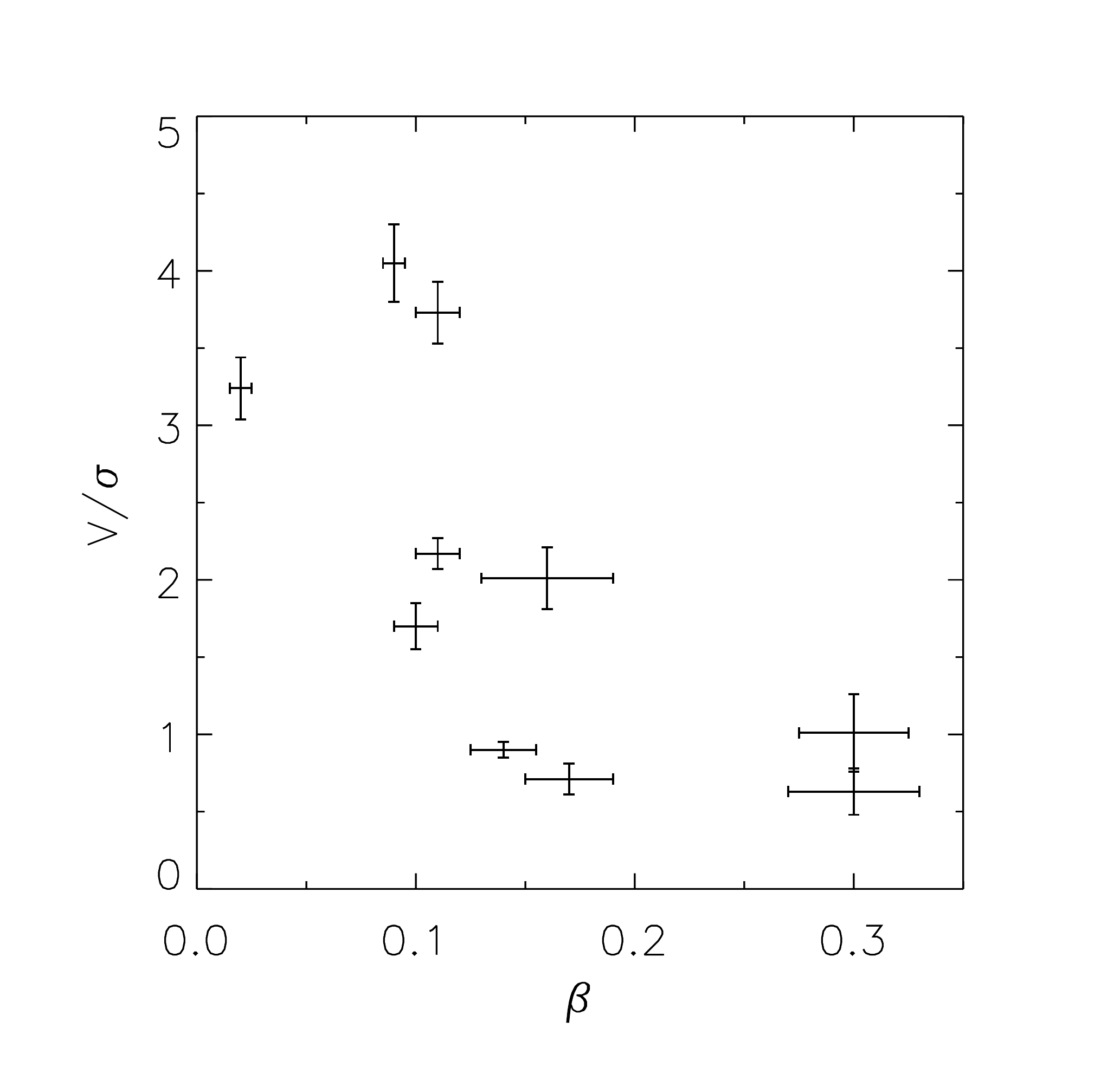}
	\caption{$\rm V/\sigma$ versus the dynamical disturbance index $\beta$ (Eq.~\ref{eq:beta}). Galaxies that deviate from a rotating disc model tend to be the more turbulent systems.}
	\label{fig:beta_turb}
\end{figure}

Finally we compare the dynamical properties $\beta$ and $\rm V/\sigma$ for our sample in Fig.~\ref{fig:beta_turb}. There is a general trend that the more a galaxy deviates from the pure rotation model, the more turbulent it is. Therefore, an important effect of minor mergers seems to be an increase in the level of turbulence in the gas. 

\subsection{Metallicity and Metallicity Gradients} 
\label{sect:metals}

Ratios between strong nebular optical emission lines can be used as tracers of gas metallicity, although they generally depend also on other gas physical parameters, such as the ionization parameter. Various direct and indirect calibrations of the strong line ratios have been proposed in the literature \cite[e.g.][]{pettini_mar04, nagao_mar06, maiolino_sep08, kewley_08}. It is beyond the scope of this paper to discuss the merit and caveats of all these calibrations, although we will shortly come back to this point in Section~\ref{sect:fgas}. Here we only emphasize that these calibrations are only valid if the gas is photoionized by stellar continua, and not by AGNs or shocks. The exclusion of regions affected by AGNs or shocks can be done by exploiting the BPT diagrams \cite[e.g.][]{kauffmann_dec03}. However, the latter involves the [OIII]/H$\beta$ ratio, which cannot be used in our case since these lines are not detected in most targets due to strong dust extinction. The [OIII] line is only clearly detected in one galaxy, CDFS2780, and H$\beta$ is only detected weakly. For this galaxy we obtain $\log(\rmn{[NII]/H\alpha})=-0.69$ and $\log(\rmn{[OIII]/H\beta})=0.30$, which comfortably place this galaxy in the HII region of the BPT diagram. However for other galaxies we only have the [NII]/H$\alpha$ ratio. We note that, even in the absence of the [OIII]/H$\beta$ ratio, AGN and shocks are characterized by [NII]/H$\alpha$ ratios larger than 0.46. Therefore, as adopted in other studies \cite[e.g.][]{mannucci_nov10}, the latter limit can be used to discriminate objects/regions affected by AGN/shock ionization from star forming HII regions.

\begin{figure*}
	CDFS13844\\
	\includegraphics[width=0.8\textwidth]{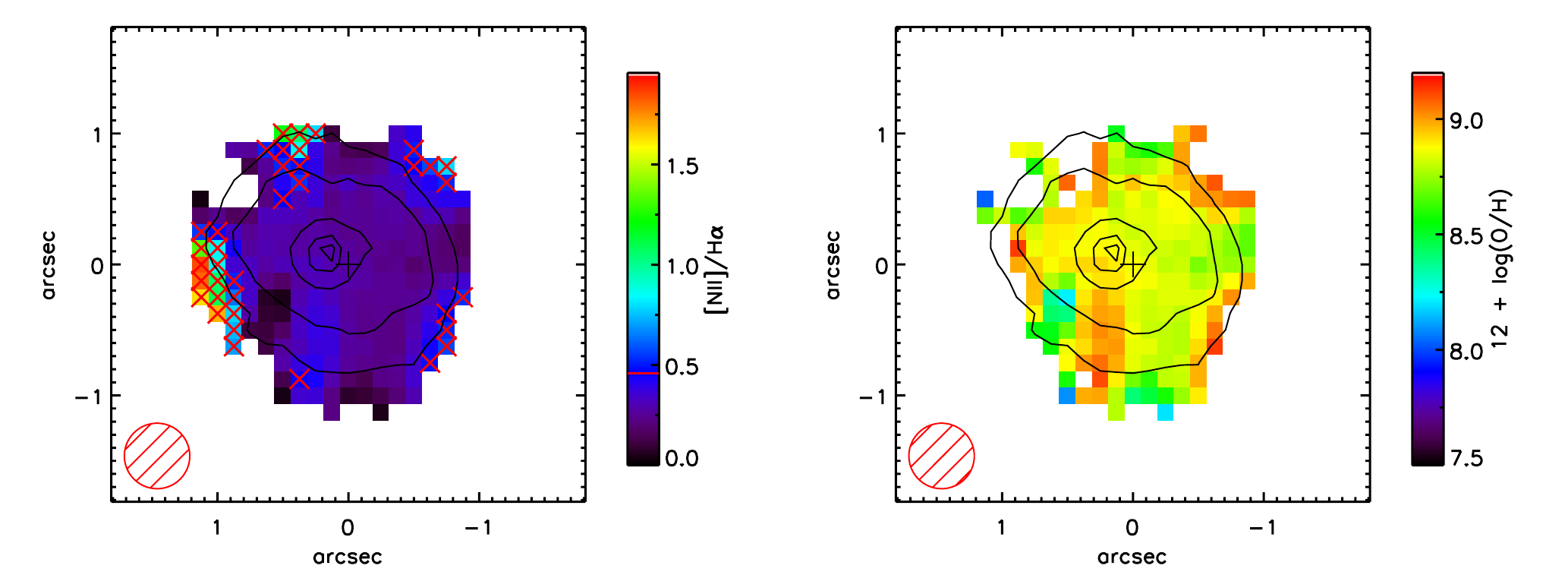}\\
	CDFS15764\\
	\includegraphics[width=0.8\textwidth]{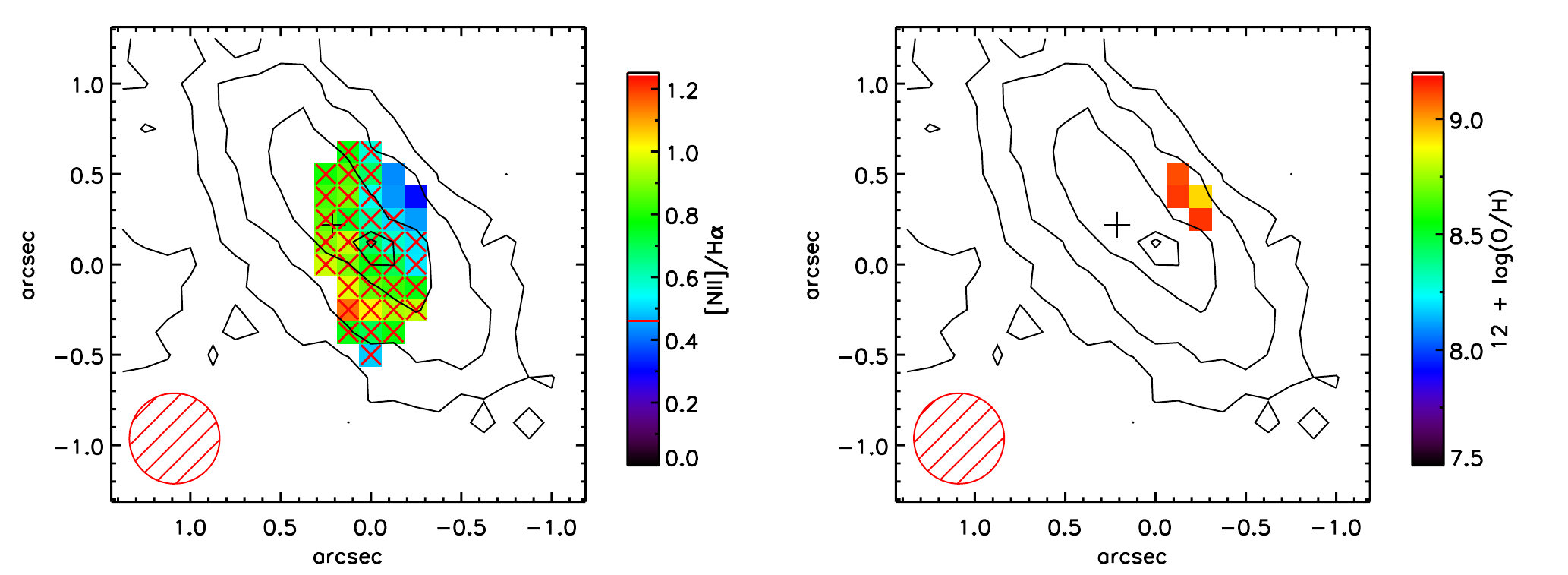}
	\caption{Metallicity gradients for two galaxies in the sample: CDFS13844 (top) and CDFS15764 (bottom). Left: Ratio of [NII]/H$\rm{\alpha}$ fluxes with red crosses indicating the regions with $R>0.46$ therefore likely to be affected by shocks or an AGN (see Section~\ref{sect:metals}) and so are excluded from the metallicity calibration. Right: The corresponding metallicity maps across the galaxy after AGN/shock removal. The black contours in both maps indicate the position of the galaxy continuum from the H-band HST images and the cross highlights the location of the peak of H$\rm{\alpha}$ emission. The size of the PSF is shown in red. The central positions (0,0) corresponds to: top R.A. = 53.0481$^\circ$, Dec. = -27.7371$^\circ$ and bottom R.A. = 53.0905$^\circ$, Dec. = -27.7122$^\circ$} 
	\label{fig:metal_grad}
\end{figure*}

The left panels of Fig.~\ref{fig:metal_grad} show the maps of the [NII]/H$\alpha$ ratios for two galaxies from the sample (maps for the entire sample available online). The red crosses indicate the regions in which [NII]/H$\rm{\alpha}$ $>0.46$, hence those regions, which are likely affected AGN/shocks, are excluded from the metallicity analysis. The right panel show the metallicity maps derived using the [NII]/H$\alpha$ ratio as a metallicity tracer from the calibration given in \cite{maiolino_sep08}, again using a S/N cut of $\sim 4$ on H$\rm \alpha$. Contours indicate the stellar light traced by H-band HST images. Metallicity are obviously shown only for the regions complying with the condition [NII]/H$\alpha$ $<0.46$. It is of interest to see that the ratio map for CDFS15764 excludes nearly the whole galaxy showing that it is almost totally dominated by shocks or an AGN. Furthermore we find that the excluded regions are in general, not coincident with the centre, therefore indicating that shocks more than AGNs are likely responsible for the high values of [NII]/H$\alpha$. This is not surprising since AGNs were accurately discarded through deep hard X-ray and mid-IR data. Interestingly, we find that the galaxies which are strongly affected by shocks according to their [NII]/H$\alpha$ ratio also seem to be those with a high $\beta$ and low $V/\sigma$ value, which represent systems with irregular and turbulent velocity fields. We have checked that the excluded regions (with high values of [NII]/H$\alpha$) are not also the regions with low S/N, however we cannot neglect that the effects of low S/N may play a role in the outer regions.

Three of these maps show a general trend of regular metallicity gradients similar to that found in local galaxies where the metallicity is higher in the central region and decreases outwards. However, six galaxies show relatively flat gradients where the metals seem to be uniformly spread across the galaxy. This is in contrast to the findings of \cite{cresci_oct10} and \cite{troncoso_13} at higher redshift ($\rm z>3$) who observe inverted gradients with inner regions more metal poor, which is attributed to the presence of pristine inflows of gas leading to a dilutions effect. This provides further evidence that inverted gradients are mostly a property typical of galaxies at $ \rm z>3$, although examples of inverted gradients are observed also at low-z and in the local universe \cite[e.g.][]{werk10, queyrel_12,sanchez13}. However an alternative interpretation for the relatively flat metallicity gradients may be due to the effects of the PSF, as discussed by \cite{yuan13b} who find that measured metallicity gradients tend to flatten with poorer angular resolution. Note that one galaxy (CDFS15764) is almost entirely dominated by shocks so it is not possible to judge the metallicity gradient in this case. Moreover, while we do not find any strong example of negative gradients, many of the galaxies showing relatively flat gradients do show points of higher metallicities located in the outer regions. This shall be discussed further in Section~\ref{sect:fgas}.

We also investigate the averaged metallicity across each galaxy in the context of the `Fundamental Metallicity Relation' (FMR), which explores how the metallicity correlates with both star formation and stellar mass. We have obtained the global metallicity of each galaxy in our sample from the integrated [NII] and H$\alpha$ fluxes, first excluding points affected by AGN-ionization or shocks.

\begin{figure}
	\flushleft{(i)}
	\includegraphics[trim=0.9cm 0.4cm 0.4cm 0.9cm, clip=true,width=84mm]{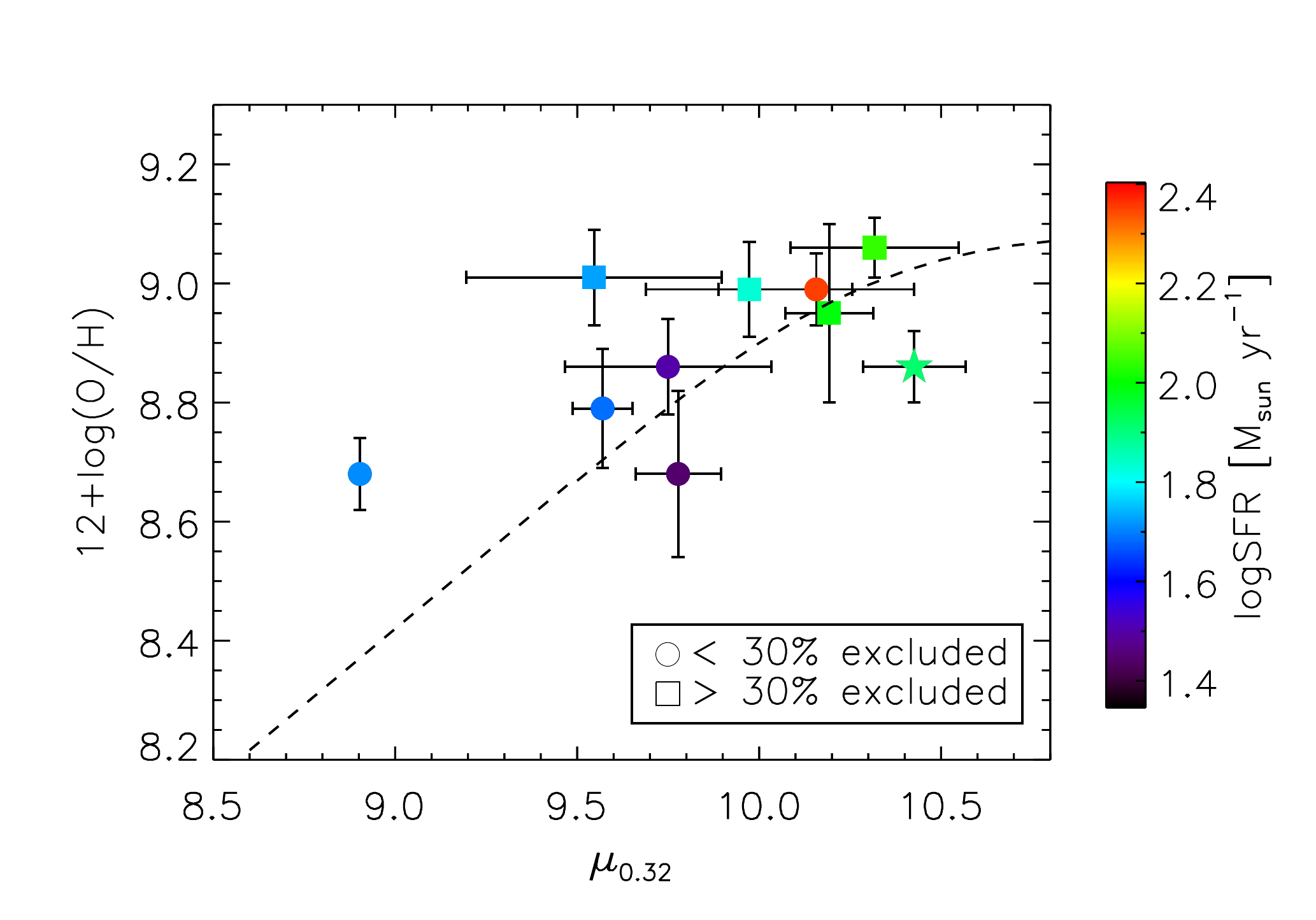} 
	\flushleft{(ii)}
	\includegraphics[trim=0.9cm 0.4cm 0.4cm 1.5cm, clip=true, width=84mm]{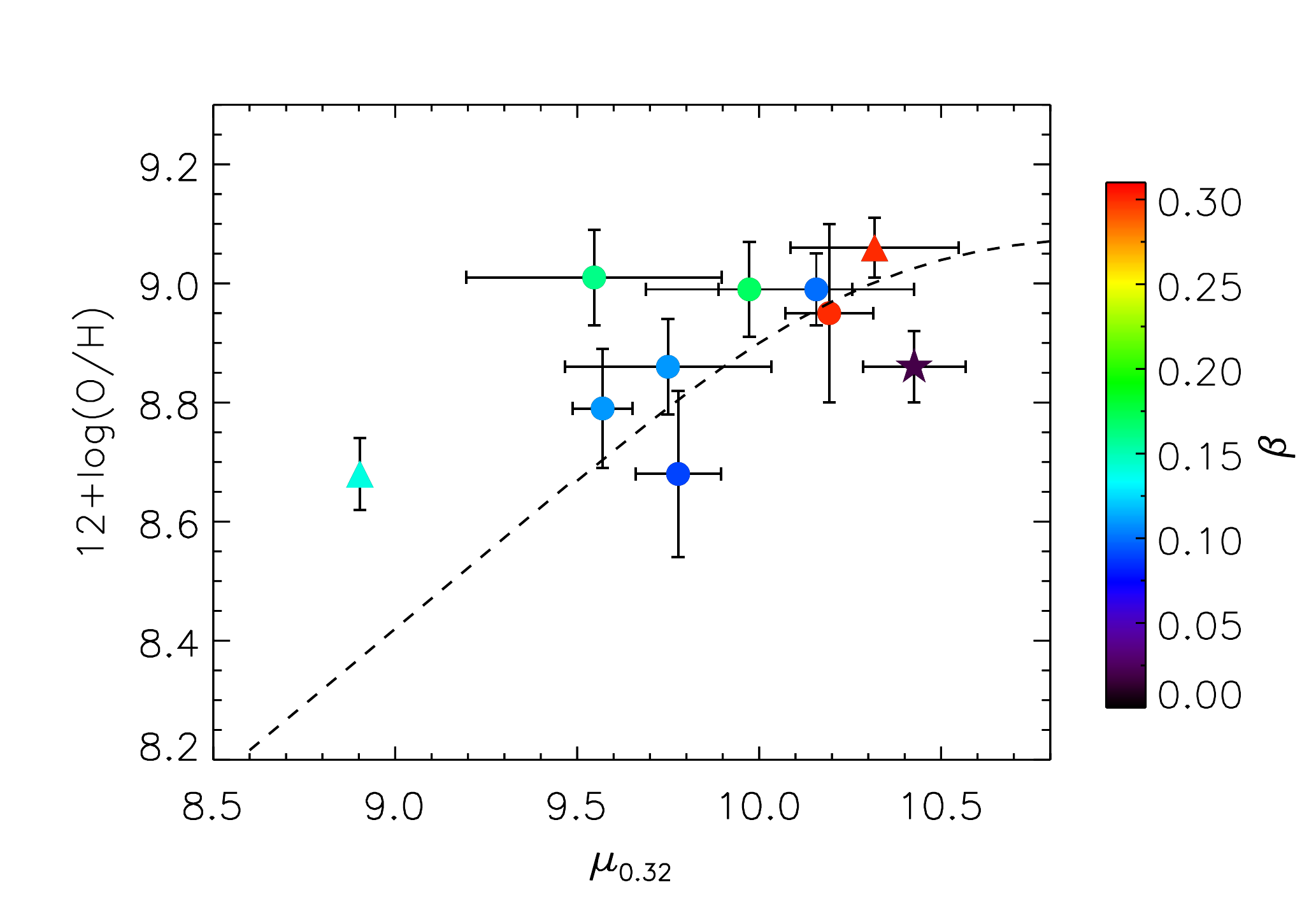}
	\flushleft{(iii)}
	\includegraphics[trim=0.9cm 0.4cm 0.4cm 1.5cm, clip=true, width=84mm]{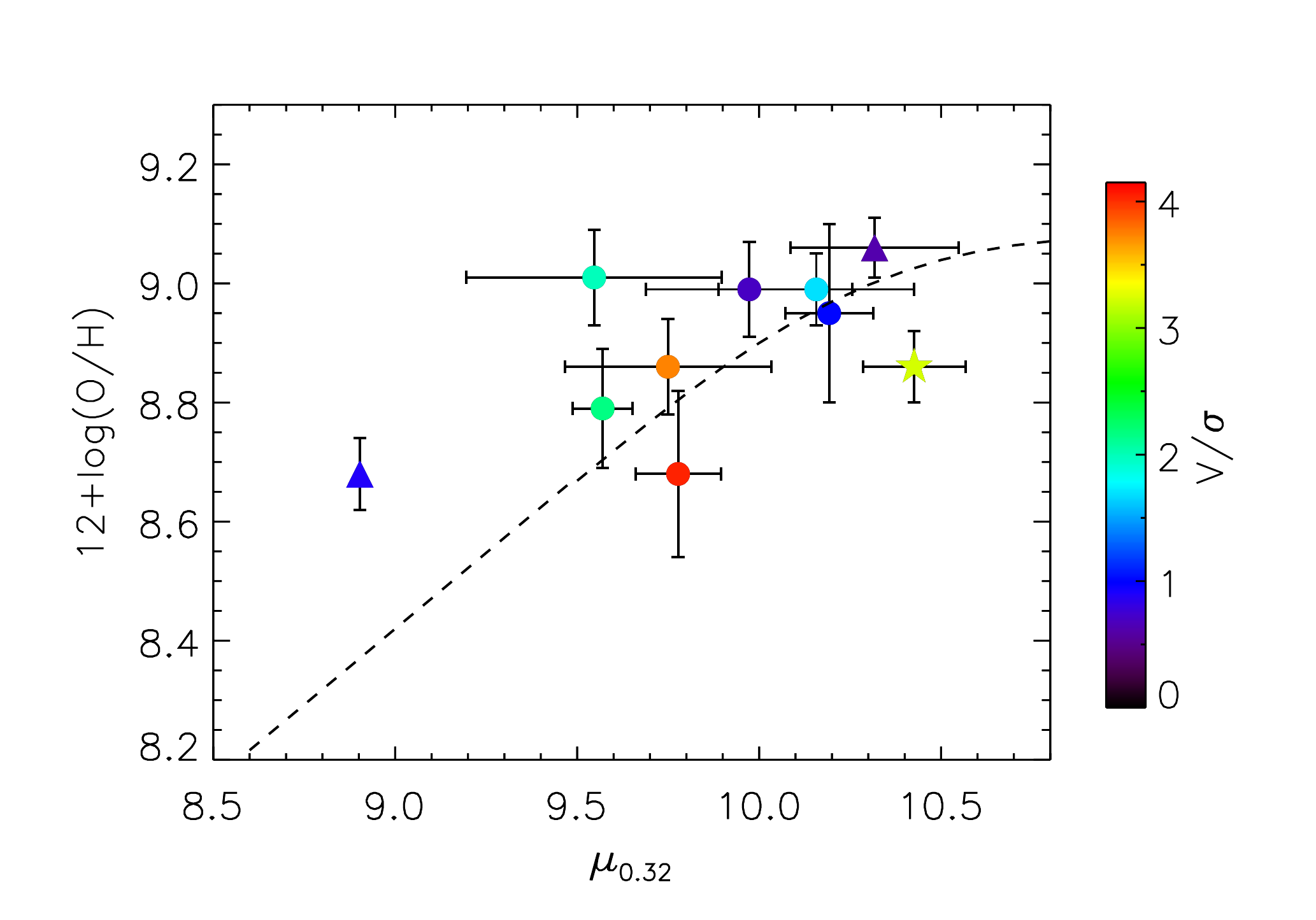}
	\caption{The FMR projection from \citet{mannucci_nov10,mannucci_jun11}	where there is a tight correlation between the parameter $\mu_{0.32} =\log(\rm M_{\rm star}-0.32~\log(\rm SFR)$) and the metallicity for local galaxies, which is illustrated with the dashed line. In (i) the galaxies are color coded by	their SFR(FIR). The circles show galaxies where $<30\%$ of the emission is excluded due to the presence of shocks and in contrast the squares show galaxies where $> 30\%$ is excluded. Beneath this plot the galaxies are color coded by: (ii) dynamical disturbance quantified by the parameter $\beta$ (Eq.~\ref{eq:beta}) and (iii) the turbulence parameter $V/\sigma$. The triangles identify the two interacting galaxies (CDFS15764 and CDFS15753), while the star symbol indicates the galaxy CDFS11583 where the H$\alpha$ emission only traces half of the galaxy. }
	\label{fig:fmr} 
\end{figure}

The results are illustrated in Fig.~\ref{fig:fmr}, which show the galaxy metallicity as a function of the parameter $\rm \mu_{0.32} = \log{(M_{star})}-0.32\log{(SFR)}$ (where $\rm M_{star}$ is in units of $\rm M_{\odot}$ and SFR in units of $\rm M_{\odot}~yr^{-1}$). According to the results presented in \cite{mannucci_nov10,mannucci_jun11}, \cite{cresci_12} and \cite{stott13}, local and intermediate redshift galaxies follow a tight relation with the parameter $\rm \mu _{0.32}$ (dispersion $\sim 0.07$ dex), which is indicated by the dashed line in Fig.~\ref{fig:fmr}. In Fig.~\ref{fig:fmr} (i), (ii) \& (iii), galaxies in our FIR selected sample at z$\sim$1.5 are color coded by their SFR, $\beta$ parameter and V/$\sigma$ respectively. In Fig.~\ref{fig:fmr} (i) we also indicate with a square symbol galaxies for which more than 30\% of the emission was discarded because affected by AGN ionization or shocks; these objects should be considered with care since the inferred average metallicity may still be subject to some AGN/shock ``pollution''.

Most galaxies follow the FMR observed locally, although with large scatter. There is no clear correlation of the deviations with the dynamical properties of the galaxy ($\beta$) or with the amount of turbulence (V/$\sigma$), supporting the general scenario in which most galaxies follow the FMR up to z$\sim$2, although with large dispersion. However, there is one remarkable exception, CDFS15753, which has a metallicity well in excess of the FMR. This is the strongly star forming satellite galaxy close to the much more massive galaxy CDFS15764, which is instead located on the FMR. Both galaxies are marked with a triangle in Fig.~\ref{fig:fmr} (ii) \& (iii). This result goes in the opposite direction with respect to the expectations of models \cite[e.g.][]{rupke_10}, which predict that the metallicity in interacting systems should decrease as a consequence of the inflow of low metallicity gas from the outer regions towards the central active region (as a consequence of the tidal disruptions and consequent loss of angular momentum of the gas). However, we note that despite being close to the massive companion, the velocity field of CDFS15753 does not look strongly disturbed ($\beta = 0.14$). This system is probably in an early stage of interaction, where the tidal processes expected by models do not yet play a major role. However, the metallicity excess observed in CDFS15753 is in nice agreement with the results obtained locally on the metallicity dependence on environment for satellite galaxies. More specifically, \cite{peng14} have shown that in the local universe satellite galaxies (defined as all galaxies which are members of clusters/groups that are {\it not} the central galaxy) follow a tight correlation between metallicity and galaxy overdensity (a correlation which is independent of stellar mass). The result is interpreted within a scenario in which the IGM is more metal enriched in dense environments and therefore, satellite galaxies accrete gas which is pre-enriched. \cite{peng14} show that this scenario can quantitatively explain the metallicity-environment correlation observed for satellite galaxies. The metallicity of central galaxies instead does not correlate with environment; this is likely because the metallicity of the central, massive galaxy is generally intrinsically high and therefore the metallicity of the inflowing IGM has a much lower effect. CDFS15753 and  CDFS15764 trace an overdense region at z$\sim$1.5 in which CDFS15753 is the low-mass satellite and CDFS15764 is the central massive galaxy. The finding that CDFS15753 is characterized by metallicity higher than the FMR strongly suggests that satellite galaxies in overdense regions at z$\sim$1.5 are subject to the same physical process as local galaxies, i.e. are over-enriched because of accreting IGM gas which is pre-enriched. This implies that the IGM in overdense regions is metal enriched already by z$\sim$1.5. We will discuss further this scenario in the next section.

The other interesting galaxy, which deviates significantly from the FMR is another massive galaxy, CDFS11583, indicated with a star in
Fig.~\ref{fig:fmr}. This galaxy is significantly below the FMR. However, this is also the galaxy for which the SFR, as traced by H$\alpha$ emission, is markedly asymmetric with respect to the (old) stellar continuum emission; more specifically, SFR is only observed on one side of the disc. This signature is likely tracing a recent minor merger event of a low metallicity, gas rich satellite on this massive galaxy; as a consequence, the metallicity is likely reflecting the one of the satellite galaxy, while the inferred mass is the one of the primary galaxy, hence resulting in an offset from the FMR. A similar scenario has been proposed by \cite{yates12}.

\subsection{Gas Fraction} 
\label{sect:fgas}
Another fundamental parameter used to study galaxy evolution at high redshifts is the gas fraction, defined as $M_{\rm{gas}}/(M_{\rm gas}+M_{\rm star})$. Local galaxies are found to have gas fractions of about 5-10\% \cite[e.g.][]{leroy_08, saintonge_jul11}, while for galaxies at higher
redshifts (z$\sim$2) these are found to be much higher, on the order of 30-40\%, \cite[e.g.][]{tacconi_may13,santini14}, revealing a clear evolution of the gas fraction over cosmic time. The Schmidt-Kennicutt (S-K) relation \citep{kennicutt_may98} defines the link between the surface density of star formation and the surface gas density and has been shown to hold out to z$\sim$2.5 \cite[e.g][]{daddi10,genzel10}. In absence of direct tracers of the gas content in high-z galaxies (which require demanding observations at millimetre wavelengths), one can obtain information on the gas surface density by inverting the S-K gas surface density relation ($\Sigma_{\rmn{SFR}} \propto \Sigma_{\rmn{gas}}^{1.4}$). This is the method we employ to compute the mass of gas in each galaxy for which we can resolve the distribution of star formation and combine this with stellar mass from the SED fitting to obtain an estimate of the gas fraction. These values are shown in Table~\ref{tab:properties}. It is interesting to note that for one of these galaxies, CDFS6758, extensive PACS+SPIRE photometric data are available, which enable us to calculate the dust mass. In this case the gas mass can be inferred from the dust-to-gas ratio, as discussed in \cite{magdis11} and \cite{santini14}. The resulting gas mass ($\log (\rm{M_{gas}/M_{\odot}}) = 10.99 \pm 0.27$) is in nice agreement with the value inferred by inverting the S-K relation ($\log (\rm{M_{gas}(SK)/M_\odot)} = 10.69 \pm 0.15$) .

\begin{figure}
	\centering
	\includegraphics[scale=0.5]{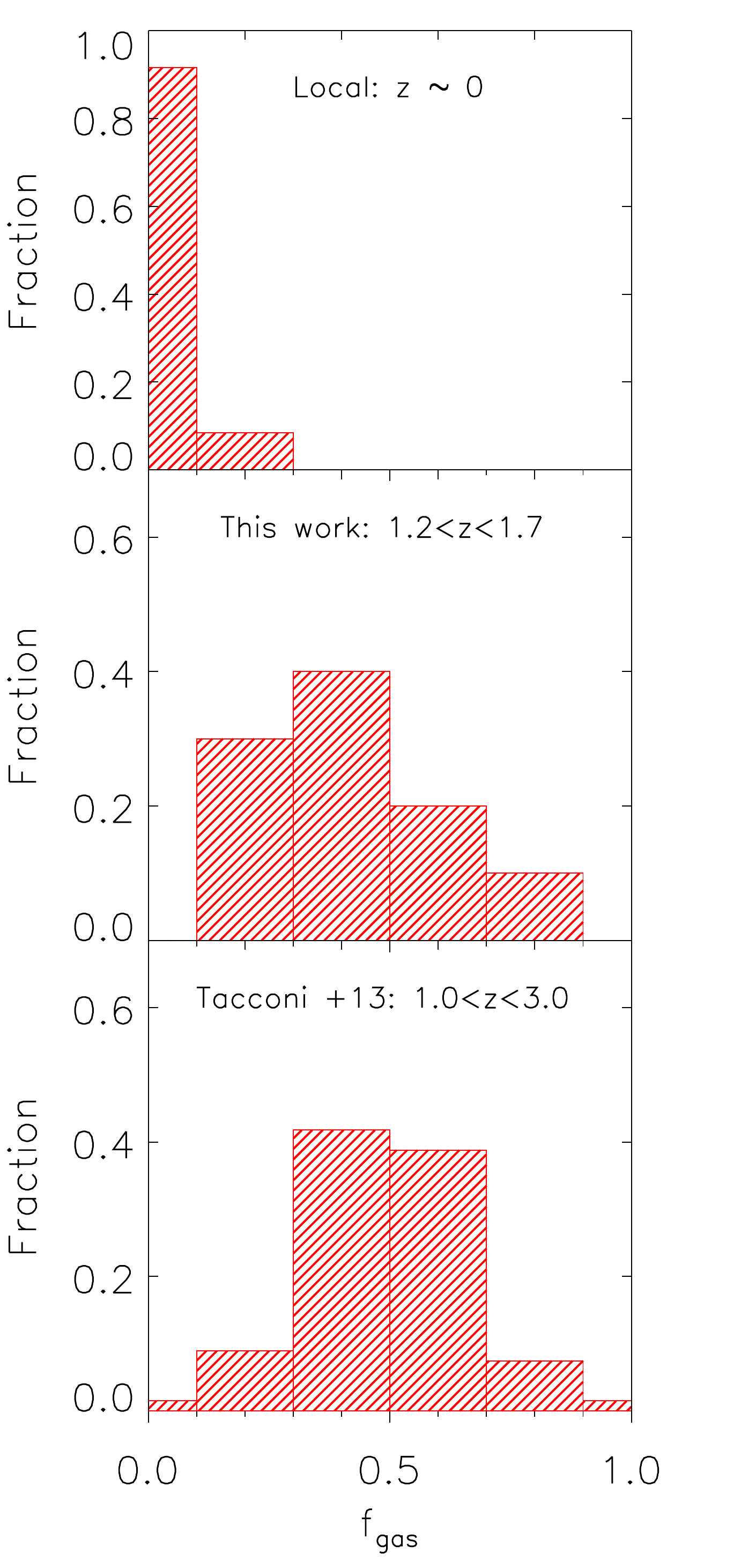} 
	\caption{Gas fractions for different samples of galaxies at different redshifts. Top: gas fraction in local galaxies inferred by \citet{saintonge_jul11} through CO observations. Middle: gas fraction for the galaxies in our sample at z$\sim$1.5, 	obtained by inverting the S-K relation. Bottom: gas fractions obtained by \citet{tacconi_may13} through CO observations of galaxies at 1.2$<$z$<$1.7.} 
	\label{fig:fgas}
\end{figure}

Fig.~\ref{fig:fgas} shows the resulting distribution of gas fractions, which we also compare with two other samples of galaxies. The top panel in
Fig.~\ref{fig:fgas} shows the distribution of gas fractions for a sample containing 250 local galaxies \citep{saintonge_jul11} spanning a mass range similar to our sample for which gas masses were measured through direct CO observations. The bottom panel shows gas fractions inferred from CO observations of galaxies at redshifts similar to ours from \cite{tacconi_may13} (50 galaxies at 1$<$z$<$1.5, 17 at 2$<$z$<$3). Our results agree nicely with those from \cite{tacconi_feb10, tacconi_may13} at similar redshift, especially if taking into account the incompleteness of the Tacconi's sample \cite[see discussion in][]{tacconi_may13}, which would shift the distribution to slightly lower values. We find that our gas fractions for galaxies at redshifts 1.2$<$z$<$1.7 are an order of magnitude higher than that of local galaxies. This is in line with our current understanding that at earlier epochs, where vigorous star formation is still ongoing, galaxies contain a much larger reservoir of gas to fuel such SFR. Whereas in the local Universe we see much lower SFRs indicating that galaxies have little fuel left for star formation.

\begin{figure*}
	\flushleft{(i)}\\
	\centering{ \includegraphics[trim=0cm -0.5cm 0cm 0.cm, clip=true, width=120mm]{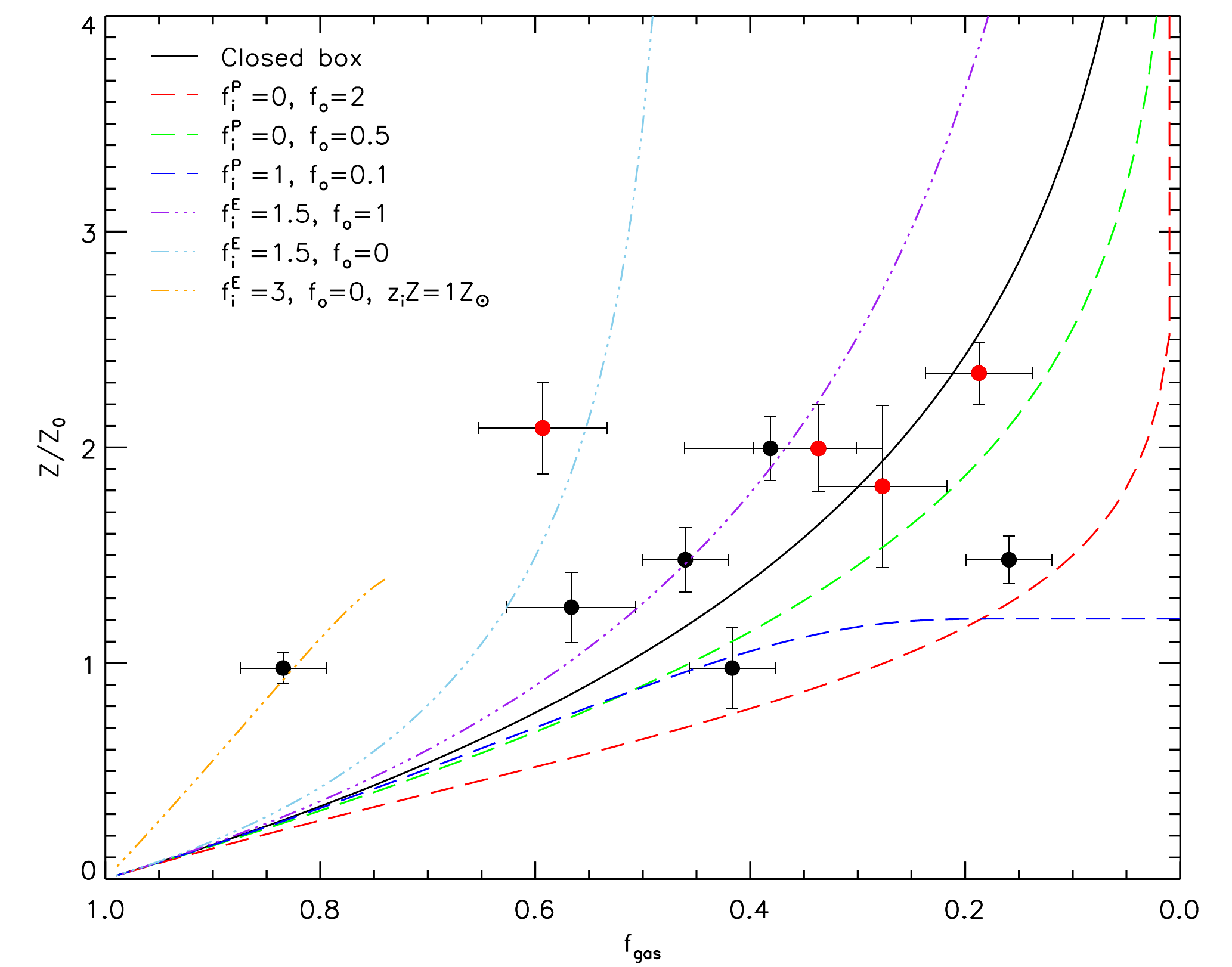} }
	\flushleft{(ii)}\\
	\centering{ \includegraphics[trim=0cm -0.5cm 0cm 0cm, clip=true, width=120mm]{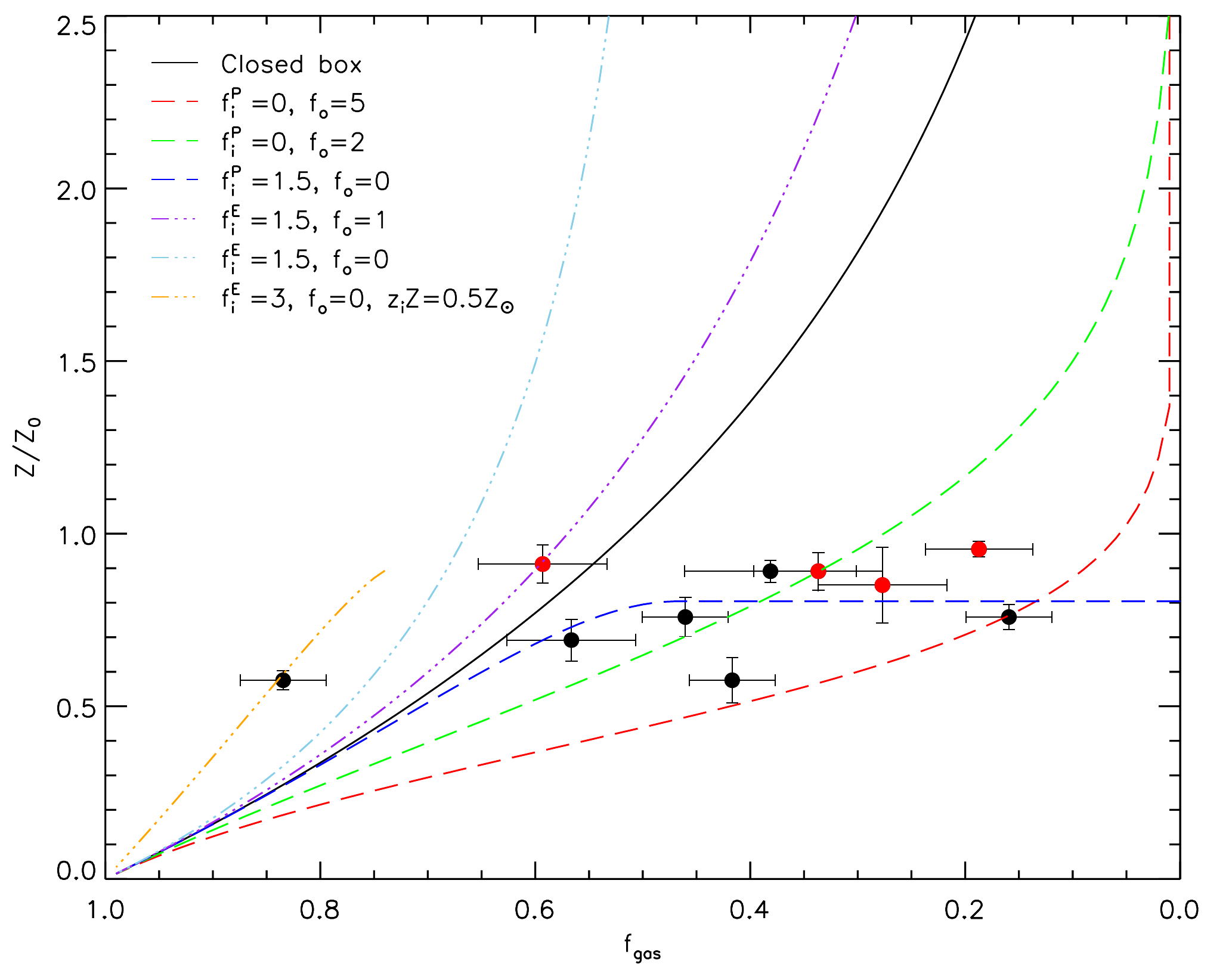} }
	\caption{Metallicity versus gas fraction for the galaxies in our sample. In panel (i) we use the metallicity calibration given in \citet{maiolino_sep08}, while in panel (ii) we use the metallicity given in \citet{pettini_mar04}. In both plots, the red points indicate galaxies where $>30\%$ of the emission is excluded due to the presence of shocks. The closed box model is over-plotted in black as well as the models for inflows and outflows from \citet{erb_feb08}. $f_{i}^{P}$ and $f_{o}$ are given as constant fractions of the star formation, where the `P' superscript denotes inflows of pristine gas. In the models the yield is set at $y=0.019$ as in \citet{erb_feb08}. The violet and light blue lines to the left of the closed box model use the secondary Erb model allowing for enriched inflows $f_{i}^{E}$ and finally the orange shows the same model but derived for enriched inflows at a fraction of solar metallicity $Z_\odot$. }
	\label{fig:fgasZ}
\end{figure*}

As discussed in \cite{erb_jun06, mannucci_oct09, troncoso_13}, the gas fraction provides a crucial piece of information to interpret the metal content in galaxies and, in particular allows us to disentangle the deviations from the closed-box case and the contribution of outflows and inflows. Fig.~\ref{fig:fgasZ} (i) shows the metallicity of each galaxy (with respect to solar) plotted against the gas fraction inferred from inverting the S-K relation. We also plot a series of models from \cite{erb_feb08} indicating different chemical evolutionary scenarios. Curves below the closed box show different models with various levels of pristine gas inflow at a rate $\rm f_i \times SFR$ and enriched outflows at a rate $\rm f_o \times SFR$, where the metallicity of the outflow is assumed to be the same as the host galaxy (see the legend for the specific values of $\rm f_i$ and $\rm f_o$ associated with each curve). The first two curves above the closed box (violet and light blue) are instead the case in which the inflowing gas is pre-enriched to the same value as the gas in the host galaxy. These are meant to approximate the case of galactic fountains in which the outer star forming regions are enriched with gas ejected from the more metal-rich central regions. The leftmost line shows the case of enriched inflows with fixed metallicity, meant to represent inflow of IGM gas pre-enriched by other galaxies.

The closed box model (solid black line) is inadequate to reproduce most of our observations as only two galaxy lies within the error bars of this relation. Two of the galaxies are found to the right of the closed box model, which indicate the presence of pristine inflows and enriched outflows of gas at constant rates between 0 -- 1 and 0.1 -- 2 of the total SFR respectively (models for such rates are shown as green, blue, red dashed lines). This result is similar to other high-z galaxies \cite[e.g.][]{mannucci_oct09, troncoso_13} and in line with other evidences that high-z galaxies are characterized by substantial outflows and inflows of gas. However, we also find that six galaxies lie to the left of the closed box model, which instead implies not only enriched outflows, but also enriched inflows of gas. Most of these six galaxies can be interpreted within the context of the ``fountain scenario'' in which enriched outflows from the galaxy centre fall back in and enrich the outer regions. Indeed, in support of this scenario, we do find that the higher gas fractions and metallicity values do lie on the edges of these galaxies.

One final noticeable and very important feature of Fig.~\ref{fig:fgasZ} (i) is the galaxy lying furthest to the left, which so far cannot be explained by the combination of models previously detailed. This represents the interacting satellite galaxy CDFS15753, which was previously found to be more metal rich than the FMR. As previously mentioned, the metallicity ``excess" in this galaxy can be interpreted within the context of the metallicity-environment relation \citep{peng14}, according to which satellite galaxies tend to have higher metallicity in dense environments as a consequence of the IGM enriched by the central massive galaxy and by the other galaxies. The result obtained for CDFS15753 implies that this IGM enrichment process was already advanced by z$\sim$1.5. To quantitatively test this scenario we have adapted the inflow models, as discussed above, to allow for a fixed metallicity of the inflowing gas. To reproduce the gas fraction and metallicity observed in CDFS15753 a solar metallicity of the inflowing gas is needed (curves orange), which is close to the metallicity of the massive central galaxy CDFS15764 of which CDFS15753 is the satellite. This would imply a nearly complete transfusion of the outflow from CDFS15764 in to CDFS15753 with little dilution on the way. This may sound somewhat extreme, but not completely unreasonable given the small distance between the two systems. Also, one should take into
account that the star-formation driven outflowing gas is generally more metal enriched than the gas in the host galaxy, as a consequence of metals freshly produced by the latest generation of supernovae \citep{ranalli08}. 

Another possible interpretation of the location of the galaxies above the closed box in Fig.~\ref{fig:fgasZ} (i) could be that the metallicities have been overestimated by the adopted calibration, i.e. \cite{maiolino_sep08}. At high metallicities the latter calibration is based on photoionization models, which tend to give metallicities higher than those inferred from the $T_\rmn{e}$ method. Recently, \cite{dopita13} have however shown that when using the K-distribution in the ionized gas, the resulting metallicities in nearby galaxies are more in agreement with the photoionization models than with the classical $T_\rmn{e}$ measurements, implying that the calibration in \cite{maiolino_sep08} may be appropriate. However, in Fig.~\ref{fig:fgasZ} (ii) we also illustrate the results when using the calibration given in \cite{pettini_mar04}, which are calibrated only with the $T_e$ method. In this case most of the galaxies lie below the closed box model (i.e. consistent with outflows and pristine inflows). However, CDFS15753 is still well above the closed box line even with this calibration, hence requiring enriched inflow, although with a metallicity lower than in Fig.~\ref{fig:fgasZ} (i).

Finally, yet another possibility is that despite our careful selection, some of these galaxies are still affected by shocks or by some AGN contribution. However in the case of CDFS15753 this effect would have to be quite dramatic and it is also unlikely given that this is a low mass galaxy and therefore unlikely to host a powerful AGN.

\section{Conclusions}

We have presented near-IR integral field spectroscopy of 12 star forming galaxies at z$\sim$1.5 detected by Herschel at 110--160 $\mu$m in GOODS-South, and whose redshifts have been accurately selected such that the optical nebular lines avoid strong OH sky lines. All of these galaxies benefit from the extensive multi-wavelength data available in this field, especially multi-band imaging from HST.

We focus on the redshifted H$\alpha$ and [NII] lines (redshifted into the H-band), since [OIII] and H$\beta$ are marginally detected only in one galaxy.

All of the selected galaxies are on the ``Main Sequence'' at z$\sim$1.5, except for the least massive galaxy,  which is well above the Main Sequence and which is the companion of the most massive galaxy in our sample.

We obtain a number of interesting results on the distribution of star formation, morphology, dynamics and metallicity, which shed light on the formation processes and observability of these star forming galaxies. The main results are summarized in the following.

\begin{itemize}
\item The distribution of star formation as traced by H$\alpha$ is generally completely different from the rest-frame UV morphology, traced by the HST V-band observations, which should also trace star formation in absence of extinction. In many cases the rest-frame UV emission is not detected despite the deep HST sensitivity. The marked differences between H$\alpha$ and UV emission and morphology are certainly associated with dust extinction and warn about the use of UV rest-frame tracers of star formation, even in Main Sequence galaxies, at high-z.

\item The morphology of star formation is also generally quite different with respect to the distribution of the old stellar population traced by the HST near-IR (H-band) continuum images. In particular, in some cases star formation extends beyond the stellar disc. In some galaxies star formation, as traced by H$\alpha$, is only present in some regions of the stellar disc; this could be the effect of minor mergers that boost star formation only in some sub-region of the galactic disc and it warns about the use of H$\alpha$ and other star formation rate tracers as a tool to infer a global view of the morphology and dynamics of galaxies. We mention that such morphological differences between H$\alpha$ and continuum of the old stellar population are unlikely due to dust extinction effects since observed at the same wavelength.

\item We find that most of these galaxies are characterized by a regular rotation curve. We quantify the presence of dynamical disturbances with a
parameter ($\beta$) that measure the average deviations from a regular rotation pattern, relative to the maximum rotational velocity. We find that most of the galaxies have small deviations from the rotation pattern, of the order of 10\%--20\% ($\beta \sim 0.10-0.15$) and only a few up to $\sim$30\%, suggesting the presence of minor merging, but are much lower than the values observed in major mergers (which typically have $\beta \sim 0.4-0.7$).

\item We find that most of these rotating discs are highly turbulent, with $\rm V/\sigma$ as low as 0.5, a feature found also in several other galaxies at similar redshifts. We find a highly significant anti-correlation between $\rm V/\sigma$ and the parameter $\beta$, suggesting that turbulence is enhanced by galaxy weak interactions and minor mergers.

\item The inferred dynamical masses are generally significantly larger than the stellar masses. Most of the excess of dynamical mass is likely associated with the large gas mass in these systems, which we have inferred by inverting the S-K relation. Yet, in a few galaxies even the large gas masses cannot account for the large dynamical mass, implying that the is some significant contribution of dark matter in the region probed by our dynamical tracers.

\item We have inferred the metallicity of the gas using [NII]/H$\alpha$ as a tracer. We find that, although with large scatter, most of the galaxies in our sample follow the FMR. The two most remarkable exceptions are the lowest mass galaxy and a high mass galaxy in the sample. The least massive galaxy has a metallicity significantly higher than expected from the FMR. This galaxy is in small group of galaxies (whose central massive galaxy is also part of our sample). The finding is in line with observations of local galaxies, whose metallicity correlate with the overdensity of galaxies, which has been ascribed to inflows of enriched gas in overdense regions. The finding that this effect is in place also at z$\sim$1.5 suggests that the IGM in overdense regions is already chemically enriched at these early epochs. The low metallicity of one of the most massive galaxies can instead be interpreted in terms of recent minor merging event with a low metallicity gas rich satellite.

\item We compare the metallicity of the galaxies with their gas fraction. The location of the galaxies on the metallicity--$\rm f_{gas}$ diagram can be explained in terms of a combination of inflows and outflows. The detailed interpretation (and the implied relative role of outflows and inflows) depends on the adopted metallicity calibration. However, regarding the lowest mass galaxy discussed above, a solid result is that the properties of this system can only be explained in terms of massive inflows of metal enriched gas, supporting the scenario presented above in which this galaxy has been accreting gas from the IGM which has been pre-enriched by other galaxies (probably mostly by the central massive galaxy) in this overdense region.

\end{itemize}

\section{Acknowledgement}
We are grateful to S. Arribas, S. Cazzoli and E. Bellocchi for providing us the cubes of the local ULIRGs used for the test in the Appendix. Based on observations made with ESO telescopes at the La Silla Paranal Observatory under programme ID 086.A-0518(A).

\bibliographystyle{mn2e}
\setlength{\labelwidth}{0pt}
\bibliography{myrefs}
\newpage

\appendix
\section{Comparison with local Major Mergers}
\label{appen1}
For our sample of galaxies at z$\sim$1.5 observed with  SINFONI we have used the $\beta$ parameter ($\beta = |V_{\rmn{residuals}}| /\Delta V$) to quantify how well we can fit the galaxy's velocity field with a rotating disc model. Low values of $\beta$ indicate a good fit, while high values of $\beta$ ($> 0.2$) indicate that the system is undergoing a major merger ot disturbances. 

\begin{figure}
	\centering
	\includegraphics[width=84mm]{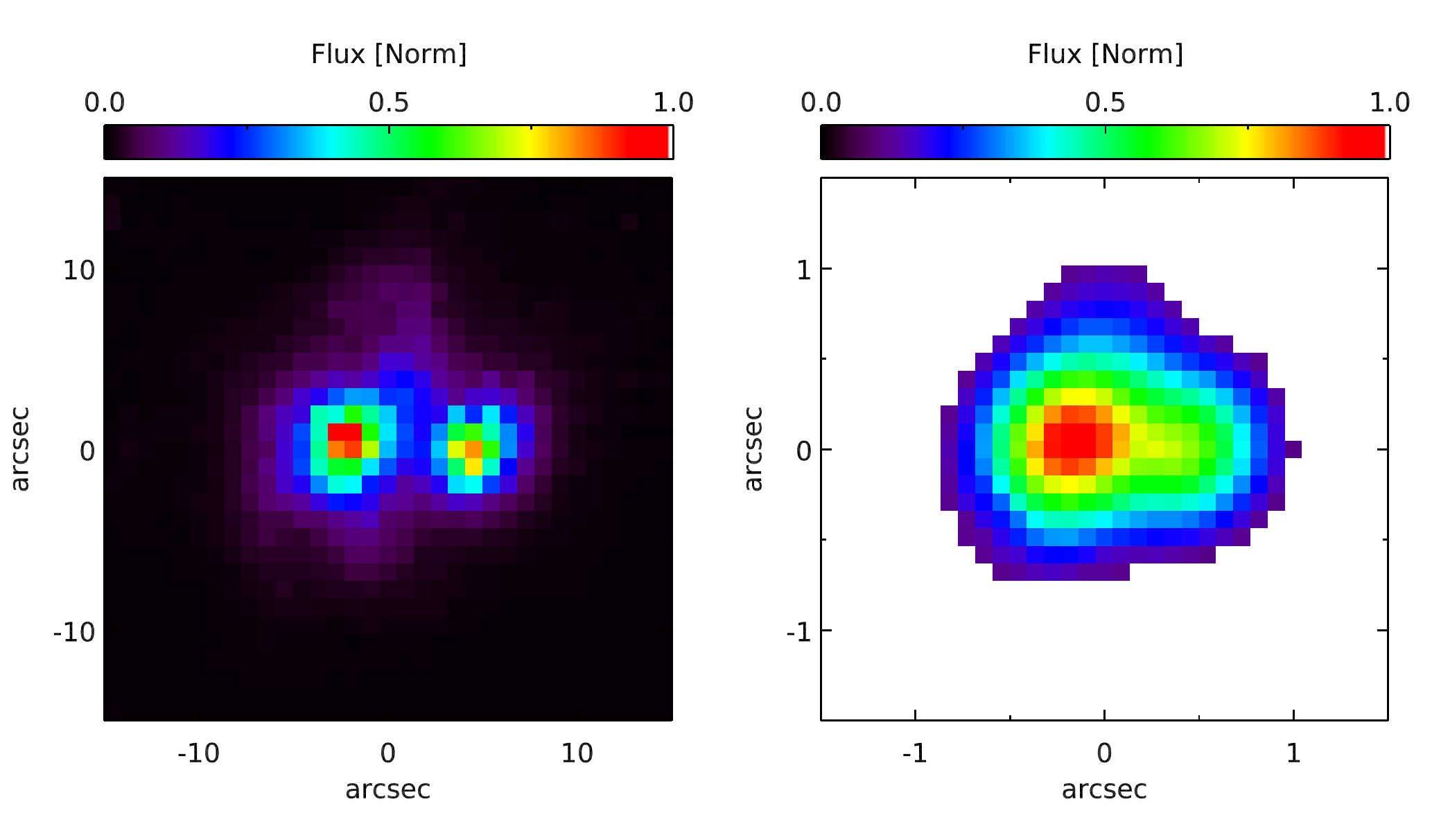} 
	\caption{An example of a local major merger taken from \citet{bellocchi13}. On the left is the original image (normalised to maximum H$\alpha$ flux) where the two separate galaxies are clearly distinguishable, whereas on the right the image has been smoothed to simulate the effect of the SINFONI beam causing the two galaxies to appear as one extended system. } 
	\label{appen:fig2}
\end{figure}

\begin{figure}
	\centering
	\includegraphics[width=88mm]{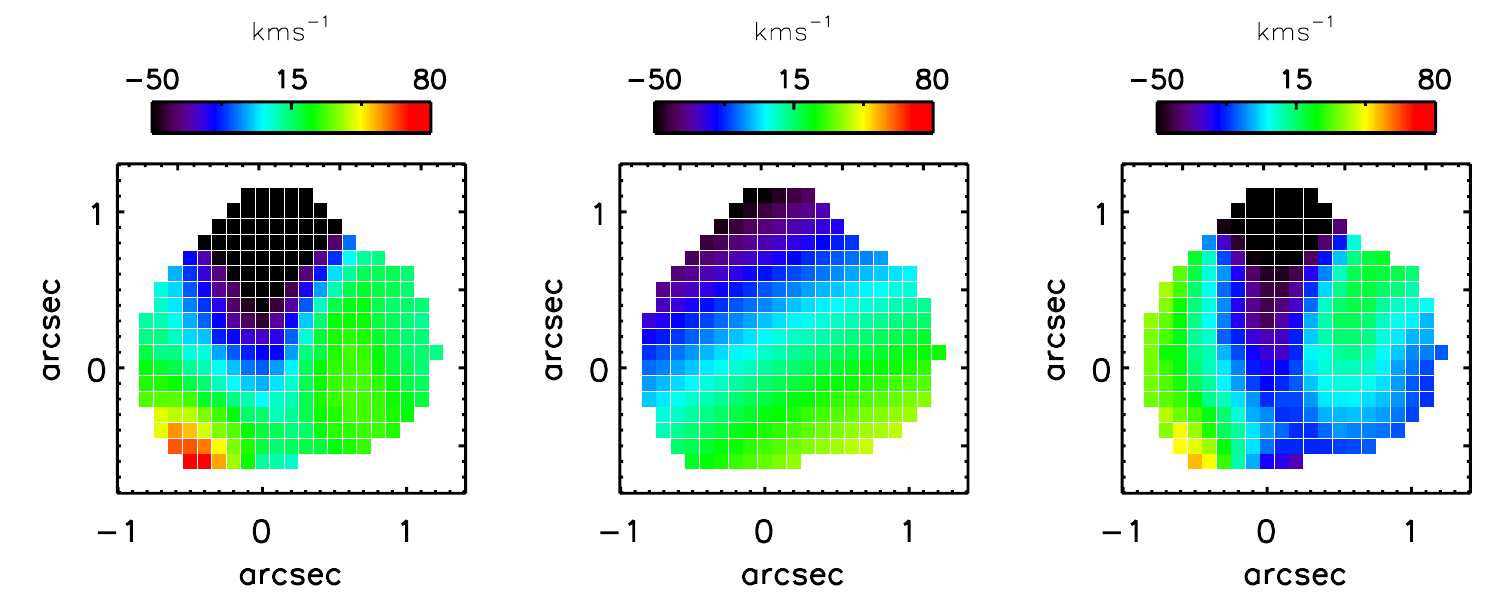} 
	\caption{Velocity model for a local major merger. From left to right the panels show the observed, modelled and residual velocity field.} 
	\label{appen:fig3}
\end{figure}

To further quantify this difference we have checked what would be the $\beta$ value of local major mergers if they were observed with our resolution at z$\sim$1.5. We have taken a small sample of six local galaxies, known to be undergoing a major merger, extracted from the sample of \cite{bellocchi13}, which includes local LIRGs and ULIRGs observed with the VLT optical IFU, focusing on H$\alpha$+[NII]. We have rescaled these galaxies to the angular resolution of our galaxies, re-sampled the cube with our spatial and spectral sampling and spatially smoothed to our PSF.
We have applied the same fitting technique to the H$\alpha$ and [NII] emission lines with three Gaussian's (see Section~\ref{sect:data}) and then attempted to fit the velocity field with a rotating disc model as in Section~\ref{sect:dynamics}. An example of the H$\alpha$ map of these galaxies is shown in Fig.~\ref{appen:fig2}, in which the left panel shows the original H$\alpha$ image (in which the two merging galaxies are still resolved), while the right panel shows the image re-sampled and smoothed to our resolution, in which the system looks like a single galaxy. Yet, Fig.~\ref{appen:fig3} shows the observed velocity field, the rotating disc fit and the residuals, from which it is clear that the dynamics of the system is highly irregular and not properly modelled by simple rotation. From the model and residuals we then obtain the $\beta$ parameter as defined in Eq.~\ref{eq:beta}.  

\begin{figure}
	\includegraphics[trim=0.9cm 0.4cm 0.9cm 0.5cm, clip=true,width=90mm]{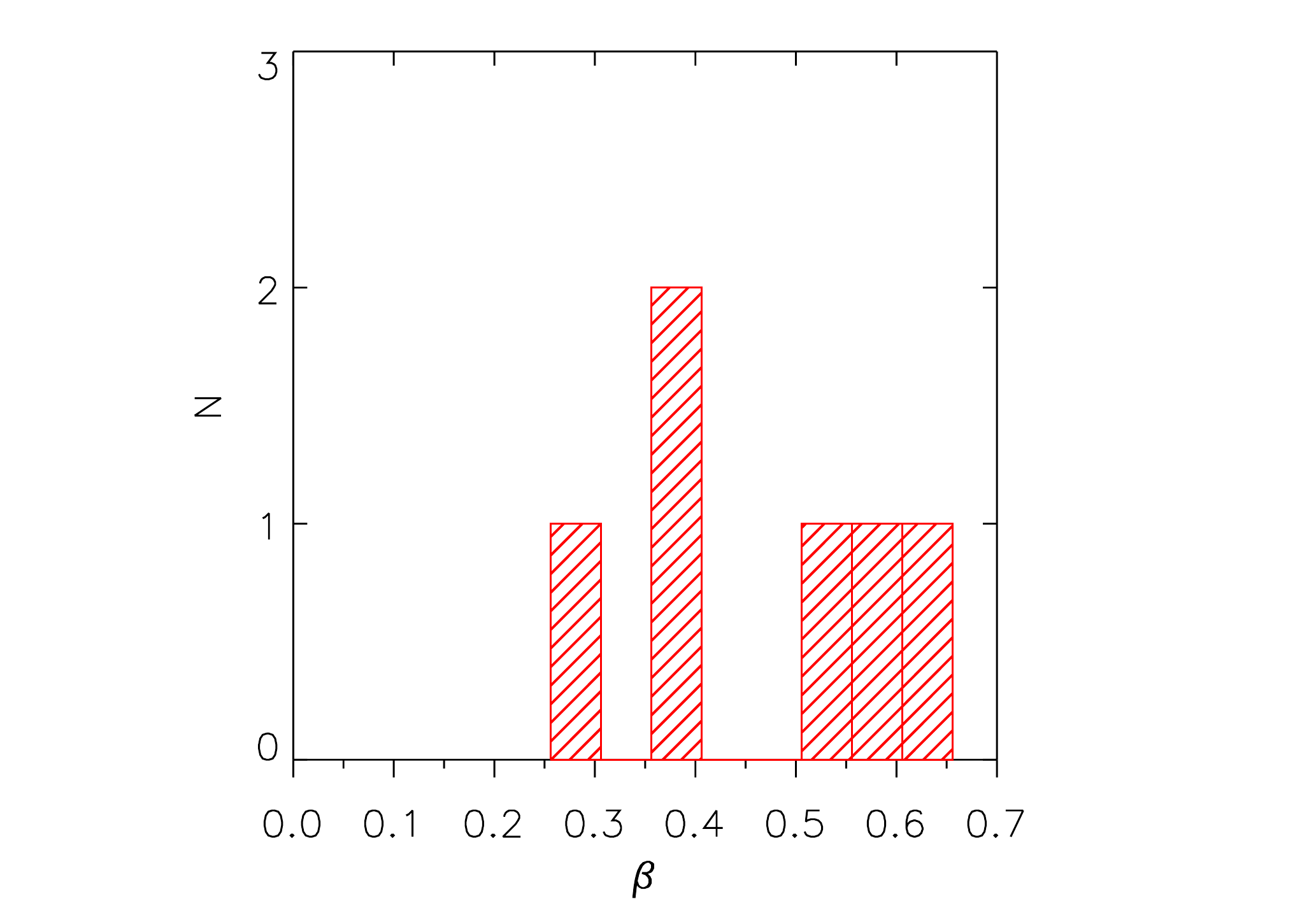} 
	\caption{The distribution of the dynamical disturbance index $\beta$ (Eq.~\ref{eq:beta}) which quantifies the level of dynamical disturbance for a small sample of six local galaxies in major mergers.} 
	\label{appen:fig1}
\end{figure}

The resulting distribution of $\beta$ is shown in Fig.~\ref{appen:fig1} In general we find, for these major mergers, high values of $\beta$, more specifically: $0.37<\beta <0.62$. Our SINFONI galaxies have much lower values of $\beta$ in most cases ($0.02<\beta<0.17$).
\newpage

\newpage

\clearpage

\section{Supplementary Material}

\begin{table}
	\caption{Central position (0,0) for the maps in Fig.~\ref{fig:sfr2}, \ref{fig:vel2}, \ref{fig:metal_grad2}.} 
	\label{tab:pos} 
	\begin{tabular}{@{}p{2cm}p{2cm}p{2cm}@{}}  
	\hline
	Galaxy  & R.A. $^\circ$ & Dec. $^\circ$\\
	&  (J2000)&  (J2000)\\
	\hline
	CDFS13844 & 53.0481 & -27.7371\\
	CDFS13784 & 53.1395 & -27.7382\\
	CDFS6758 & 53.0714 & -27.8341\\
	CDFS14224 & 53.1323 & -27.7323\\
	CDFS16485 & 53.1293 & -27.7014 \\
	CDFS2780 & 53.2092 & -27.8885\\
	CDFS11583 & 53.1642 & -27.7698\\
	CDFS15753 & 53.0910 & -27.7123\\
	CDFS15764 & 53.0905 & -27.7123\\
	CDFS10299 & 53.0750 & -27.7860\\
	\hline
	\end{tabular}\\
\end{table}

\begin{figure*}
	CDFS13844\\
	\includegraphics[scale=0.7]{sfr0}

	CDFS13784\\
	\includegraphics[scale=0.7]{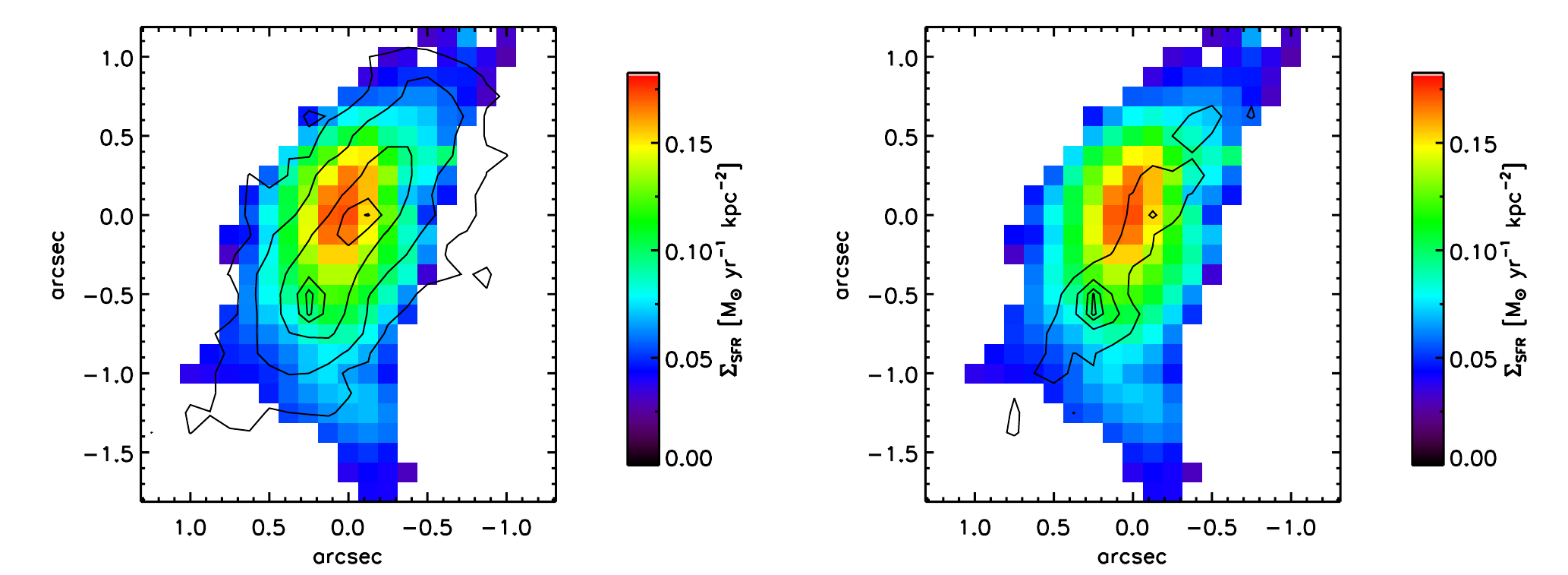}

	CDFS6758\\
	\includegraphics[scale=0.7]{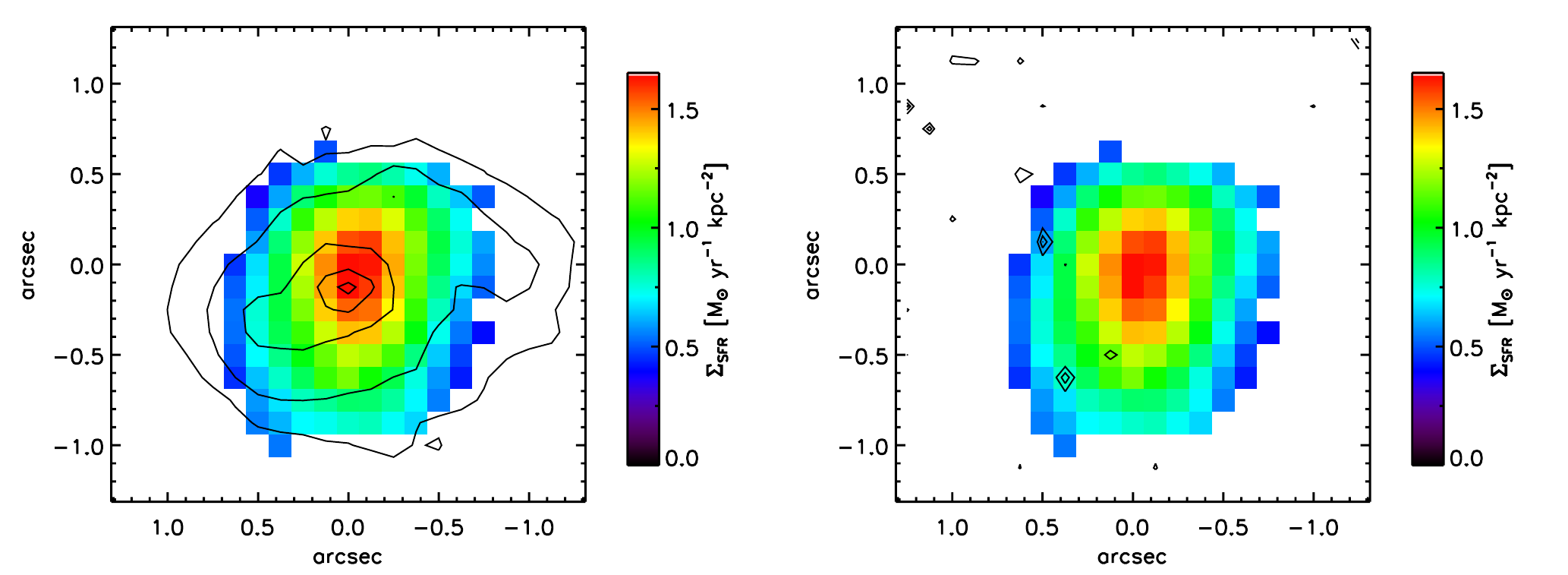}
	\caption{Star formation rate surface density maps inferred from the H$\rm{\alpha}$ surface brightness distribution, scaled to match the total
	SFR inferred from the FIR. The $\rm \Sigma _{SFR}$ maps are compared with ({\it left}) the near-IR (160W) HST images (contours) and ({\it right}) the optical V-band HST image. See Table~\ref{tab:pos} for central coordinates. }
	\label{fig:sfr2}
\end{figure*}

\begin{figure*}
	\contcaption{}
	CDFS14224\\
	\includegraphics[scale=0.7]{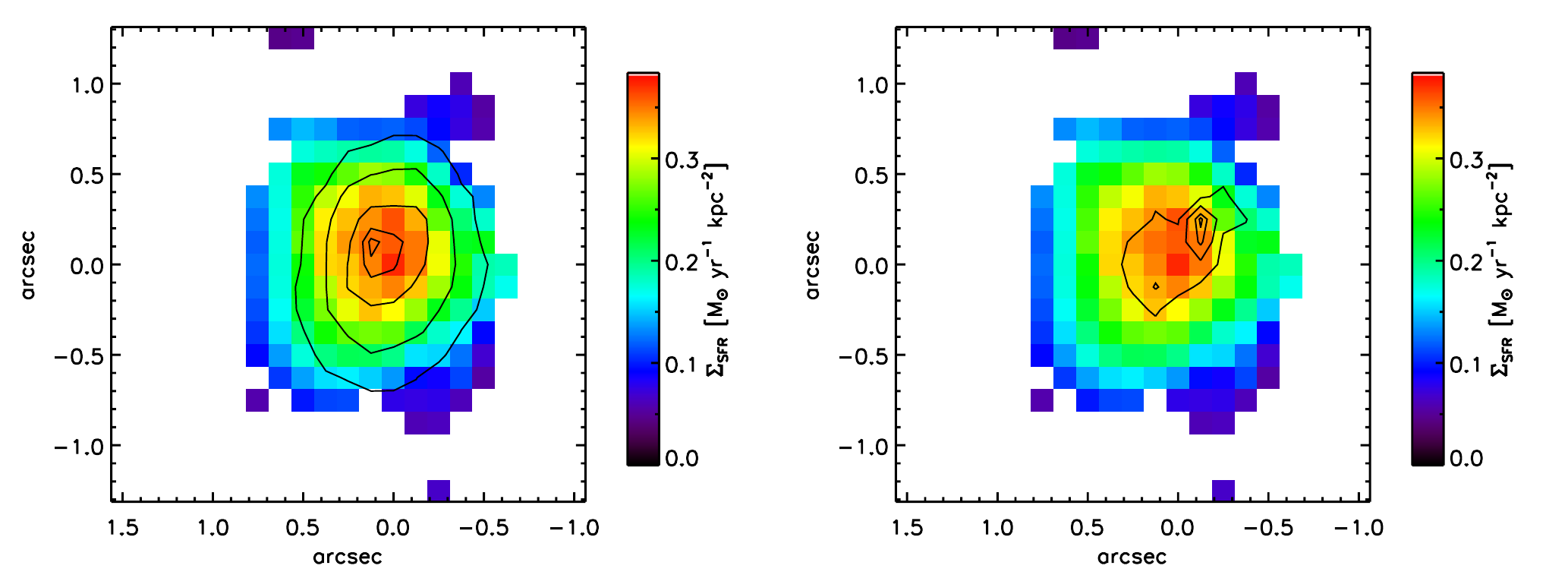}
	
	CDFS16485\\
	\includegraphics[scale=0.7]{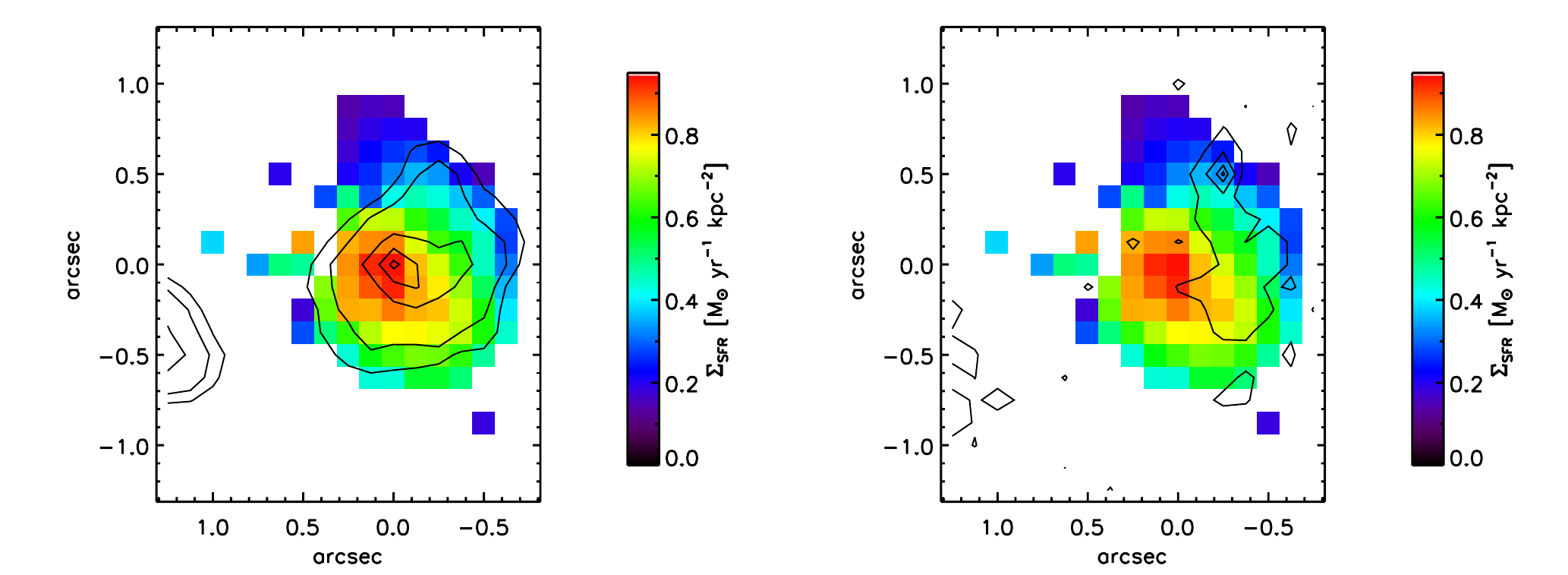}

	CDFS2780\\
	\includegraphics[scale=0.7]{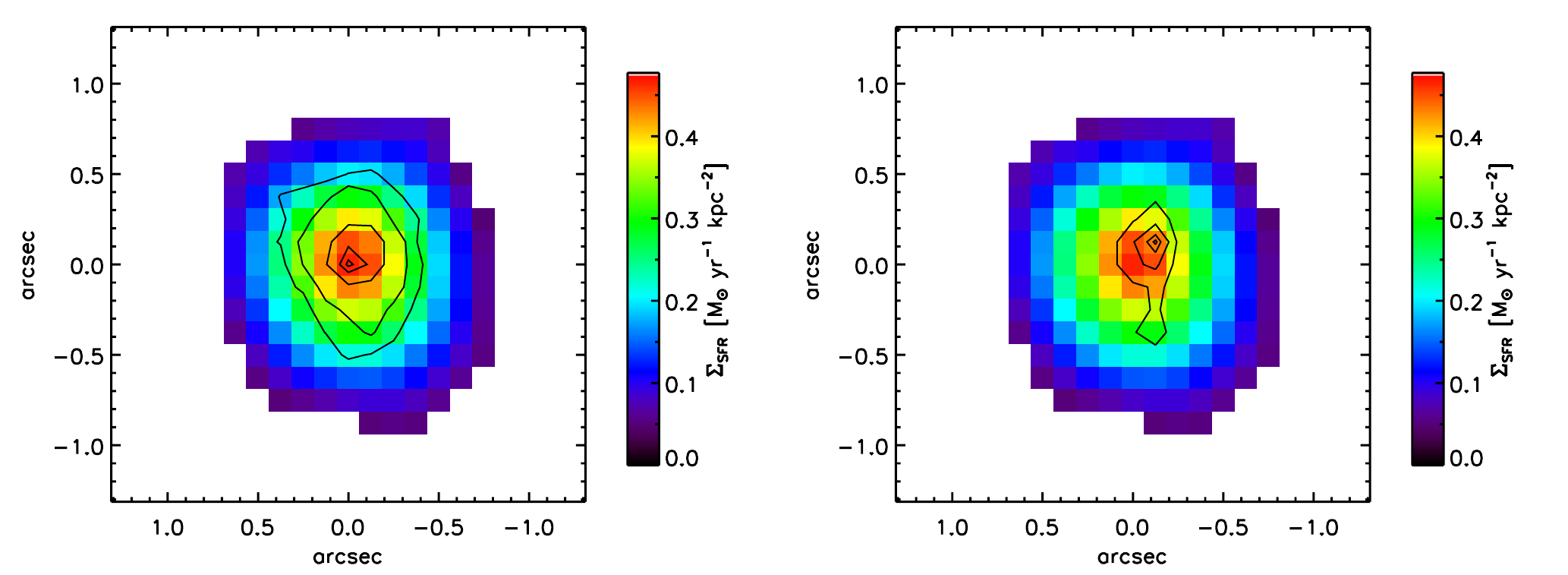}

	CDFS11583\\
	\includegraphics[scale=0.7]{sfr6}

\end{figure*}

\begin{figure*}
	\contcaption{}
	CDFS15753\\
	\includegraphics[scale=0.7]{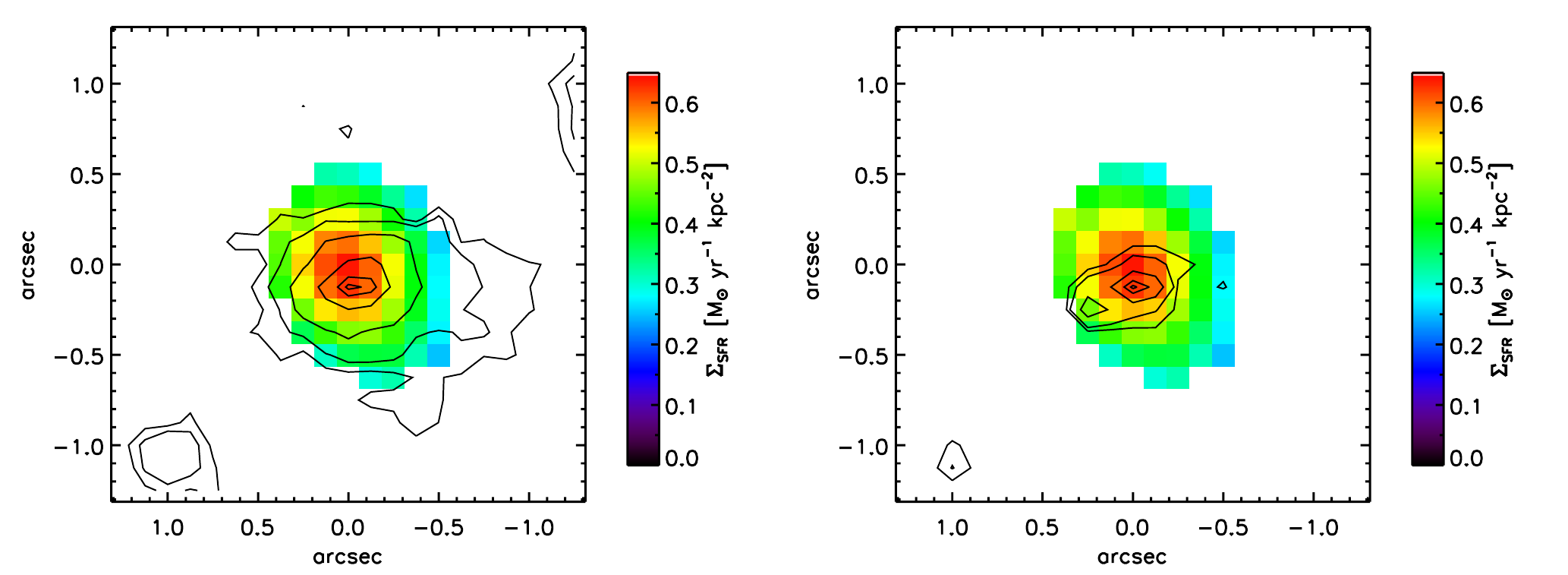}
	
	CDFS15764\\
	\includegraphics[scale=0.7]{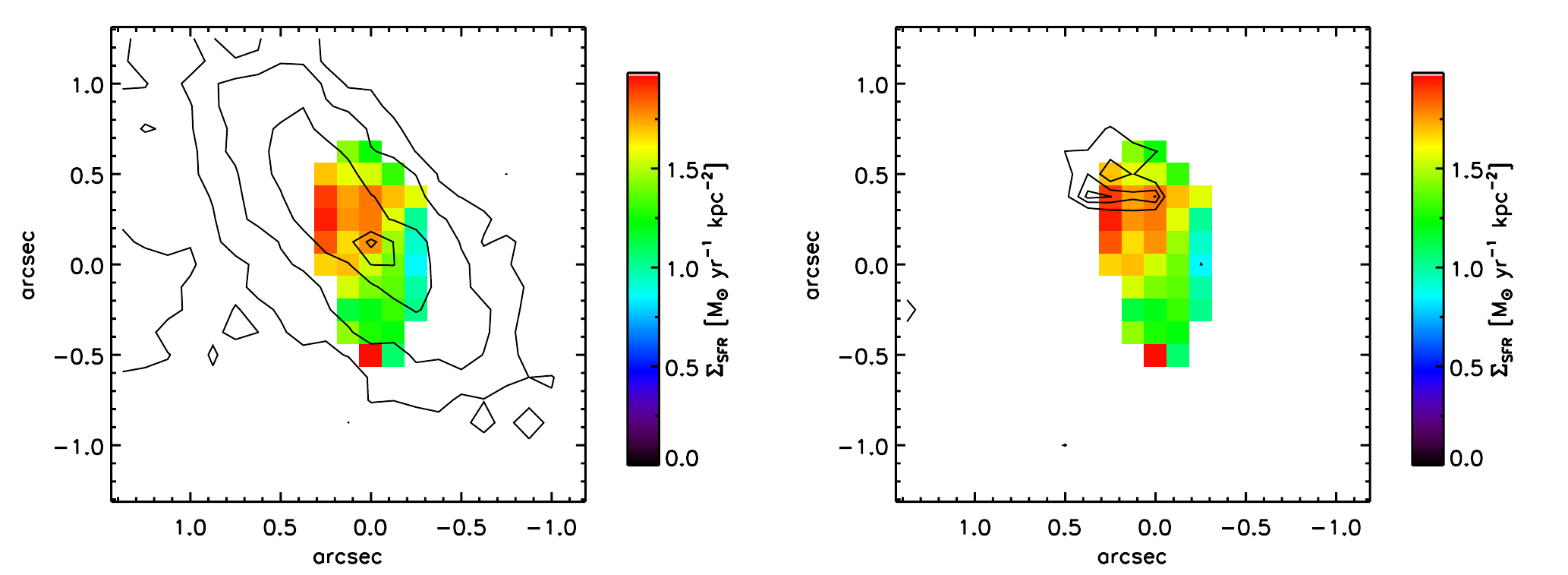}

	CDFS10299\\
	\includegraphics[scale=0.7]{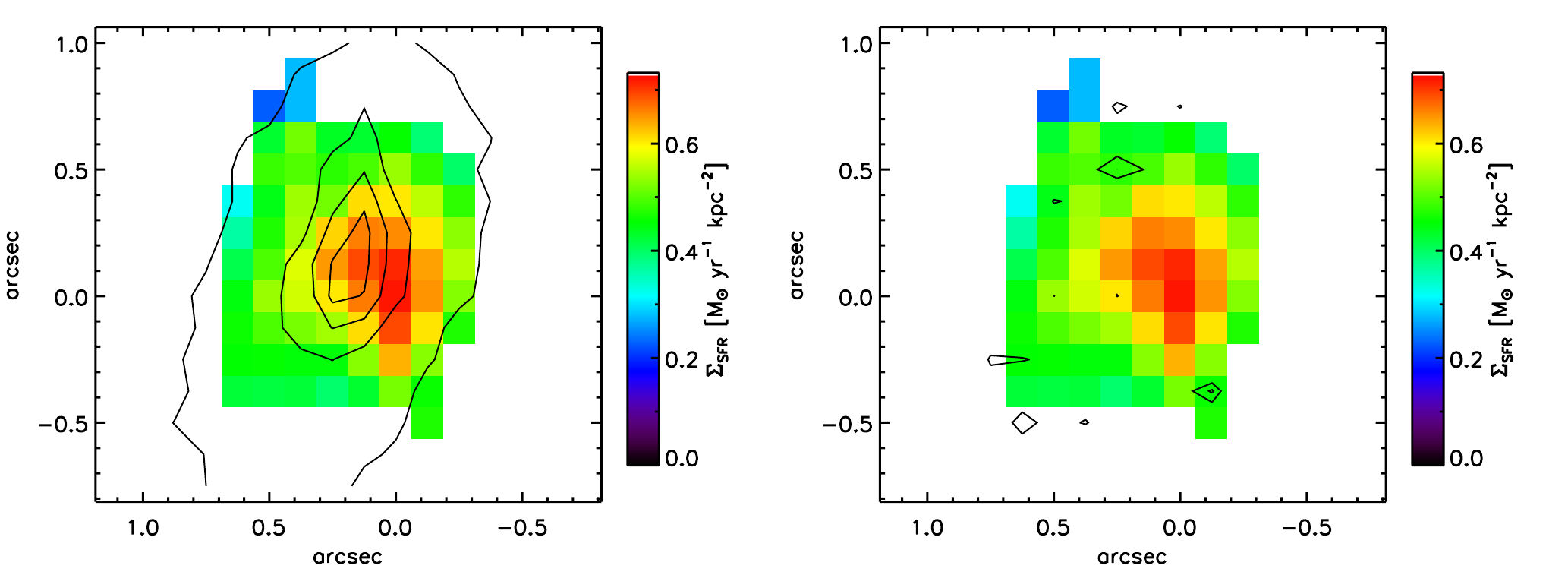}
\end{figure*}
\clearpage

\begin{figure*}
	\textbf{CDFS13844}
	\includegraphics[scale=0.8]{model_13844}
	\vspace{0.5cm}

	\textbf{CDFS13784}
	\includegraphics[scale=0.8]{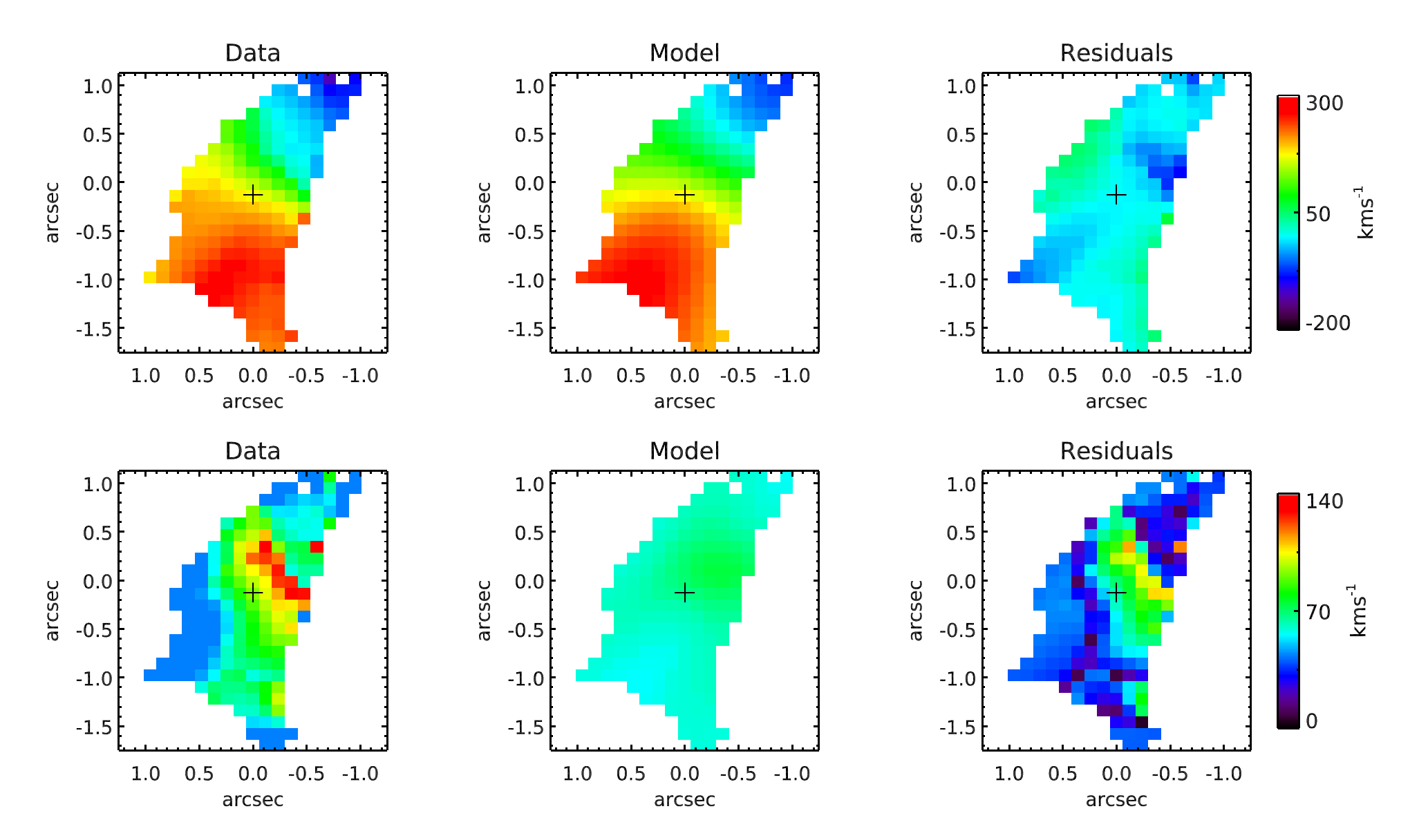}

	\caption{Velocity field and associated modelling inferred from the H$\alpha$+[NII] lines (using a signal-to-noise cut of 4). For each galaxy the top row shows the velocity field, where the left panel shows the observed velocity pattern, the central panel shows the best fit with a rotating disc model and the right panel shows the residuals. The bottom row shows the velocity dispersion, where the left panel shows the observed velocity dispersion, the central panel the velocity dispersion due to the PSF smearing of the velocity field and the right panel shows the intrinsic velocity dispersion obtained by the difference in quadrature of the previous two quantities. The cross marks the position of the peak H$\alpha$ emission,  see Table~\ref{tab:pos} for central coordinates. }	
	\label{fig:vel2}
\end{figure*}

\begin{figure*}
	\contcaption{}
	\textbf{CDFS6758}
	\includegraphics[scale=0.8]{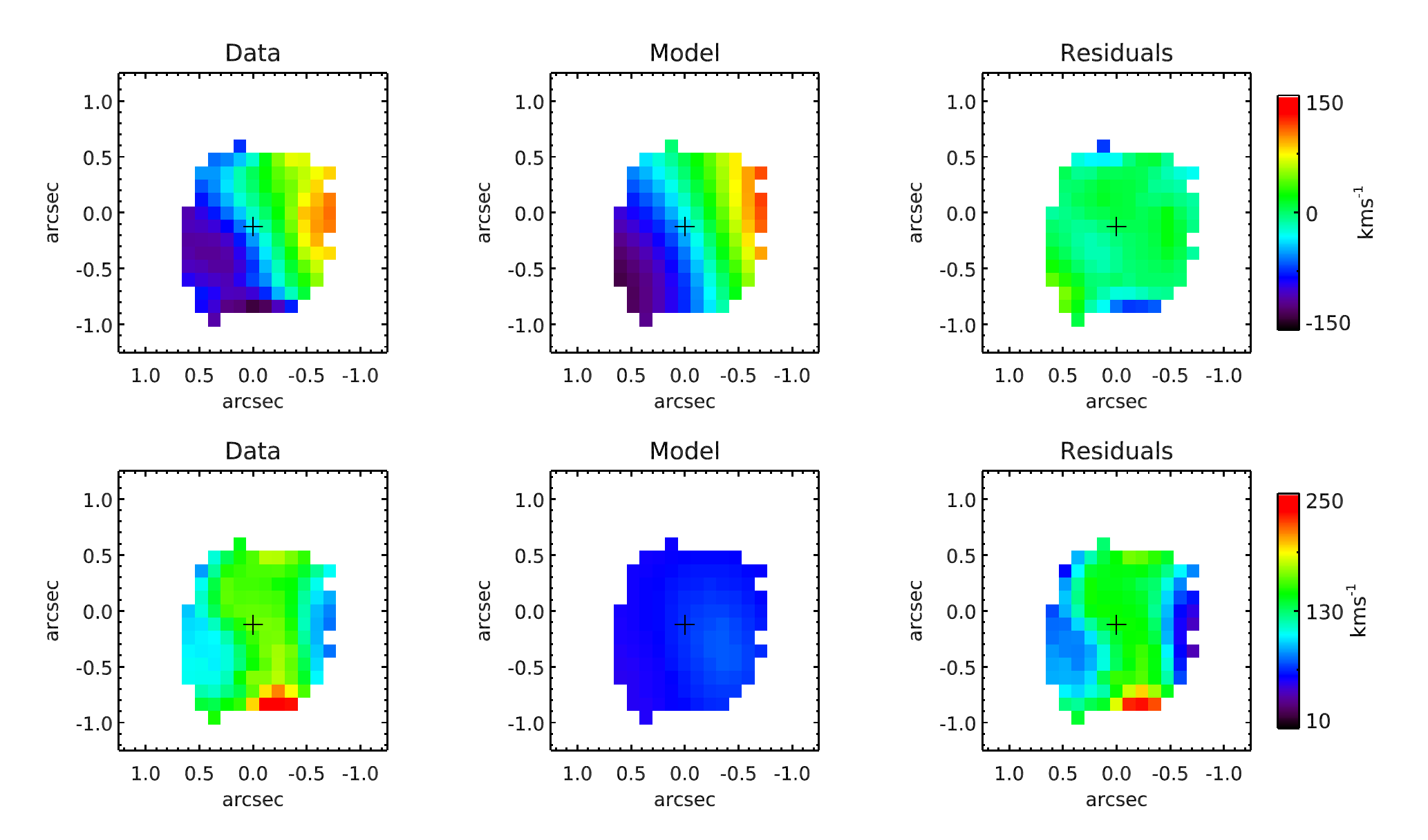}
	\vspace{0.5cm}

	\textbf{CDFS14224}
	\includegraphics[scale=0.8]{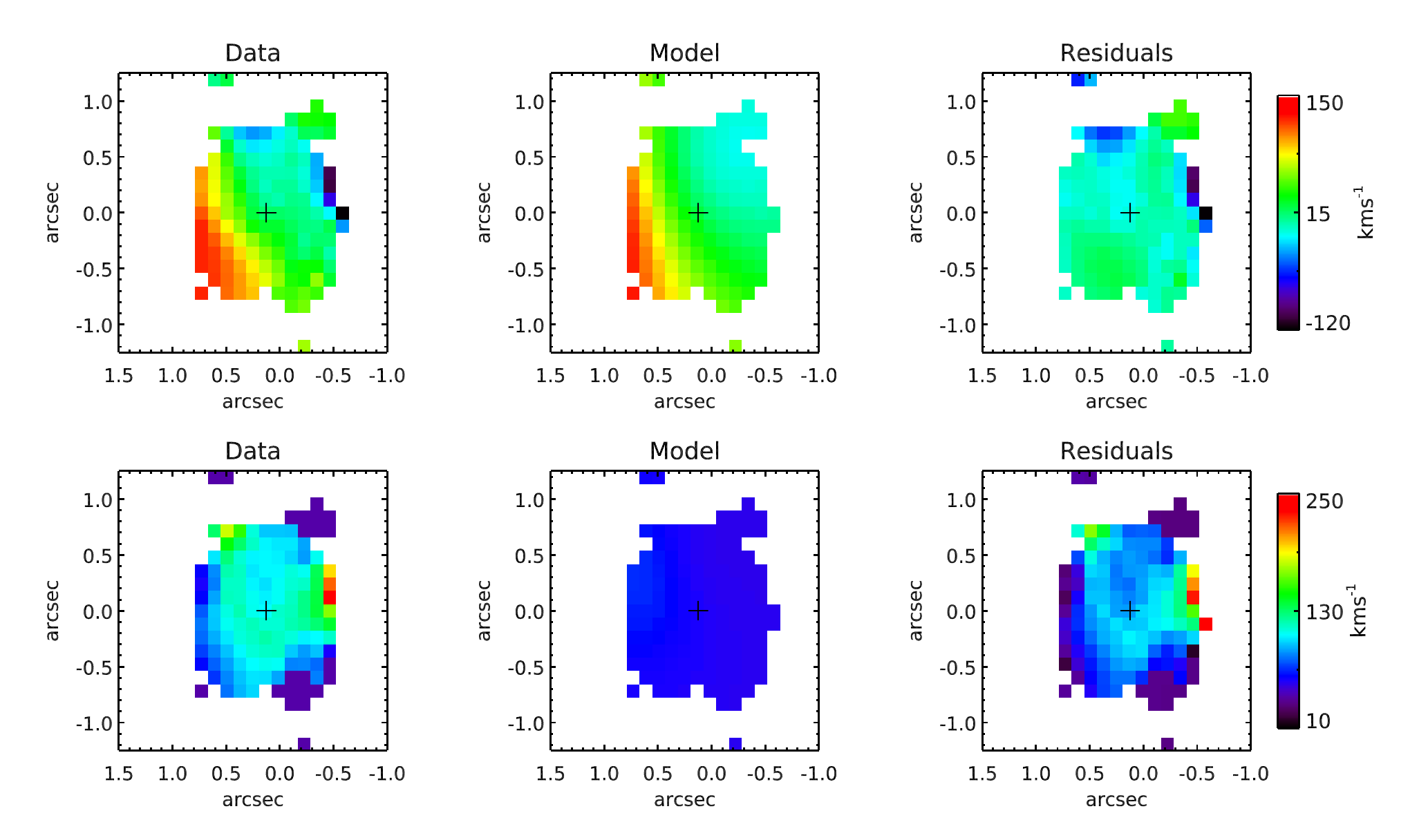}
\end{figure*}

\begin{figure*}
	\contcaption{}
	\textbf{CDFS16485}
	\includegraphics[scale=0.8]{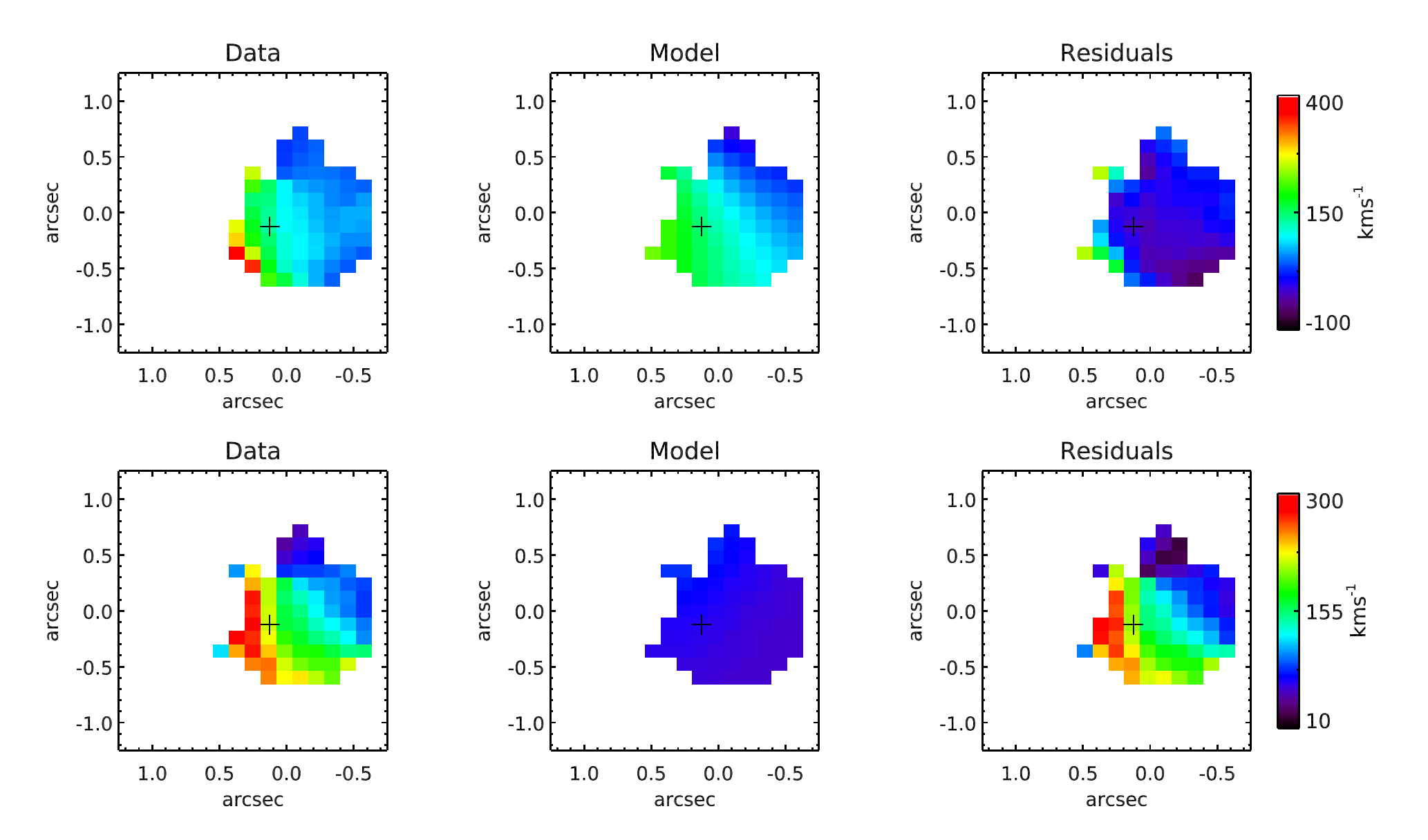}
	\vspace{0.5cm}

	\textbf{CDFS2780}
	\includegraphics[scale=0.8]{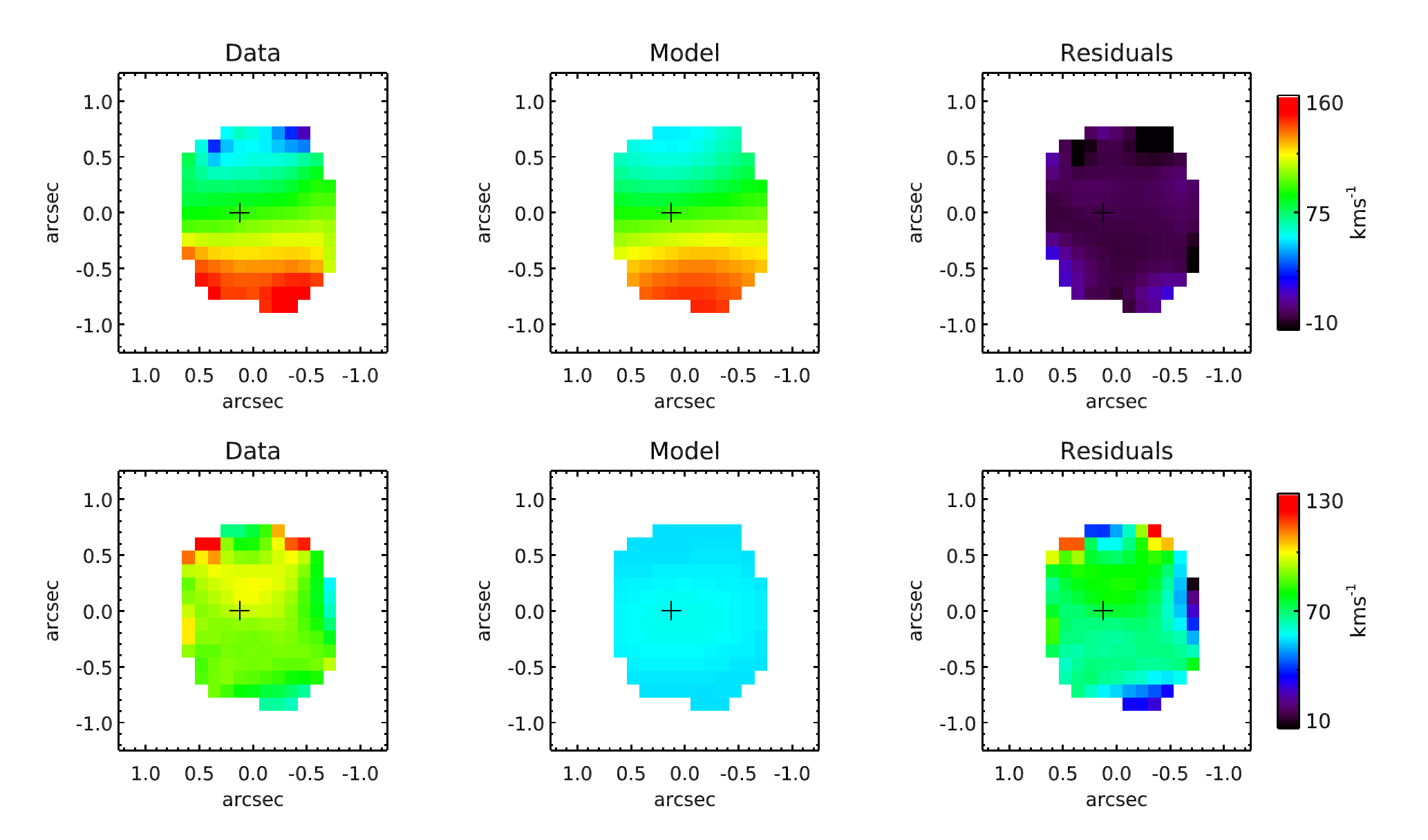}
\end{figure*}

\begin{figure*}
	\contcaption{}
	\textbf{CDFS11583}
	\includegraphics[scale=0.8]{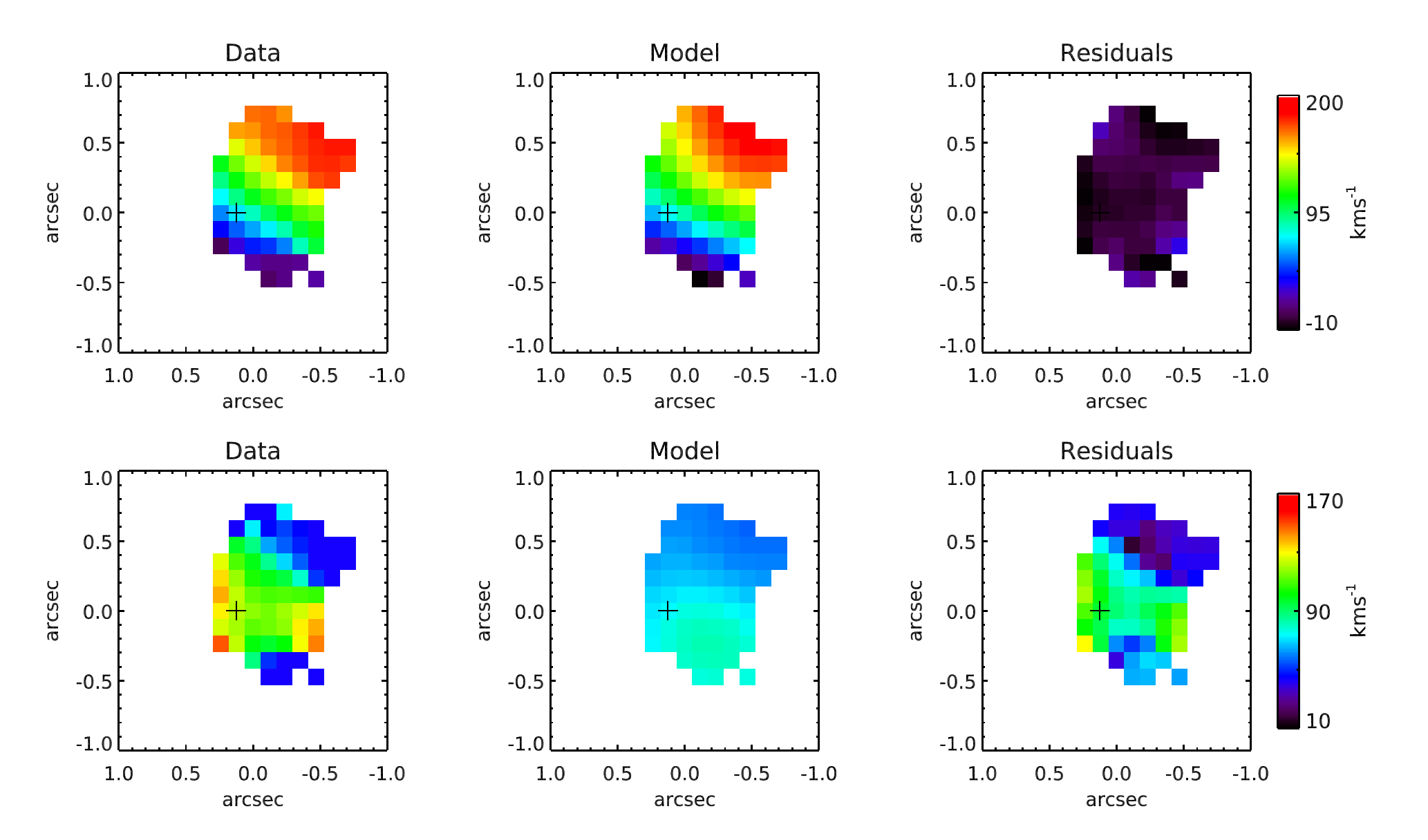}
	\vspace{0.5cm}

	\textbf{CDFS15753}
	\includegraphics[scale=0.8]{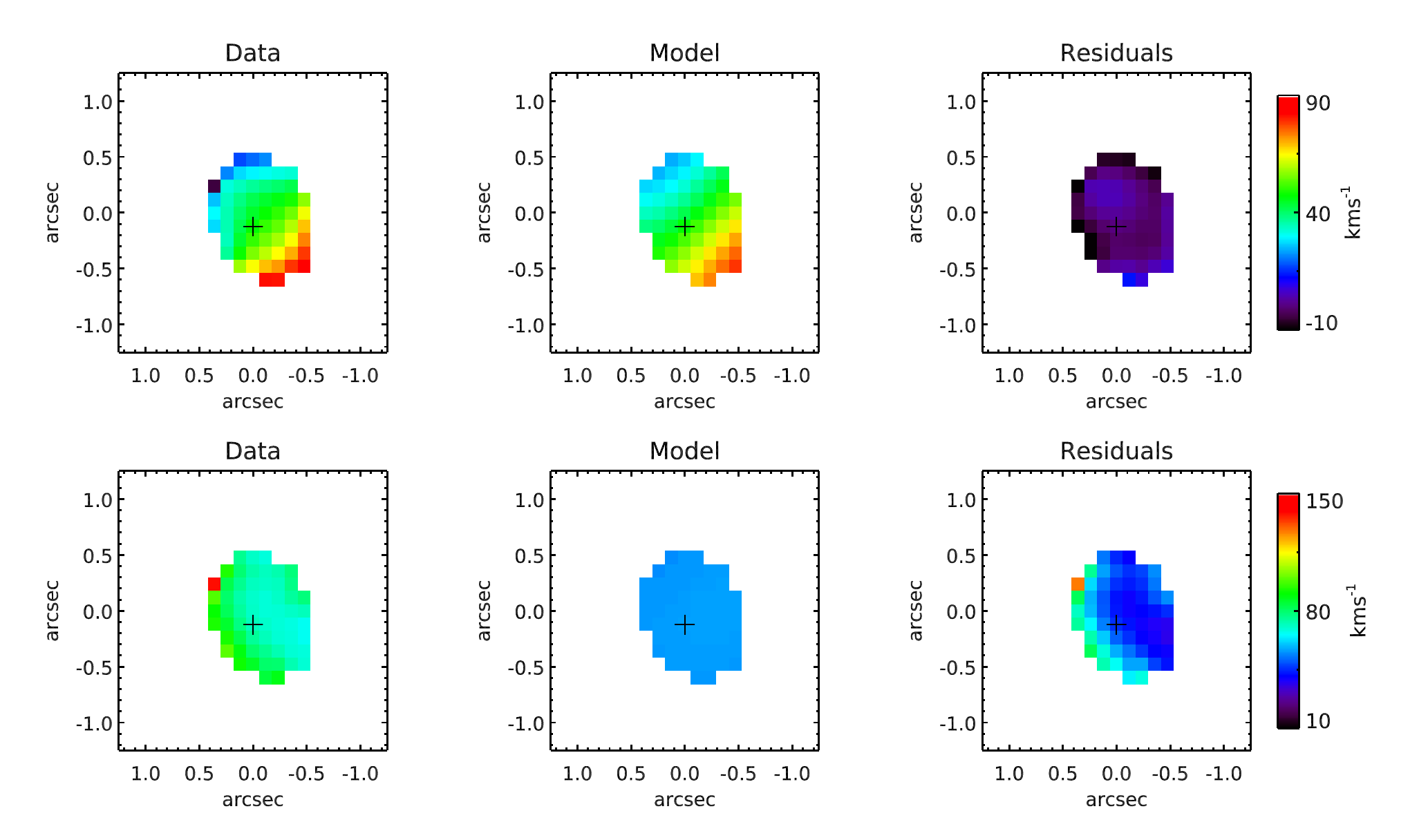}
\end{figure*}

\begin{figure*}
	\contcaption{}
	\textbf{CDFS15764}
	\includegraphics[scale=0.8]{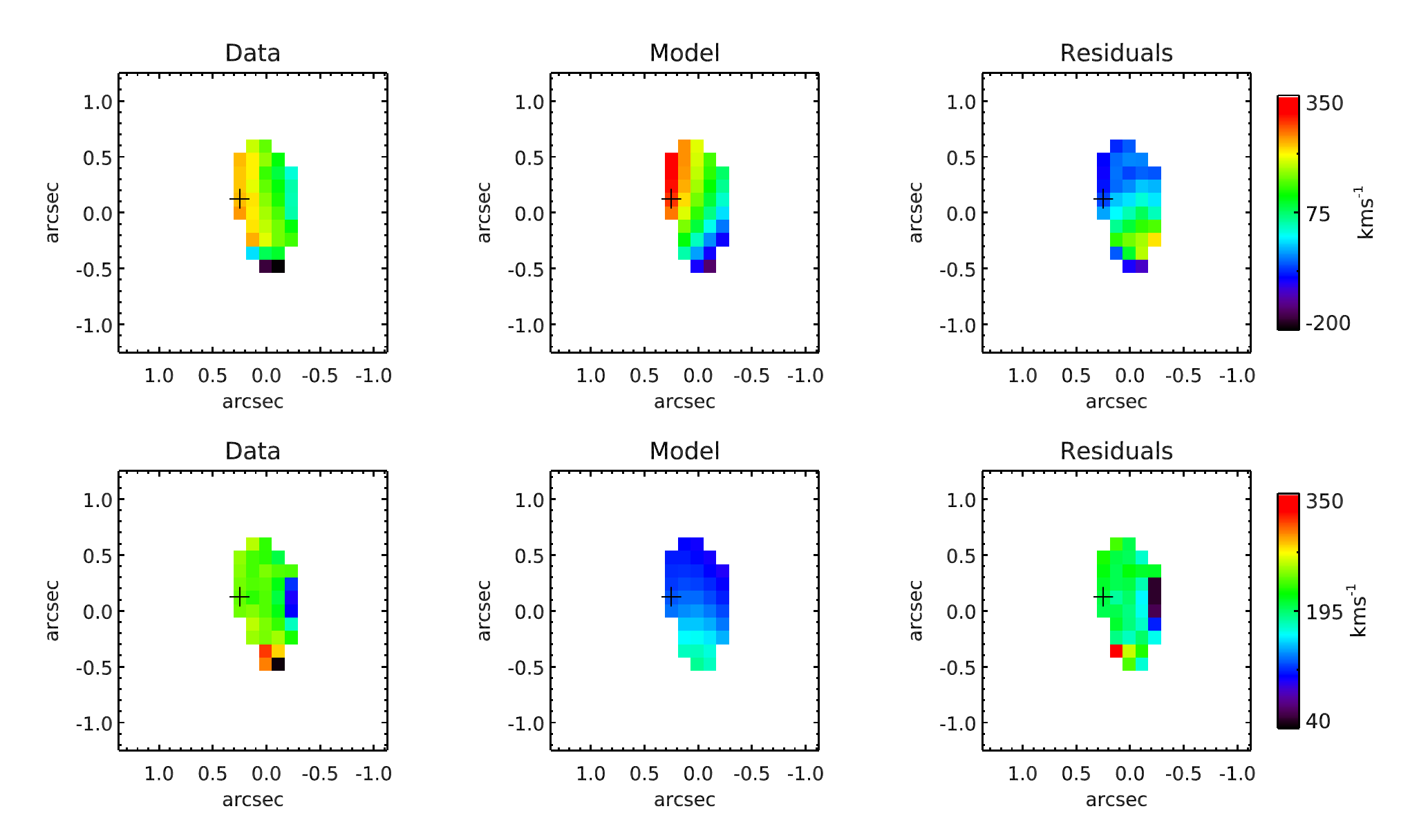}
	\vspace{0.5cm}

	\textbf{CDFS10299}
	\includegraphics[scale=0.8]{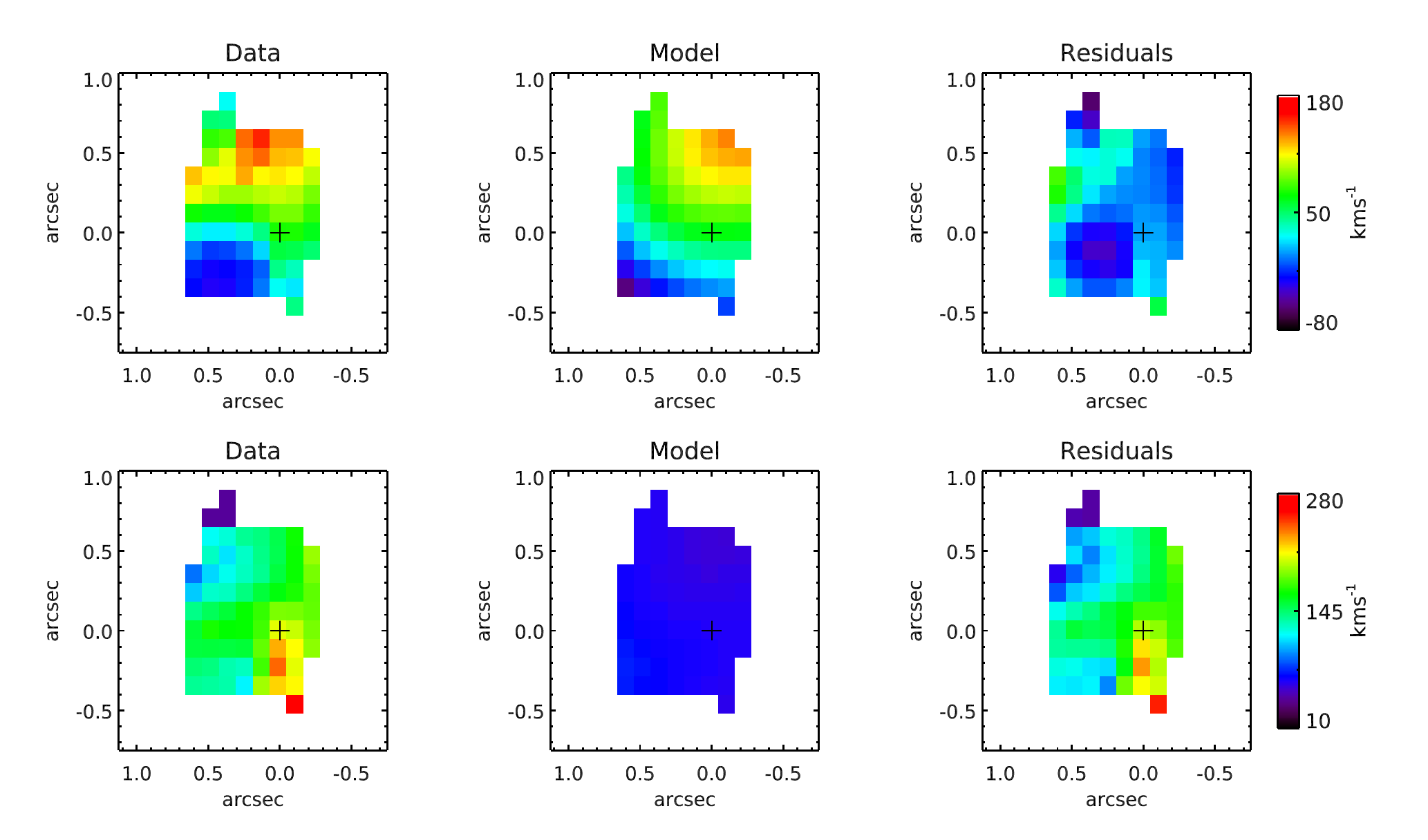}
\end{figure*}
\clearpage

\begin{figure*}
	CDFS13844\\
	\includegraphics[scale=0.7]{metallicity0}

	CDFS13784\\
	\includegraphics[scale=0.7]{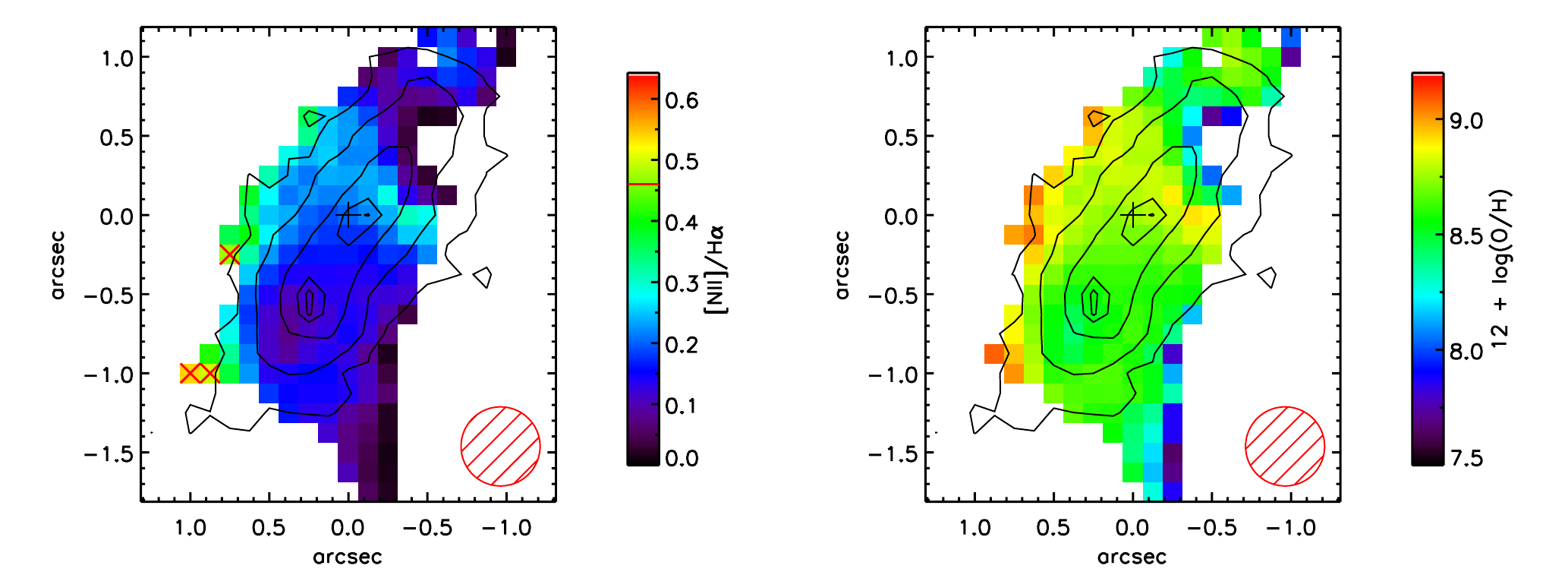}

	CDFS6758\\
	\includegraphics[scale=0.7]{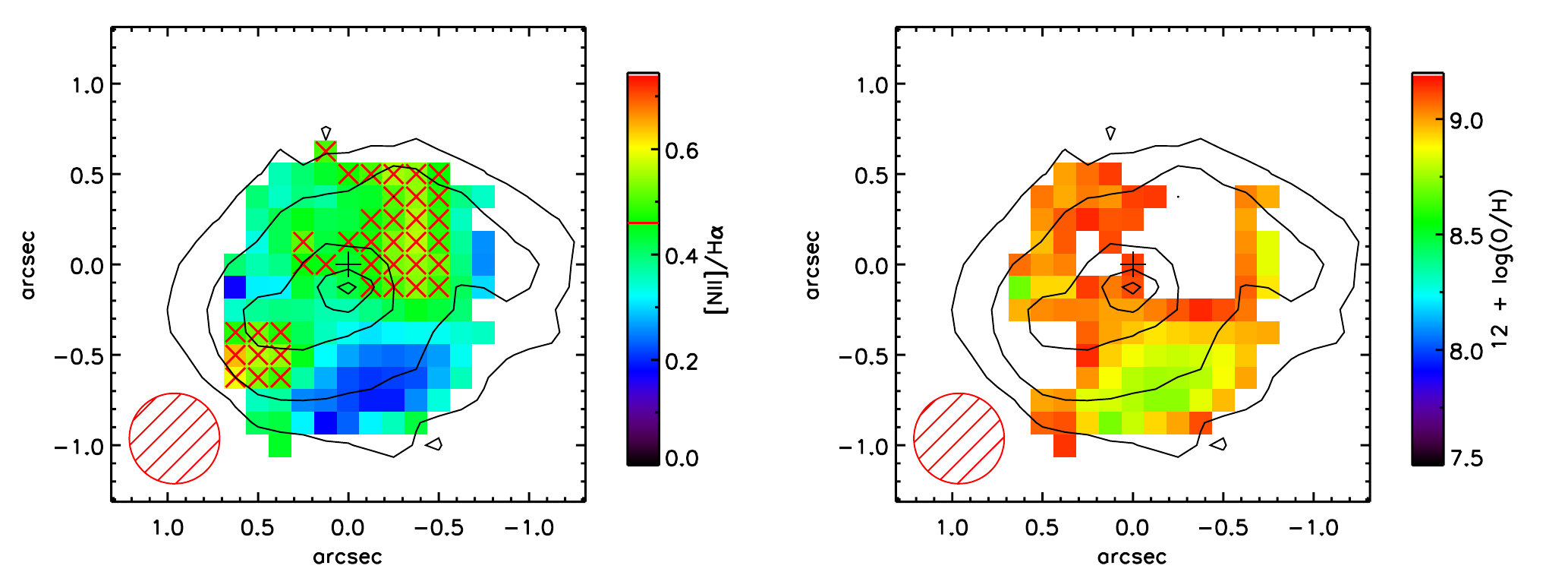}
	\caption{Metallicity gradients. Left: Ratio of [NII]/H$\rm{\alpha}$ fluxes with red crosses indicating the regions with $R>0.46$ therefore likely to be affected by shocks or an AGN (see Section 4.3 in main text) and so are excluded from the metallicity calibration. Right: The corresponding metallicity maps across the galaxy after AGN/shock removal. The black contours in both maps indicate the position of the galaxy continuum from the H-band HST images and the cross highlights the location of the peak of H$\rm{\alpha}$ emission. The size of the PSF is shown in red. See Table~\ref{tab:pos} for central coordinates.}
	\label{fig:metal_grad2}
\end{figure*}

\begin{figure*}
	\contcaption{}
	CDFS14224\\
	\includegraphics[scale=0.7]{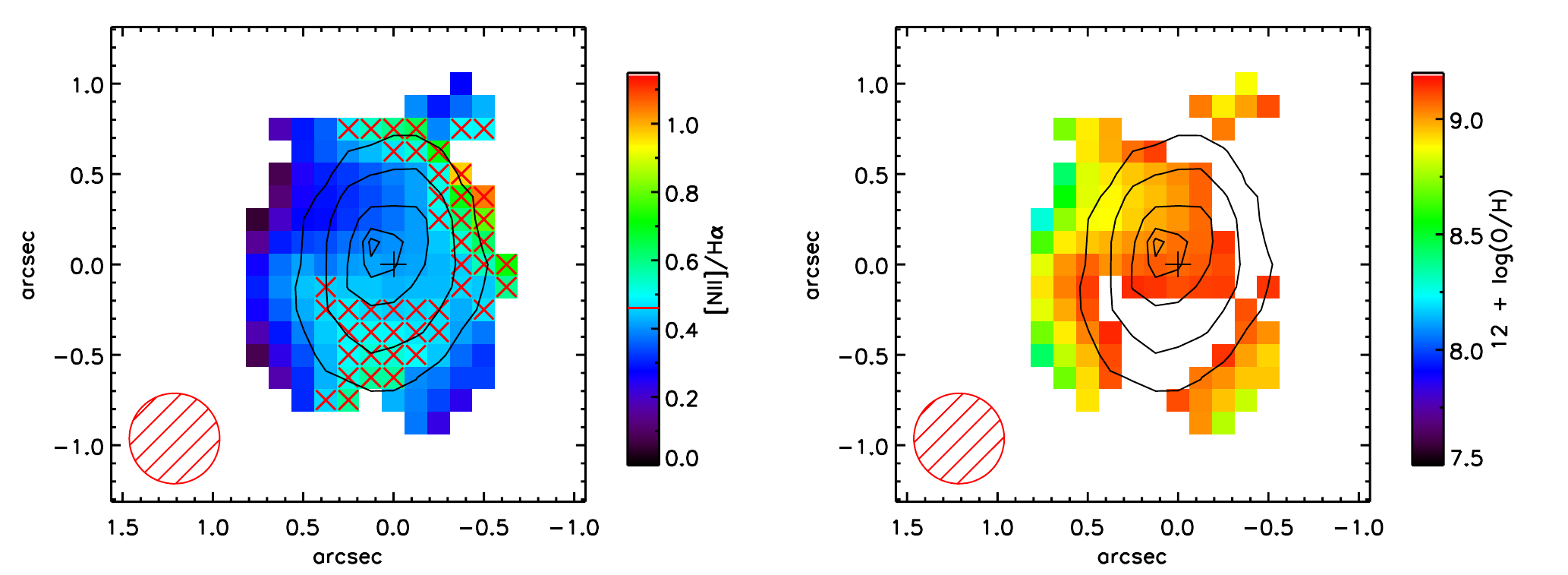}

	CDFS16485\\
	\includegraphics[scale=0.7]{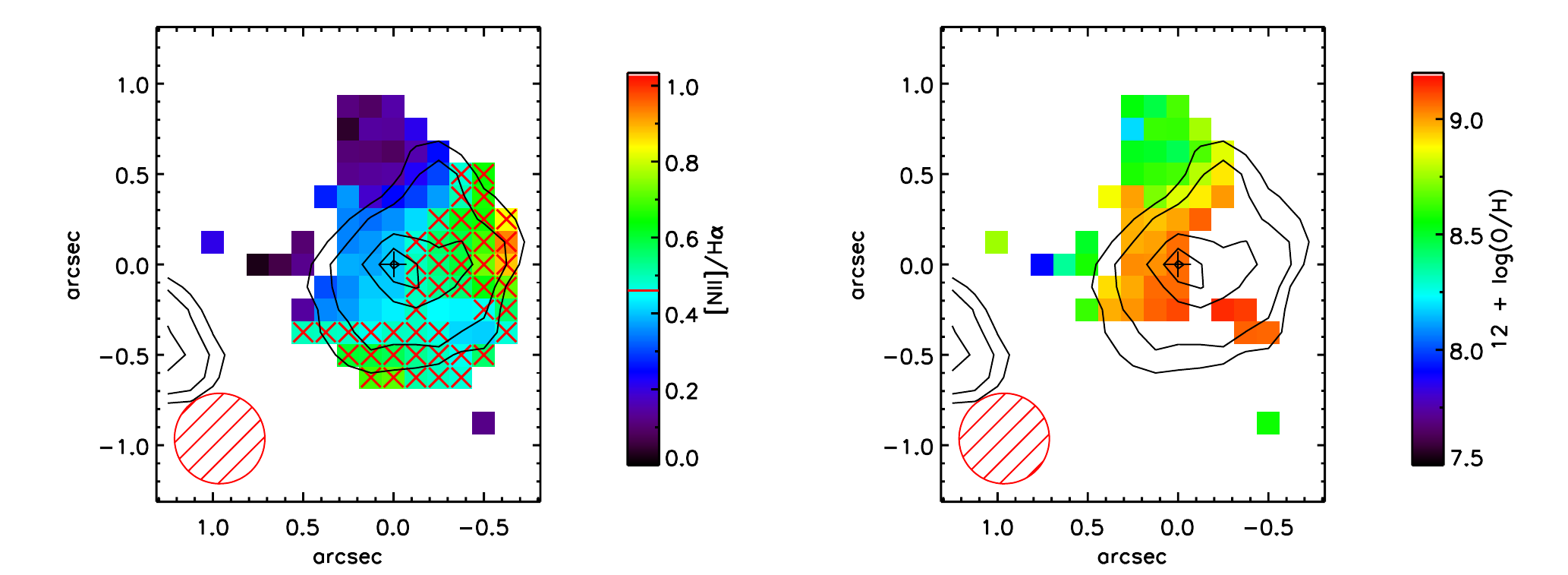}

	CDFS2780\\
	\includegraphics[scale=0.7]{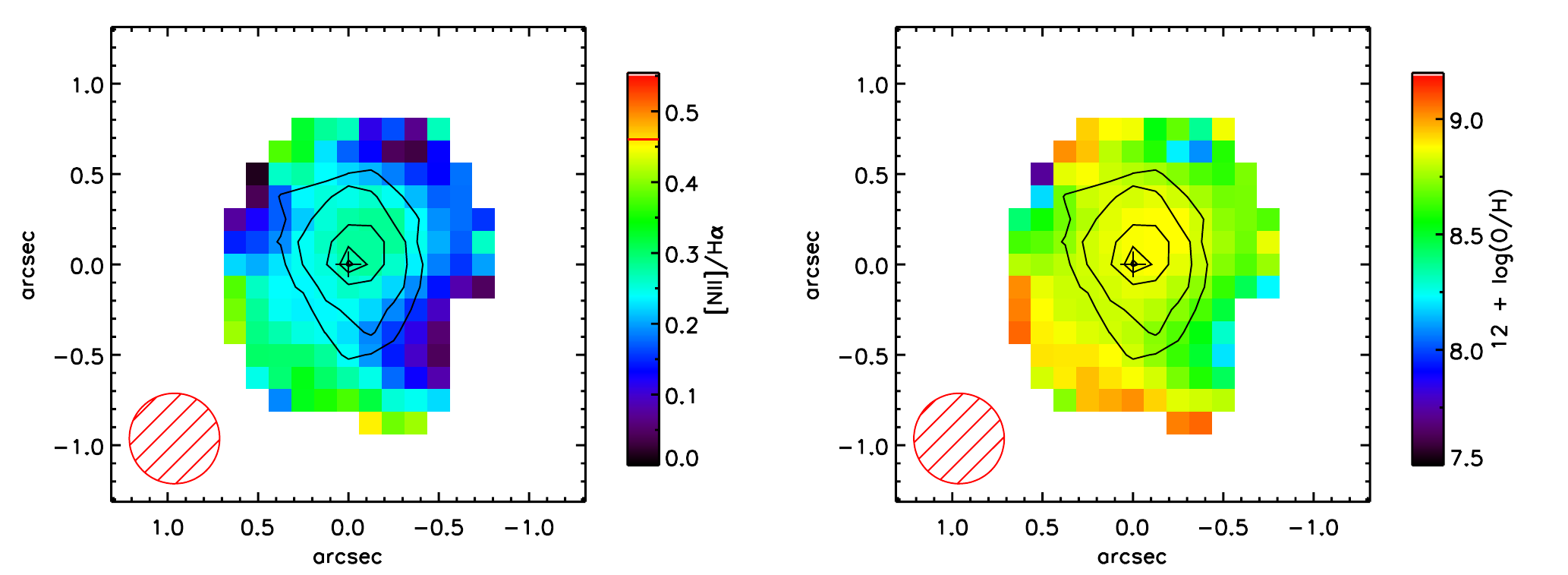}

	CDFS11583\\
	\includegraphics[scale=0.7]{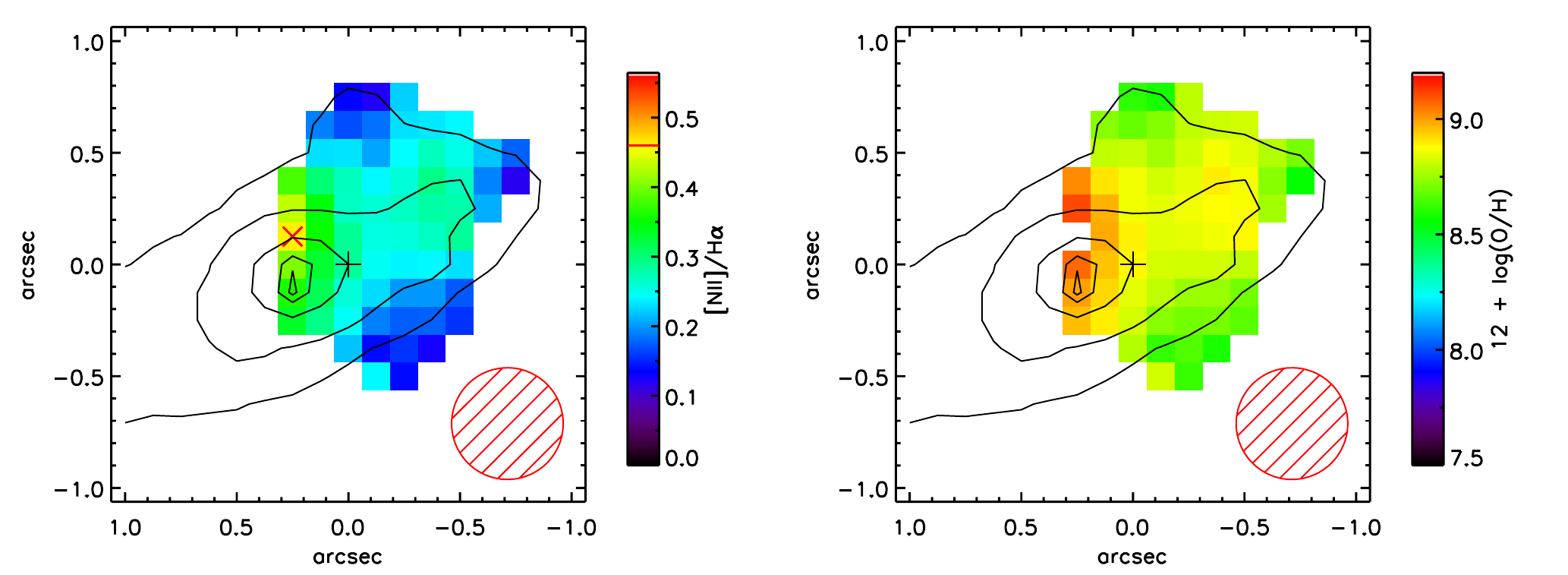}
\end{figure*}

\begin{figure*}
	\contcaption{}
	CDFS15753\\
	\includegraphics[scale=0.7]{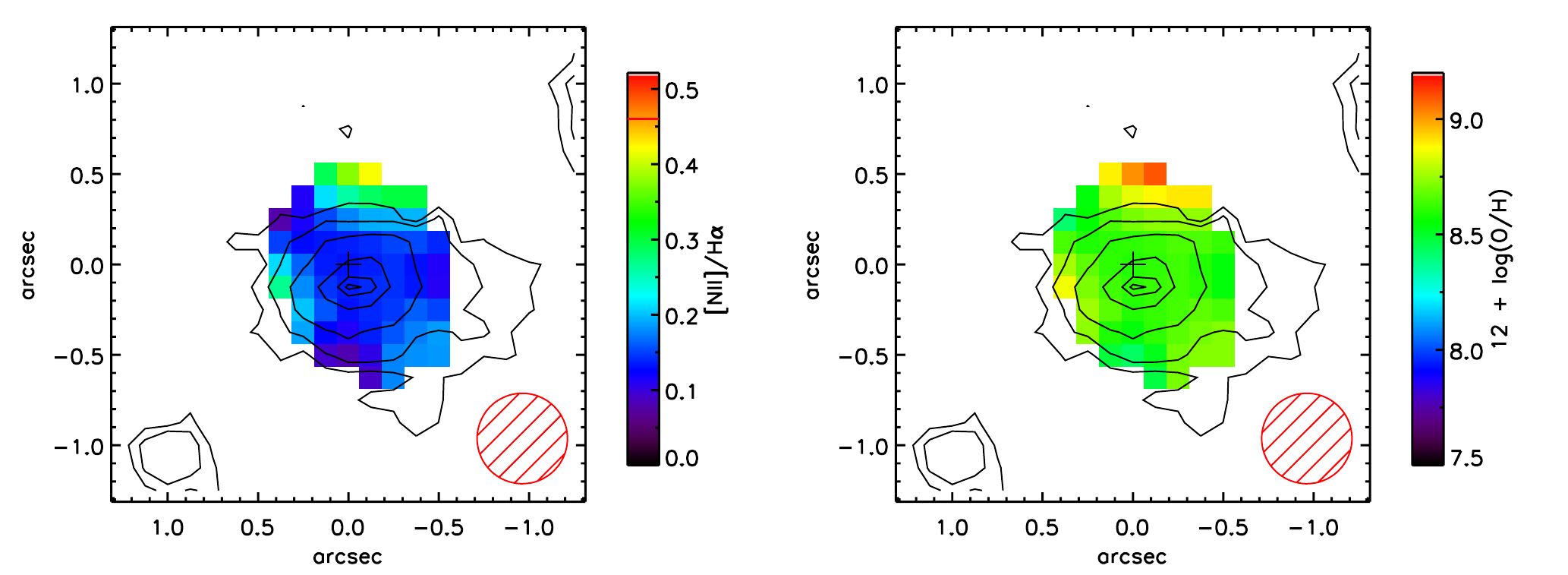}

	CDFS15764\\
	\includegraphics[scale=0.7]{metallicity8}

	CDFS10299\\
	\includegraphics[scale=0.7]{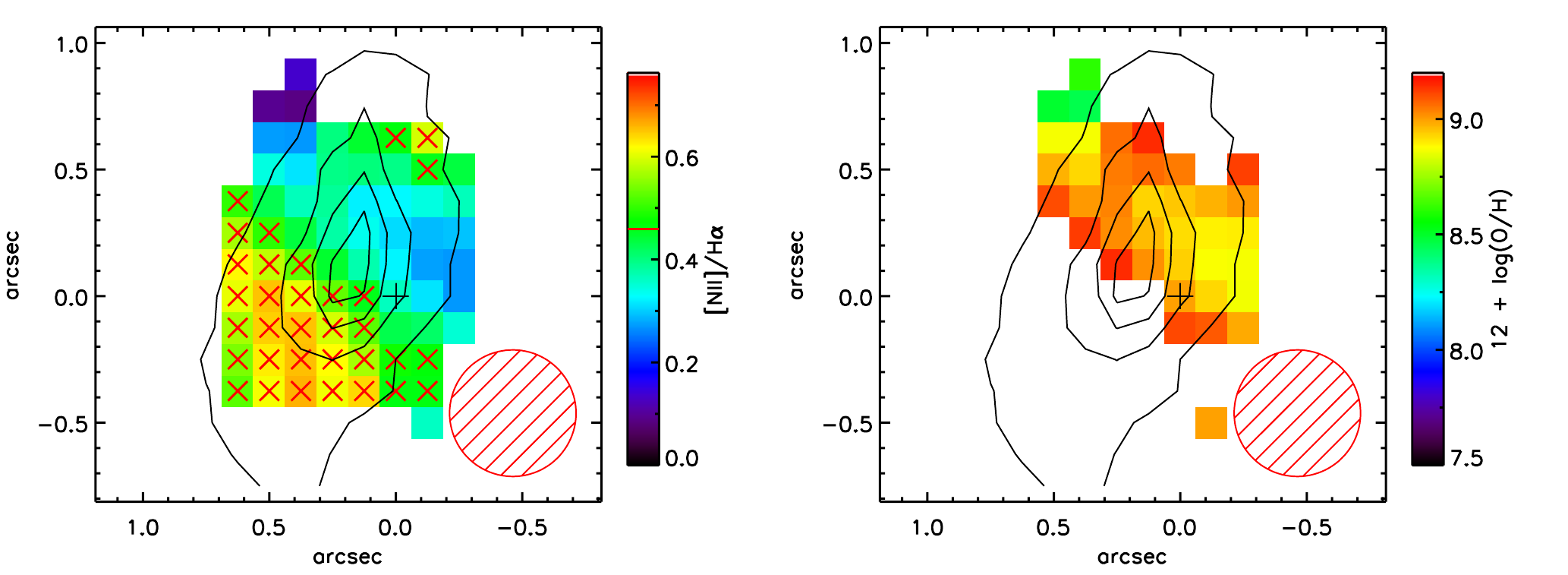}
\end{figure*}
\clearpage

\label{lastpage}
\end{document}